\documentclass[longauth]{aa}

\usepackage[separate-uncertainty=true, range-phrase=-, range-exponents=combine-bracket, range-units=single]{siunitx}
\usepackage[version=4]{mhchem}
\usepackage[varg]{txfonts}
\usepackage{xcolor}
\usepackage{hyperref}
\usepackage{url}
\hypersetup{colorlinks, 
 linkcolor = blue,
 urlcolor = blue,
 citecolor = blue,
 anchorcolor = blue}

\DeclareSIUnit{\parsec}{pc}
\DeclareSIUnit{\Lsun}{L_\odot}
\DeclareSIUnit{\Msun}{M_\odot}
\DeclareSIUnit{\micron}{\um}

\AddToHook{begindocument/before}{\usepackage{hyperref}}

\nolinenumbers

\begin{document}

\title{The Rich JWST Spectrum of the Western Nucleus of Arp~220: Shocked Hot Core Chemistry Dominates the Inner Disk}

\titlerunning{Shocked Hot Core Chemistry in Arp~220 W}

\author{Victorine A. Buiten \inst{\ref{inst:Leiden}}
\and Paul P. van der Werf \inst{\ref{inst:Leiden}}
\and Serena Viti \inst{\ref{inst:Leiden},\ref{inst:Bonn},\ref{inst:ucl}}
\and Daniel Dicken \inst{\ref{inst:UKATC}}
\and Almudena Alonso Herrero \inst{\ref{inst:CAB}}
\and Gillian S. Wright \inst{\ref{inst:UKATC}}
\and Maarten Baes \inst{\ref{inst:Gent}}
\and Torsten B\"oker \inst{\ref{inst:ESA-STScI}}
\and Bernhard R. Brandl \inst{\ref{inst:Leiden}, \ref{inst:Delft}}
\and Luis Colina \inst{\ref{inst:CAB-Torrejon}}
\and Macarena Garc\'ia Mar\'in \inst{\ref{inst:ESA-STScI}}
\and Thomas R. Greve \inst{\ref{inst:DAWN},\ref{inst:DTU}}
\and Pierre Guillard \inst{\ref{inst:Sorbonne},\ref{inst:IUF}}
\and Olivia C. Jones \inst{\ref{inst:UKATC}}
\and Laura Hermosa Mu\~noz \inst{\ref{inst:CAB}}
\and \'Alvaro Labiano \inst{\ref{inst:ESAC}}
\and G\"oran \"Ostlin \inst{\ref{inst:Stockholm}}
\and Lara Pantoni \inst{\ref{inst:Gent}}
\and Fabian Walter \inst{\ref{inst:MPIA}}
\and Martin J. Ward \inst{\ref{inst:Durham}}
\and Michele Perna \inst{\ref{inst:CAB-Torrejon}}
\and Ewine F. van Dishoeck \inst{\ref{inst:Leiden}}
\and Thomas Henning \inst{\ref{inst:MPIA}}
\and Manuel G\"udel \inst{\ref{inst:Vienna},\ref{inst:Zurich}}
\and Thomas P. Ray \inst{\ref{inst:DIAS}}
}

\institute{Leiden Observatory, Leiden University, PO Box 9513, 2300 RA Leiden, The Netherlands \label{inst:Leiden}
\and Transdisciplinary Research Area (TRA) ‘Matter’/Argelander-Institut für Astronomie, University of Bonn, Bonn, Germany \label{inst:Bonn}
\and Physics and Astronomy, University College London, London, UK \label{inst:ucl}
\and UK Astronomy Technology Centre, Royal Observatory, Blackford Hill Edinburgh, EH9 3HJ, Scotland, UK \label{inst:UKATC}
\and Centro de Astrobiolog\'{\i}a (CAB), CSIC-INTA, Camino Bajo del Castillo s/n, 28692 Villanueva de la Ca\~nada, Madrid, Spain \label{inst:CAB}
\and Sterrenkundig Observatorium, Universiteit Gent, Krijgslaan 281 S9, 9000 Gent, Belgium \label{inst:Gent}
\and European Space Agency, c/o Space Telescope Science Institute, 3700 San Martin Drive, Baltimore, MD 21218, USA \label{inst:ESA-STScI}
\and Faculty of Aerospace Engineering, Delft University of Technology, Kluyverweg 1, 2629 HS Delft, The Netherlands \label{inst:Delft}
\and Centro de Astrobiolog\'ia (CAB), CSIC–INTA, Cra. de Ajalvir Km. 4, 28850 Torrej\'on de Ardoz, Madrid, Spain \label{inst:CAB-Torrejon}
\and Cosmic Dawn Center (DAWN), Denmark \label{inst:DAWN}
\and DTU Space, Technical University of Denmark, Elektrovej, Building 328, 2800 Kgs. Lyngby, Denmark \label{inst:DTU}
\and Sorbonne Universit\'e, CNRS, UMR 7095, Institut d’Astrophysique de Paris, 98bis bd Arago, 75014 Paris, France \label{inst:Sorbonne}
\and Institut Universitaire de France, Minist\`ere de l’Enseignement Sup\'erieur et de la Recherche, 1 Rue Descartes, 75231 Paris Cedex 05, France \label{inst:IUF}
\and Telespazio UK for the European Space Agency (ESA), ESAC, Camino Bajo del Castillo s/n, 28692 Villanueva de la Cañada, Spain \label{inst:ESAC}
\and Department of Astronomy, Stockholm University, The Oskar Klein Centre, AlbaNova 106 91, Stockholm, Sweden \label{inst:Stockholm}
\and Max Planck Institute for Astronomy, Konigstuhl 17, 69117 Heidelberg, Germany \label{inst:MPIA}
\and Centre for Extragalactic Astronomy, Durham University, South Road, Durham DH1 3LE, UK \label{inst:Durham}
\and Dept. of Astrophysics, University of Vienna, T\"urkenschanzstr. 17, 1180 Vienna, Austria \label{inst:Vienna}
\and ETH Z\"urich, Institute for Particle Physics and Astrophysics, Wolfgang-Pauli-Str. 27, 8093 Z\"urich, Switzerland \label{inst:Zurich}
\and Dublin Institute for Advanced Studies, 31 Fitzwilliam Place, D02 XF86, Dublin, Ireland \label{inst:DIAS}
}

\abstract
{We present full \qtyrange{3}{28}{\micron} JWST MIRI/MRS and NIRSpec/IFU spectra of the western nucleus of Arp~220, the nearest ultraluminous infrared galaxy. This nucleus has long been suggested to possibly host an embedded Compton-thick AGN. millimetre observations of the dust continuum suggest the presence of a distinct \qty{20}{\parsec} core with a dust temperature of $T_\mathrm{d} \gtrsim \qty{500}{\kelvin}$, in addition to a \qty{100}{\parsec} circumnuclear starburst disk. However, unambiguously identifying the nature of this core is challenging, due to the immense obscuration, the nuclear starburst activity, and the nearby eastern nucleus. With the JWST integral field spectrographs, we can, for the first time, separate the two nuclei across this full wavelength range, revealing a wealth of molecular absorption features towards the western nucleus. We analyse the rovibrational bands detected at \qtyrange[range-units=single]{4}{22}{\micron}, deriving column densities and rotational temperatures for 10 distinct species. Optically thick features of \ce{C2H2}, \ce{HCN} and \ce{HNC} suggest that this molecular gas is hidden behind a curtain of cooler dust, and indicate that the column densities of \ce{C2H2} and \ce{HCN} are an order of magnitude higher than previously derived from \textit{Spitzer} observations. We identify a warm \ce{HCN} component with rotational temperature $T_\mathrm{rot} = \qty{330}{\kelvin}$, which we associate with radiative excitation by the hot inner nucleus. We propose a geometry where the detected molecular gas is located in the inner regions of the starburst disk, directly surrounding the hot \qty{20}{\parsec} core. The chemical footprint of the western nucleus is reminiscent of that of hot cores, with additional evidence for shocks. Despite the molecular material's close proximity to the central source, no evidence for the presence of an AGN in the form of X-ray-driven chemistry or extreme excitation is found.
}

\date{Received date / Accepted date}


\maketitle

\nolinenumbers

\section{Introduction}
Arp~220, the nearest ultra-luminous infrared galaxy (ULIRG), has been the subject of extensive observational studies across the electromagnetic spectrum. Its enormous infrared luminosity ($L_\mathrm{IR} = \qty{2e12}{\Lsun}$), first discovered by the \textit{Infrared Astronomical Satellite} (IRAS) \citep{Soifer1984}, is comparable to the bolometric luminosity of a quasar. This extreme power led to a yet-unresolved debate about its nature as either a ``superstarburst'' or host to a deeply buried active galactic nucleus (AGN). Optical photometry, as well as HI \SI{21}{\centi\meter} imaging, revealed tidal tails suggesting that Arp~220 is an advanced-stage major merger of two gas-rich galaxies \citep{JosephWright1985, Sanders1988, Hibbard2000}. A prominent dust lane, oriented NW to SE, hides two remnant nuclei only $1\arcsec$ apart, first resolved by radio observations \citep{Norris1988} and later in ground-based infrared (IR) images \citep{Graham1990, Soifer1999}.

The two nuclei of Arp~220, hereafter denoted as the western nucleus (WN) and the eastern nucleus (EN), are intriguing structures on their own, with the WN being the most luminous and obscured of the two. Radio-VLBI observations have revealed 97 compact point sources, interpreted as supernovae (SNe) and supernova remnants (SNRs) and distributed in ellipses centred on the two nuclei \citep[e.g.][]{Smith1998, Lonsdale2006, Varenius2019}. These SNe and SNRs are well-matched to the nuclear disk structures seen in high-resolution ALMA dust continuum observations. The disks are counter-rotating, with the eastern disk approximately aligned with the surrounding \SI{}{\kilo\parsec}-scale rotating structure \citep{Sakamoto1999, Scoville2015, Scoville2017, Sakamoto2017}. These findings all point towards a recent nuclear starburst.

Analyses of the millimetre continuum suggest the presence of a distinct inner core component for the WN, with column density $N(\ce{H2}) \sim \SI{e26}{\centi\meter^{-2}}$, in addition to the starburst disk \citep{Scoville2017, Sakamoto2017, Sakamoto2021a}. If this core harbours a hidden AGN, it is very Compton-thick, and all of its associated X-ray emission will be absorbed before reaching an outside observer. Indeed, \citet{Teng2015} conclude that their Fe~K$\mathrm{\alpha}$ detection is consistent with arising from the starburst. Furthermore, to date no high-excitation mid-infrared (MIR) AGN lines have been detected \citep{Goldberg2024, Perna2024}. The extreme optical depths towards the WN impede the search for an AGN, but they also make it difficult to confirm its absence. Indirect probes of AGN activity must therefore be explored.

The nuclear region contains large amounts of chemically rich molecular gas. The earliest interferometric line observations already showed a concentration of 70\% of the total \ce{CO} (1-0) emission in the inner $< \SI{1500}{\parsec}$ \citep{Scoville1986}. Later submillimetre spectral scans revealed complex line forests full of broad emission lines \citep[e.g.][]{Martin2011, Aladro2015, Sakamoto2021b}. At the highest spatial resolutions achieved with ALMA, sub-continuum absorption appears towards the compact inner core of the WN in a wide variety of molecular lines \citep{Scoville2017, Sakamoto2021b}.

Beyond (sub)millimetre rotational lines, the molecular gas of Arp~220 has also been probed with centimetre radio observations and \textit{Herschel} far-infrared (FIR) spectroscopy. At centimetre wavelengths, lines of \ce{NH3}, \ce{H2CO}, \ce{CH2NH}, have been detected \citep{Araya2004, Takano2005, Ott2011, Salter2008}. \textit{Herschel} detected high-excitation rotational lines of \ce{CO} and \ce{HCN}, as well as \ce{H2O}, \ce{OH}, \ce{NH3}, and several molecular ions \citep{Rangwala2011, Gonzalez-Alfonso2012, Gonzalez-Alfonso2013}. Arp~220 also hosts the first known \SI{18}{\centi\meter} \ce{OH} megamaser \citep{Baan1982, BaanHaschick1984, Baan2023}.

Evidence for molecular outflows is plentiful. Blueshifted absorption and $P$-Cygni profiles have been found in a variety of lines across a wide spectral range \citep[e.g.][]{Baan1989, Sakamoto2009, Rangwala2011, Gonzalez-Alfonso2012, Tunnard2015, Perna2020}. \citet{BarcosMunoz2018} spatially resolved a \SI{200}{\parsec}-scale collimated bipolar outflow at the WN in \ce{HCN} (1-0) and \ce{CO} (1-0), reaching line-of-sight velocities of \SI{500}{\kilo\meter\per\second}. \citet{Wheeler2020} and \citet{Sakamoto2021b} confirmed this collimated high-velocity outflow in \ce{HCN} (4-3) and several other \ce{CO} lines. A less prominent polar outflow has been suggested for the EN, but only the blueshifted absorbing lobe has been firmly detected \citep{Sakamoto2021b}. Furthermore, recent JWST NIRSpec results reveal \SI{}{\kilo\parsec}-scale hot molecular outflows from both nuclei, traced by \ce{H2} S(1) 1-0 emission \citep{Perna2024, Ulivi2024}.

Given the numerous molecular species detected in Arp~220, a characterisation of its chemistry has been proposed as a potential avenue to find a buried AGN. While much of its chemical footprint is reminiscent of Galactic hot cores and other starburst galaxies \citep[e.g.][]{Lahuis2007, Martin2011}, evidence for substantial X-ray-driven chemistry or AGN-driven mechanical heating would point towards AGN activity deep inside the nucleus. Such scenarios have been proposed based on overluminous \ce{HNC} emission \citep{Aalto2007}, elevated [\ce{HCN}]/[\ce{HCO+}] abundance ratios \citep[e.g.][]{Tunnard2015}, and high inferred column densities of \ce{H2O+}, \ce{OH+} and \ce{OH} \citep{Rangwala2011, Gonzalez-Alfonso2012}, but none of these diagnostics are without alternative interpretation.

Another angle that has been explored observationally is through lines of vibrationally-excited molecules. \citet{Salter2008} already detected several direct $l$-type transitions of vibrationally-excited \ce{HCN} (\ce{HCN}-vib) at centimetre wavelengths in absorption; later submillimetre observations also detected rotational \ce{HCN}-vib emission lines, as well as lines of vibrationally-excited \ce{HC3N} \citep{Martin2011, Aalto2015, Martin2016, Sakamoto2021b}. These lines have been used to probe the hot inner envelopes of Galactic hot cores \citep[e.g.][]{VanDerTak1999, Boonman2001}. They only arise in regions where a strong mid-infrared continuum \citep[$T_\mathrm{d} > \SI{100}{\kelvin}$;][]{Sakamoto2010} can efficiently pump the molecules into a vibrationally-excited mode, and are therefore not affected by contamination or self-absorption of colder gas in the same way their ground-state counterparts are. \citet{Sakamoto2021b} confirmed this notion by comparing emission sizes of a wide variety of lines in a high-resolution spectral scan; they found that the \ce{HCN}-vib emission was very compact.

For \ce{HCN}, the pumping mechanism in question involves absorption in the \SI{14}{\micron} $\nu_2$ band, exciting its bending mode. This absorption band has been observed in Arp~220 and several other U/LIRGs with \textit{Spitzer} \citep{Lahuis2007}. With the advent of JWST and its integral field spectroscopy (IFS) capabilities, several nuclear regions of other U/LIRGs without reported \ce{HCN}-vib emission have been added to this list \citep{Buiten2024a, Gonzalez-Alfonso2024, Garcia-Bernete2024}. Rovibrational absorption bands provide insight into local excitation conditions. Depending on the size of the effective background continuum source, they can probe a narrower line of sight than the point spread function (PSF) of the telescope. Thus, they are an excellent probe of the inner nuclear regions. Additionally, molecules like \ce{C2H2}, \ce{CO2} and \ce{CH4}, which are key components of the interstellar chemistry of high-mass star-forming regions, but lack any allowed rotational transitions, can be detected through their rovibrational bands \citep[e.g.][]{Lahuis2000, Sonnentrucker2007, Barr2020, Francis2024, VanGelder2024}. Mid-infrared molecular signatures can therefore serve as a useful chemical probe through a direct comparison to spectra of Galactic hot cores.

In this work, we present the detection and analysis of a large number of rovibrational molecular bands in the WN of Arp~220 with JWST MIRI/MRS and NIRSpec/IFU. With JWST, the two nuclei of Arp~220 are, for the first time, separated in near- and mid-infrared spectroscopic observations, and the strong absorption towards the WN is no longer contaminated by continuum emission from the EN. The bands of \ce{HCN}, \ce{C2H2} and \ce{CO2}, previously detected at low spectral resolution with \textit{Spitzer} \citep{Lahuis2007}, are now seen in unprecedented detail. This work complements a study of the mid-infrared emission lines in the MIRI/MRS cube (Van der Werf et al. in prep) and an extensive search for high-excitation fine structure lines by \citet{Goldberg2024}, as well as two studies on the emission lines seen with NIRSpec \citep{Perna2024, Ulivi2024}. In the present paper, we only consider the WN, where the absorption lines are strongest and allow for a highly detailed analysis.

In Section \ref{sec:data}, we summarise the JWST observations and data reduction. In Section \ref{sec:results}, we describe the spectral fitting procedure used, and present the resulting models and inferred gas properties for the detected bands. A discussion on the excitation conditions and chemistry of the observed molecular gas components, and their implications for the nature of the nucleus, follows in Section \ref{sec:discussion}. Finally, we summarise our findings in Section \ref{sec:conclusions}. Throughout this work we assume a flat Planck cosmology with $H_0 = \qty{67.4}{\kilo\meter\second^{-1}\mega\parsec^{-1}}$ and $\Omega_m = 0.315$ \citep{planck2020}. Under these cosmological parameters and assuming a systemic redshift of $z = 0.0188$, Arp~220 has a luminosity distance of \qty{84.8}{\mega\parsec}, and a projected angular scale of \qty{396}{\parsec\per\arcsec}.

\section{Observations and data reduction} \label{sec:data}
\subsection{MIRI}
A full description of the observations and data reduction is given in Van der Werf et al. (in prep). Here we provide a summary.

Arp~220 was observed by the MIRI MRS as part of the Mid-Infrared Characterisation of Nearby Iconic galaxy Centers (MICONIC) Guaranteed Time Observations (GTO) program 1267 (PI: D. Dicken; \href{http://dx.doi.org/10.17909/t6c5-ks25}{DOI 10.17909/t6c5-ks25}). This program targets the nuclear regions of the Milky Way and several iconic nearby galaxies: Arp~220, Mrk~231 \citep{AlonsoHerrero2024}, NGC~6240 \citep{HermosaMunoz2024}, Cen~A, and SBS~0335-052. Dedicated background exposures and simultaneous images were taken to enable the subtraction of thermal telescope emission and zodiacal light, and to allow for astrometric calibration. A four-point dither pattern was used for the science exposures, and two-point dithering for the backgrounds. The uncalibrated data products were downloaded through the MAST portal and reduced with the JWST Science Calibration pipeline v1.13.4, using CRDS context file \texttt{jwst\_1200.pmap}. Several non-standard settings were used for the Stage 1 outlier detection, most notably the use of the cosmic ray shower detection step. Pixel-based ``image-from-image'' background subtraction was performed in Stage 2, making use of the dedicated backgrounds. In Stage 2, we also used the pixel replacement option. Finally, in the Stage 3 outlier detection, we used a larger kernel size of $11 \times 11$ pixels. The MIRI MRS cubes cover a spectral range of $\qtyrange{4.9}{27.9}{\micron}$ and have a resolving power varying between 1330 at the longest wavelengths and 3710 at the shortest \citep{Jones2023, Wright2023, Argyriou2023}.

\begin{figure}[ht]
    \centering
    \includegraphics[width=\linewidth]{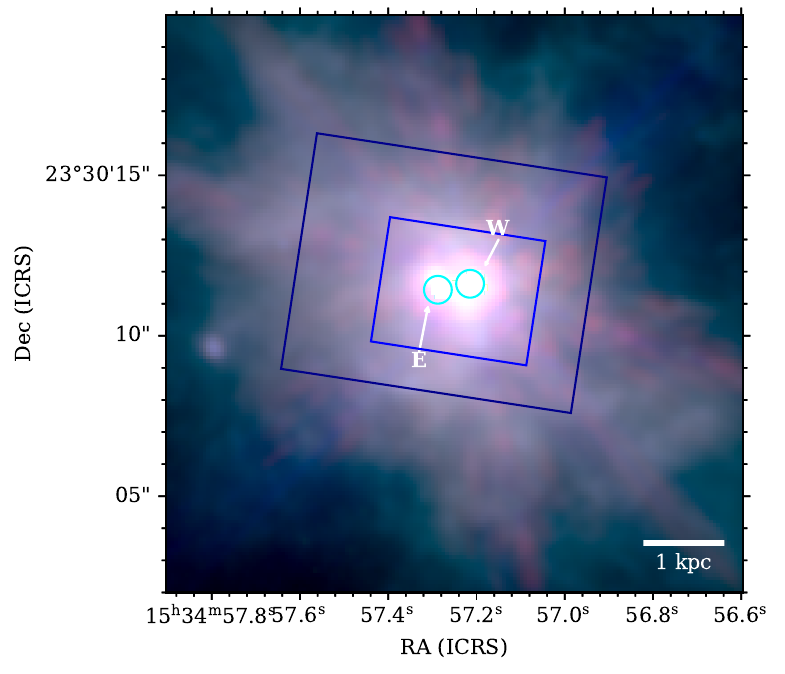}
    \caption{MIRI false-colour image of Arp~220, using the F1280W (red), F1130W (green) and F770W (blue) filters. The blue rectangles indicate the smallest (channel 1) and largest (channel 4) fields of view of the MRS observations. The cyan circles indicate extraction apertures for the western and eastern nuclei, both with a radius of 0\farcs{435}.}
    \label{fig:miri_image}
\end{figure}

\begin{figure*}[ht]
    \centering
    \includegraphics[width=\linewidth]{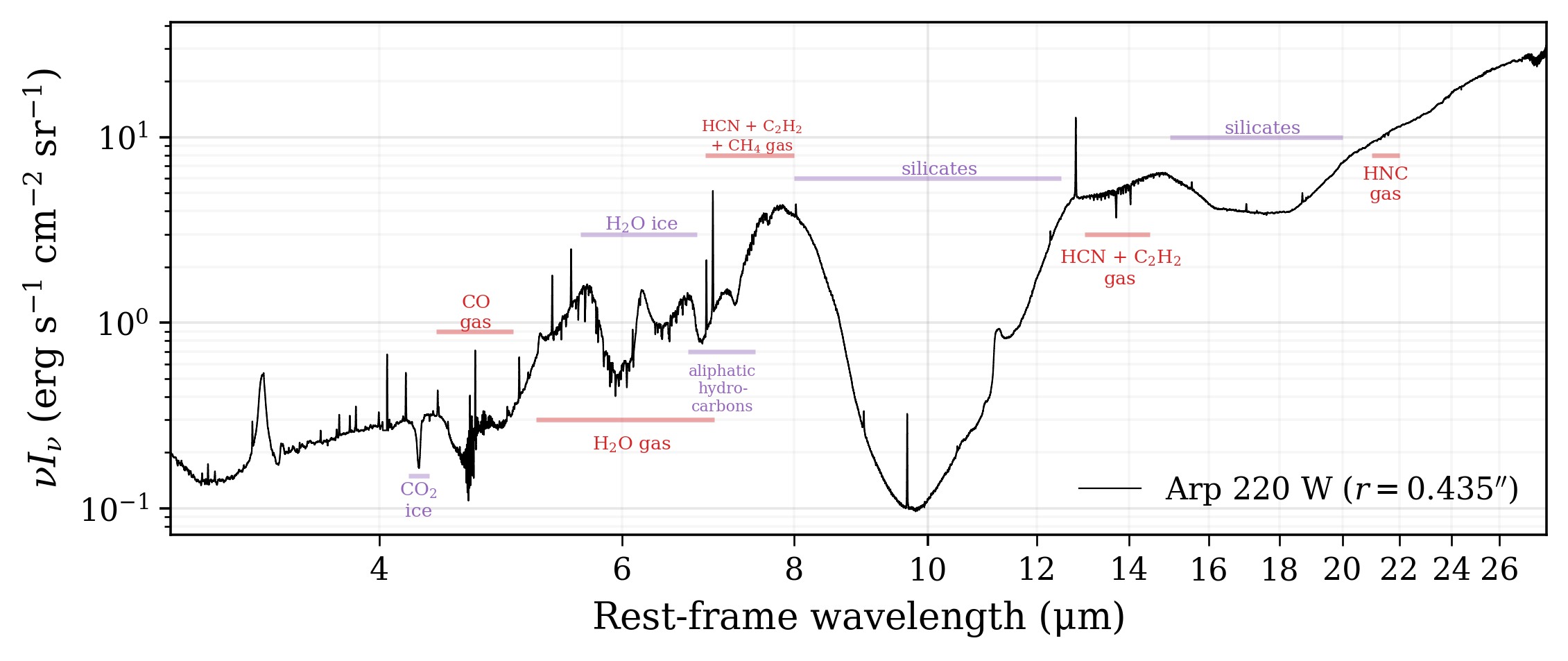}
    \caption{Combined NIRSpec/IFU G395H/F290LP + MIRI MRS spectrum of the western nucleus of Arp~220, extracted from a circular aperture of 0\farcs435 in radius. Major absorption features are indicated in purple (solids) and red (gas).}
    \label{fig:fullspec}
\end{figure*}

In post-processing, the astrometry of the data cubes was corrected based on Gaia Data Release 3 stars \citep{GaiaDR3} in the simultaneous imaging data. An aperture of 0\farcs{435} in radius centred on the western nucleus was selected for spectral extraction (see Fig. \ref{fig:miri_image}). This aperture size was chosen to reduce the effect of ``wiggles'' due to spatial undersampling of the PSF \citep{Law2023}, in order to reveal weak molecular absorption lines. We note that in channel 4, the PSF becomes too wide to completely separate the two nuclei, and the continuum emission from the western nucleus becomes dominant. However, as we exclusively use continuum-normalised spectra to study absorption features, the increasing PSF size does not affect our analysis, and we do not need to apply an aperture correction. For each channel, the redshift was manually calibrated using \ce{H2} lines to ensure that velocities are measured with respect to the bulk of the warm molecular gas. The redshifts derived from \ce{H2} lines generally correspond well to the value of $z_\mathrm{WN} = 0.01786 \pm 0.00005$ derived from rotational \ce{CO} and \ce{HCO+} lines by \citet{Sakamoto2009}. A more detailed analysis of the \ce{H2} lines detected with the MIRI MRS is presented by Van der Werf et al. (in prep). The full extracted MIRI MRS spectrum of the western nucleus is presented in Fig.~\ref{fig:fullspec}.

\subsection{NIRSpec}
In coordination with the MIRI observations, NIRSpec observations were taken as well. These are presented by \citet{Perna2024} and \citet{Ulivi2024}; here we summarise the data acquisition and reduction. The nuclear region was observed with each of the three high-resolution grating settings, using a four-point dither pattern. Data reduction was performed using a modified version of the JWST Science Calibration pipeline v1.8.2 and CRDS context file \texttt{jwst\_1063.pmap}. Modifications included corrections for $1/f$ noise to the count-rate images, and outlier detection as implemented by \citet{D'Eugenio2024}. Of the three final data cubes, in this work we only use the one from the G395H/F290LP grating setting. This cube covers a wavelength range $\qtyrange{2.9}{5.2}{\micron}$ with resolving power $R \approx 2700$ \citep{Boeker2023}.

We extract the spectrum from an aperture of 0\farcs{15} in radius, centred on the WN. Given the higher angular resolution of NIRSpec, a smaller aperture than for the MIRI data was used to avoid contaminating emission from the nearby bright star cluster \citep{Perna2024}. Experimentation with aperture size led us to conclude that, indeed, the CO absorption is deeper in the smaller aperture, suggesting that the continuum emission is not fully dominated by the WN at these wavelengths. {This is substantiated by comparatively faint, somewhat extended emission seen in narrow-band images extracted from NIRSpec cubes \citep{Perna2024}.} A spectrum extracted from an aperture 0\farcs{435} in radius, to match that used for the MIRI data, is shown in Fig. \ref{fig:fullspec}.

{In addition to the GTO program, MIRI and NIRCam imaging observations were taken as part of Director's Discretionary program 2739, available at \href{http://dx.doi.org/10.17909/94a9-h925}{DOI 10.17909/94a9-h925}. In the present work, MIRI imaging in three filters is used exclusively for the purpose of visualising the MIRI/MRS and NIRSpec fields of view, and the aperture selection, shown in Fig.~\ref{fig:miri_image}.}

\section{Results and analysis} \label{sec:results}

\begin{table*}
    \centering
    \caption{Summary of Derived Gas Properties}
    \begin{tabular}{lcccccccccr}
        \hline\hline
        Species & Wavelength & Component\tablefootmark{a} & $N$ & $T_\mathrm{rot}$ & $f_\mathrm{bg}$ & $V_\mathrm{rad}$ & $\sigma_V$\\
         & \SI{}{\micron} & & \SI{}{\centi\meter^{-2}} & \SI{}{\kelvin} & & \SI{}{\kilo\meter\second^{-1}} & \SI{}{\kilo\meter\second^{-1}} \\
         \hline
        CO & 4.7 & c & $\numrange[]{8e17}{2e18}$ & $\numrange{15}{19}$ & $\numrange{1}{0.77}$ & $+60$ & $\sim 90$ \\
        CO & 4.7 & mw & $\numrange{1.4e18}{2.3e18}$ & $\numrange{88}{120}$ & $\numrange{1}{0.77}$ & $-60$ & $\sim 140$ \\
        CO & 4.7 & h & $\numrange{9.5e18}{1.4e19}$ & $\numrange{730}{690}$ & $\numrange{1}{0.77}$ & $0$ & $\sim 190$ \\
        \ce{H2O}\tablefootmark{b} & 6.2 & w & $\numrange{2.4e18}{7.0e18}$ & $\numrange{300}{270}$ & $\numrange{1}{0.35}$ & 0 & $\sim 90$ \\
        \ce{CH4} & 7.7 & w & $\numrange{4.2e18}{4.7e18}$ & $\numrange{333}{291}$ & $\numrange{0.10}{0.12}$\tablefootmark{c} & 0 & $\sim 70$ \\
        \ce{C2H2}\tablefootmark{d} & 13.7 & mw ($\mathrm{\nu_5}$) & $\SI{{2.29\pm0.05}e17}{}$ & $149\pm2$ & $0.28\pm0.003$\tablefootmark{e} & $-55$ & $\sim 90$ \\
        \ce{C2H2} & 7.5 & mw ($\mathrm{\nu_4 + \nu_5}$) & $\numrange{4.3e17}{2.6e18}$ & $\numrange{179}{157}$ & $\numrange{1}{0.2}$ & 0 & $\sim 70$ \\
        \ce{HCN} & 14.0 & c ($\mathrm{\nu_2}$) & $\SI{{2.45\pm0.11}e17}{}$ & $52\pm2$ & $0.28\pm0.003$\tablefootmark{e} & $-75$ & $\sim 90$\\
        \ce{HCN} & 14.0 & w ($\mathrm{\nu_2}$) & $\SI{{1.15\pm0.08}e17}{}$ & $328 \pm 22$ & $0.28\pm0.003$\tablefootmark{e} & $-75$ & $\sim 90$\\
        \ce{HCN} & 7.1 & mw ($\mathrm{2\nu_2}$) & $\numrange{5e17}{7e18}$ & $\numrange{110}{130}{}$ & $\numrange{1}{0.1}$ & $-15$ & $\sim 75$ \\
        \ce{CO2} & 15.0 & c & $\numrange{5.8e15}{2.6e17}$ & $\numrange{40}{59}$ & $\numrange{1}{0.05}$ & $\SI{-75}{}$ & $\sim 90$ \\
        \ce{N2H+} & 14.6 & c/mw & $\numrange{1.3e15}{4.4e16}$ & $\numrange{132}{192}$ & $\numrange{1}{0.05}$ & $-75$ & $\sim 90$ \\
        \ce{C2H} & 5.4 & mw & $\numrange{2.6e17}{3.7e18}$ & $\numrange{138}{107}$ & $\numrange{1}{0.1}$ & $-50$ & $\sim 70$ \\
        \ce{NO} & 5.3 & c & $\numrange{1.3e17}{1.8e18}$ & $\numrange{60}{48}$ & $\numrange{1}{0.1}$ & $-50$ & $\sim 70$ \\
        \ce{HNC} & 21.6 & c & $\SI{{3.2\pm1.2}e16}{}$ & $38\pm4$ & $0.06 \pm 0.02$ & $0$\tablefootmark{f} & $\sim 90$ \\
        \ce{HCO+} & 12.1 & c? & $\lesssim \SI{5e15}{}$ & & & & \\
        \ce{CS} & 7.9 & & & & & & \\
        \ce{HC3N} & 15.1 & c? & $\lesssim \SI{e16}{}$ & & & \\
        \ce{HCNH+} & 12.5 & c? & & & & \\
        \hline
    \end{tabular}
    \tablefoot{
    Background fractions were fit directly when constrained by optically thick features. In these cases, the uncertainties quoted are statistical 68\% confidence intervals from the MCMC sampling, and do not reflect the larger systematic uncertainties. In all other cases, parameter ranges are given based on fits with the maximum and minimum possible background fraction. The radial velocities are either based on direct measurements of individual lines (e.g. \ce{CO}, \ce{C2H2}, \ce{HCN}) or estimated from a visual comparison of the observed and modelled spectra (e.g. \ce{N2H+}, \ce{CO2}).
    \tablefoottext{a}{Cold (c), moderately warm (mw), warm (w) or hot (h); see Section \ref{sec:categories}.}
    \tablefoottext{b}{The \ce{H2O} ortho-to-para ratio inferred from several fits was consistent with the equilibrium value of 3; it was therefore fixed to this value in the final run.}
    \tablefoottext{c}{Not well-constrained in the fit. Ranges are taken between $f_\mathrm{bg}(\ce{C2H2}) = \numrange{1}{0.2}$.}
    \tablefoottext{d}{An ortho-to-para ratio of $1.71 \pm 0.05$ was inferred from the \SI{13.7}{\micron} band; it was fixed to this value in the \SI{7.5}{\micron} fit.}
    \tablefoottext{e}{Coupled in the fit. {If the warm \ce{HCN} is located in a deeper layer of the region, its background fraction may be overestimated and, consequently, its actual column density could be higher than reported here.}}
    \tablefoottext{f}{With respect to the systemic redshift $z_\mathrm{WN} = 0.01786$ \citep{Sakamoto2009}.}
    }
    \label{tab:properties_summary}
\end{table*}

The extracted JWST spectra of the WN of Arp~220 contain a large number of molecular bands. Here we present these detections, alongside model spectra of the bands. The inferred properties of the gas are summarised in Table \ref{tab:properties_summary}. In the following we outline the general method used to model the molecular bands.

The first step in the analysis of each band is to extract the background continuum. We model the local continuum by fitting a basis spline with manually placed nodes, and divide the observed spectrum by this continuum model. Throughout this work we assume that the lines are pure absorption lines, and that therefore the absolute fluxes are irrelevant. We reflect on this assumption in Section \ref{sec:discussion_line_emission}.

Given the large line widths and the chemical complexity of the nucleus, almost no isolated individual lines are present in the spectrum, and therefore a rotation diagram analysis \citep{GoldsmithLanger1999} is insufficient to describe the observed molecular signatures. Instead, throughout this work we adopt a spectral fitting procedure with local thermodynamic equilibrium (LTE) models and Markov Chain Monte Carlo (MCMC) sampling, implemented through the \texttt{emcee} Python package \citep{emcee}. We estimate radial velocities and dispersions from direct measurements where possible, or from visual comparison to model spectra otherwise, and fix these kinematics in the models\footnote{In specific cases (Sections \ref{sec:res_6um} and \ref{sec:res_HNC}) we deviate from this approach and fit the velocity dispersion as well to account for line broadening in the optically thick regime.}. For each gas component, the rotational excitation temperature $T_\mathrm{rot}$ and column density $N$ are left as free parameters. In many cases, bands of several species overlap, and they must be fit together.

The dust continuum observed towards the nuclei of U/LIRGs is typically produced not by a single uniform emission source, as we implicitly assume our background to be, but rather by a distribution of dust with a temperature gradient \citep[e.g.][]{Armus2007, Donnan2024}. The detected molecular gas could be located anywhere in this dust distribution. If, at any wavelength, a significant fraction of the observed continuum is produced in the foreground, the absorption signal produced in a deeper layer will be diluted by this foreground emission. In this work we do not model the spectral energy distribution of the continuum, remaining deliberately agnostic about the location of the molecular gas and avoiding the complexity and degeneracies involved in SED models. Instead, we parametrise the fraction of observed continuum flux that passes through the absorbing gas through the \textit{background fraction} $f_\mathrm{bg}$. This parameter is mathematically equivalent to the covering factor, which describes the projected overlap between a uniform background source and a uniform absorbing cloud. However, we do not consider this a correct interpretation here, as high-resolution ALMA data of the Arp~220 nuclei show extended molecular gas traced by many lines \citep[e.g.][]{Sakamoto2021b}.

The background fraction is largely degenerate with column density unless (some of) the features become optically thick. {If the deepest features are no longer optically thin, the absorption depth ratio between optically thick and optically thin features becomes a probe of the background fraction.} In particular, the detection of both a {deep, blended} $Q$-branch and the {shallower} $P$- and $R$-branch lines can break {the degeneracy between background fraction and column density, if the $Q$-branch is shallower than expected from the $R$- and $P$-branch lines}. In our analysis, we try to simultaneously fit the background fraction along with the temperature and column density. We use a scalar background fraction under the assumption that, across the spectral range of the fit, the spectral shape of the unabsorbed continuum does not vary greatly from that of the assumed total continuum. For the narrow range of a single rovibrational band, this assumption is typically valid. In cases where the background fraction is unconstrained, we present a range of inferred column densities based on the minimum and maximum possible background fraction. In the absence of obviously saturated features, the minimum possible background fraction is equal to the absorption depth of the deepest feature: $f_\mathrm{bg,min} = 1 - I_\mathrm{obs,min}/I_\mathrm{obs,cont}$.

To construct the spectral models, we adopt line lists and partition functions primarily from the HITRAN database \citep{HITRAN}. CDMS \citep{CDMS} is used instead for \ce{N2H+}, \ce{HC3N} (Section~\ref{sec:res_14um}), \ce{HCO+} (Section~\ref{sec:res_HNC}), and \ce{C2H} (Section~\ref{sec:res_C2H}); for \ce{HNC} (Section~\ref{sec:res_HNC}) the line list is taken from GEISA \citep{GEISA} and the partition function from CDMS. For each spectral fit, the final best-fit model is constructed by taking the median of the sampled posterior for each parameter, and modelling the spectrum under those conditions. Details for each individual spectral region are provided in the relevant sections. All inferred gas properties are summarised in Table \ref{tab:properties_summary}.

\subsection{\ce{C2H2} and HCN fundamental bands} \label{sec:res_14um}
\begin{figure*}[ht]
    \centering
    \includegraphics[width=\linewidth]{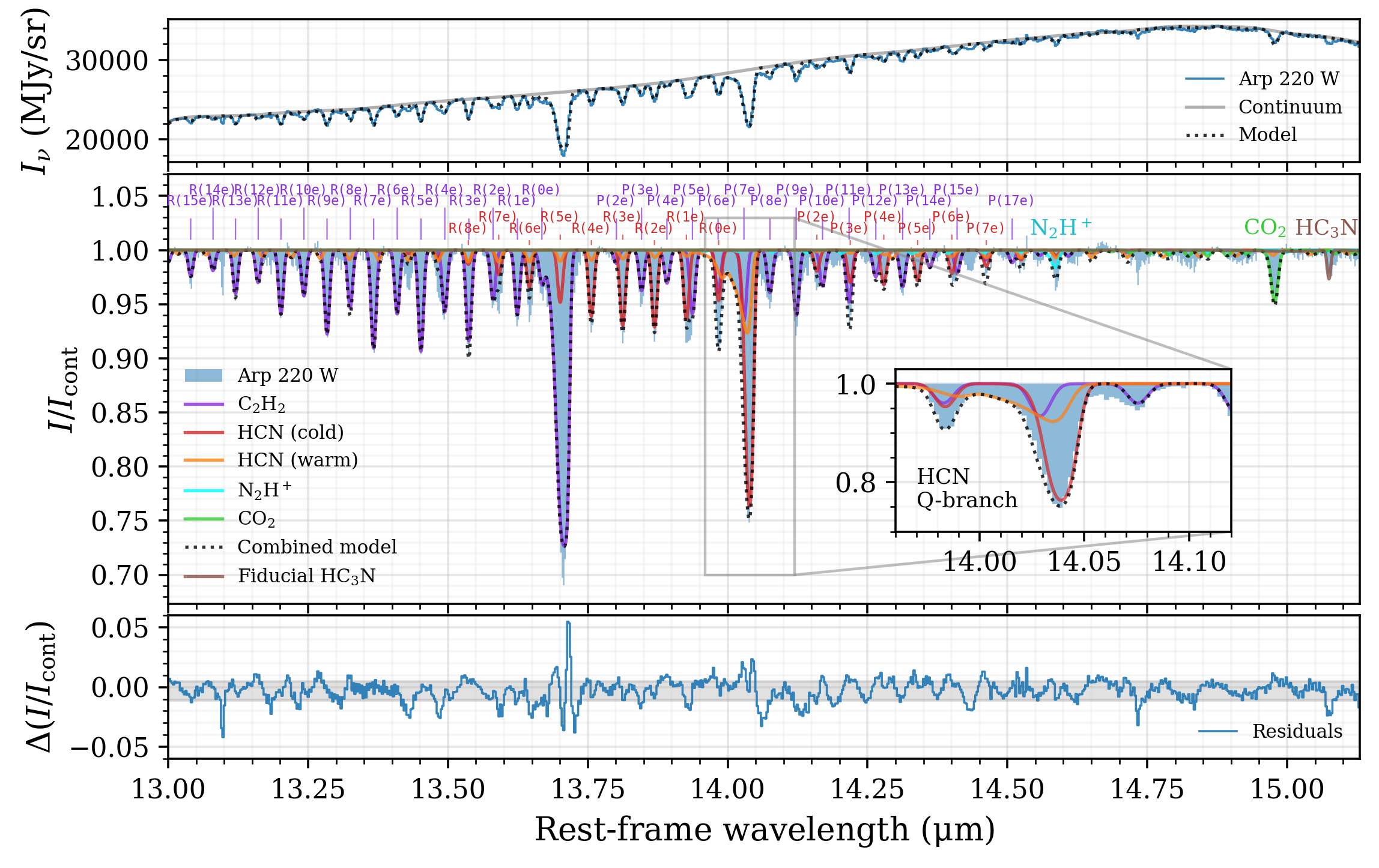}
    \caption{Top panel: observed spectrum and total model of the \SI{14}{\micron} region. Middle panel: {continuum-normalised} spectrum and model absorption spectrum. The model includes \ce{C2H2}, two \ce{HCN} components, \ce{N2H+} and \ce{CO2}. A fiducial model of \ce{HC3N} with $N = \SI{3e15}{\centi\meter^{-2}}$, $T = \SI{50}{\kelvin}$ is plotted as well. All components are shown with $f_\mathrm{bg} = 0.28$, following the value inferred from \ce{C2H2} and \ce{HCN}. The inset shows the model of the \ce{HCN} $Q$-branch in detail, illustrating that a $\mathbf{\SI{330}{\kelvin}}$ component is needed to explain the asymmetric shape of the $Q$-branch. Strong lines in the $R$- and $P$-branches of \ce{C2H2} and \ce{HCN} are labelled for reference. Bottom panel: the residual spectrum. The grey shaded area indicates the 68\% interval of the residuals.}
    \label{fig:14micron_model_spec}
\end{figure*}

The MIRI spectrum around \SI{14}{\micron} shows prominent absorption features of \ce{HCN} and \ce{C2H2} (Fig. \ref{fig:14micron_model_spec}). The $Q$-branches, previously detected with Spitzer \citep{Lahuis2007} and now seen at higher spatial and spectral resolution, reach absorption depths of 25\% and 30\% respectively. Additionally, a large number of $R$- and $P$-branch lines of both species are clearly detected. Three weaker but clearly detected features can be attributed to the $Q$-branches of \ce{CO2}, \ce{N2H+}, and \ce{HC3N}.

We estimate the radial velocity and velocity dispersion of \ce{C2H2} and \ce{HCN} from their $R$-branch lines, considering only those that are relatively unaffected by contamination from the other species. We then take the median radial velocity and dispersion of each species separately, and use these as fixed parameters in the spectral fits. We note that, by using directly measured line widths, we implicitly assume that the lines are optically thin. This assumption can be verified by considering the maximum optical depth of the lines. For \ce{C2H2}, the background fraction must exceed $0.31$ to produce the observed $Q$-branch depth. Its deepest $R$-branch line reaches an absorption depth of $10\%$. Therefore, the true peak optical depth of this line is at most $\tau_0 < 0.4$, and we can safely use the directly measured line widths in our fits.

Initial fits exhibited an asymmetry between lines arising from ortho and para levels of \ce{C2H2}, where the ortho-\ce{C2H2} lines were well-fit but the para-\ce{C2H2} lines were underpredicted. For this reason, we implemented the ortho-to-para ratio (OPR) for this species as a free parameter. Furthermore, we find that a single \ce{HCN} component cannot explain the shape of the $Q$-branch at \SI{14.0}{\micron}, and incorporate a second, warmer \ce{HCN} component. We note that attempts to fit the background fraction of each component separately led to very similar values for \ce{C2H2} and the cold \ce{HCN} component, but failed to constrain that of the warm \ce{HCN} component. This latter background fraction may be smaller if the warm \ce{HCN} is located in a deeper layer of the nucleus, in which case its implied column density would be larger.

After constraining the \ce{C2H2} and \ce{HCN} properties, we combine the best-fit model spectrum with models of \ce{CO2} and \ce{N2H+} to simultaneously fit for the properties of the latter two species. For these species we only detect the $Q$-branches, and therefore we run the fit for both $f_\mathrm{bg} = 1$ and $f_\mathrm{bg} = 0.05$ to determine the full range of possible column densities. The corresponding total model spectrum is shown in Fig.~\ref{fig:14micron_model_spec}. A fiducial model of \ce{HC3N}, which has a $Q$-branch at \SI{15.07}{\micron}, is presented as well. The inferred column densities and temperatures of \ce{C2H2}, \ce{HCN}, \ce{CO2} and \ce{N2H+} are summarised in Table~\ref{tab:properties_summary}.

ALMA observations of the WN previously found a collimated, compact molecular outflow perpendicular to the disk at radial velocities of up to \SI{-520}{\kilo\meter\per\second} \citep{BarcosMunoz2018}. This outflow was clearest in \ce{HCN} (1-0), and therefore we would expect to see evidence of this outflow in cold \ce{HCN}. Although our mid-IR absorption lines of \ce{HCN} are blueshifted, the derived radial velocities are only $\sim \SI{-75}{\kilo\meter\per\second}$. We find no blueshifted features of \ce{C2H2} or \ce{HCN} at the $\qtyrange{280}{520}{\kilo\meter\per\second}$ outflow velocities reported by \citet{BarcosMunoz2018}, but note that one of relatively low column density $\lesssim \SI{e16}{\per\centi\meter\squared}$ or much smaller background fraction could be buried beneath the more prominent bulk features. We also note that we do not detect any clear hot band absorption (i.e. from a higher vibrational state) for either \ce{HCN} or \ce{C2H2}.

\subsection{Water vapour} \label{sec:res_6um}
\begin{figure*}[ht]
    \centering
    \includegraphics[width=\linewidth]{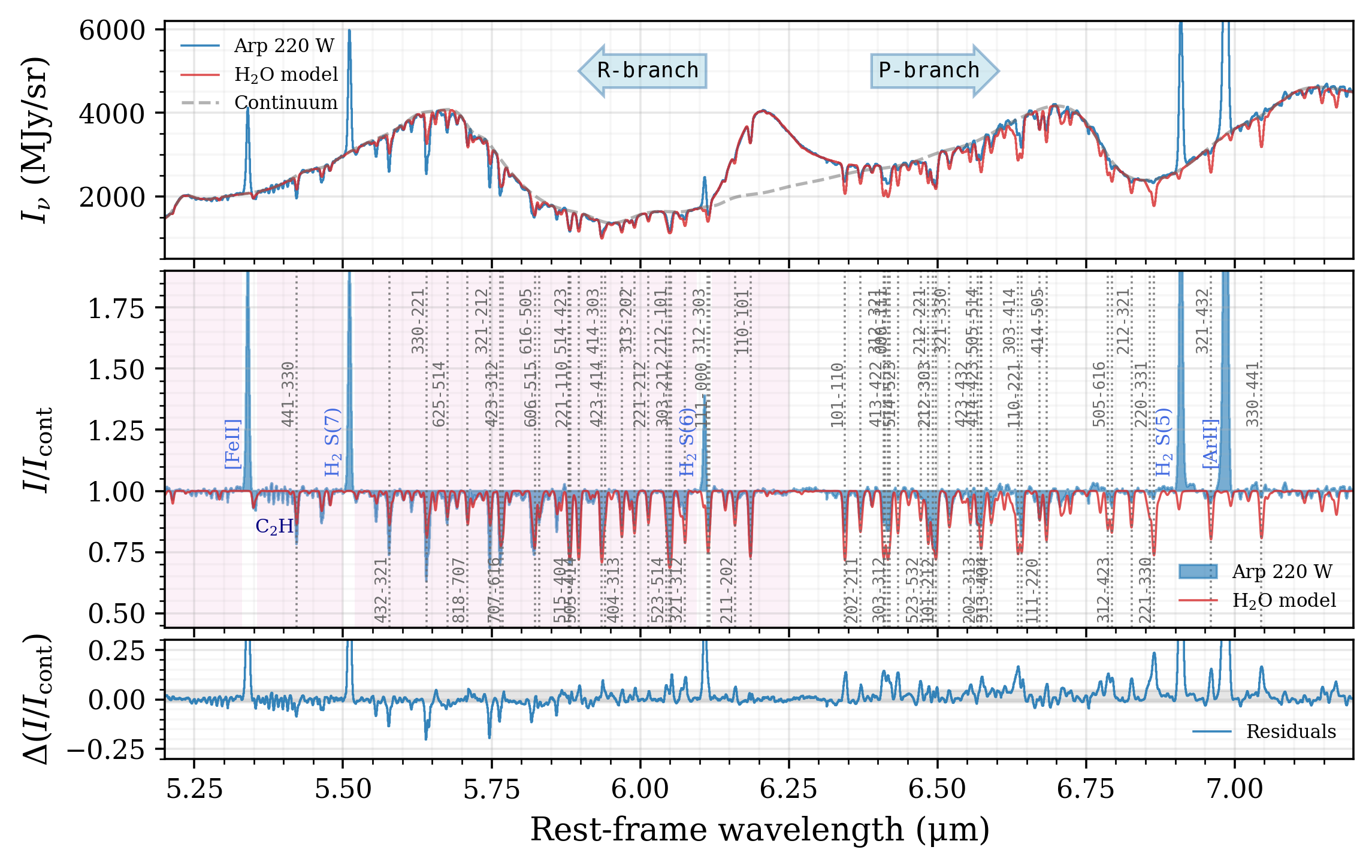}
    \caption{The spectrum around \SI{6}{\micron} and the best-fit \ce{H2O} model for $f_\mathrm{bg} = 0.35$. The top panel shows the absolute flux spectrum and continuum model. The middle panel shows the {continuum-normalised} spectrum. Strong \ce{H2O} lines are labelled by their upper- and lower-level quantum numbers as $J_{u}, K_{a,u}, K_{c,u} - J_l, K_{a,l}, K_{c,l}$. Several strong emission lines and absorption lines of \ce{C2H} are labelled as well. The pink shaded area indicates the spectral region considered in the fit. The bottom panel shows the residuals on the {continuum-normalised} spectrum; the grey shaded area indicates their 68\% interval.}
    \label{fig:H2O_model_spec}
\end{figure*}

We detect prominent {rovibrational} water absorption between \SI{5}{\micron} and \SI{7}{\micron}, as shown in Fig.~\ref{fig:H2O_model_spec}. As the water lines dominate the absorption over a wide spectral region, we do not simultaneously model other species. The $P$-branch lines at wavelengths $\lambda > \SI{6.3}{\micron}$ appear significantly weaker than the $R$-branch lines, possibly indicative of $P$-/$R$-branch asymmetry \citep[e.g.][]{Gonzalez-Alfonso2002, Buiten2024a}. This asymmetry arises from the fact that absorption lines into a particular vibrationally excited level are typically much stronger in the $R$-branch than in the $P$-branch. The $P$-branch can thus more easily be filled in, if some emission is also present. Therefore, in all cases where such an asymmetry is found, we restrict our fits to the $R$-branch only.

{\citet{Martin2011} and \citet{Gonzalez-Alfonso2012} detected lines from the \ce{H^{18}O} isotopologue in submillimetre and far-infrared rotational lines respectively.}
We do not unambiguously detect any lines of \ce{H2^{18}O}, but note that even at the elevated \ce{H2^{18}O} abundances that have been reported for Arp~220 \citep{Gonzalez-Alfonso2012}, these isotopologue features would be at most 5\% deep (if $f_\mathrm{bg} = 1$) and largely indistinguishable from the much stronger \ce{H2^{16}O} lines. {The} rotational water lines at \qtyrange{12}{26}{\micron}, commonly seen towards Galactic protostars \citep{Francis2024, VanGelder2024}, are not detected. {This non-detection is not unexpected as these mid-infrared rotational lines trace very highly excited gas and are predicted to be much weaker than the far-infrared \ce{H2O} lines, assuming the rotational temperature derived from the rovibrational band.}

A complicating factor in the water spectrum is the strong contribution of PAH emission in this spectral range, in particular the feature at \SI{6.2}{\micron}. If this emission arises in the foreground, it fills in the absorption lines with a particular spectral shape that a scalar background fraction cannot account for. \citet{Spoon2004} noted that, in the ISO spectrum of Arp~220, the PAH features are not strongly deformed by the \qty{9.7}{\micron} silicate absorption, implying that the PAH emission indeed originates from in front of most of the obscuring dust. Therefore, in our analysis, we assume that the PAH emission is produced in the foreground. As such, we fit a basis spline to the apparent \SI{6.2}{\micron} PAH feature, and subtract it before dividing out the continuum. We note that we also detect weaker PAH features at \SI{5.25}{\micron}, \SI{5.43}{\micron} and \SI{6.02}{\micron} \citep{Chown2024}; we do not make any correction for these features in the \ce{H2O} fit. The shape of the underlying \ce{H2O} ice feature is treated as continuum, as both the gas and ice features are in absorption.

The fact that the strongest \ce{H2O} lines have similar absorption depths ($\sim 30\%$ in the $R$-branch) suggests that some of these lines may be optically thick. We therefore estimate the velocity dispersion of the lines by leaving both the background fraction and the line width as free parameters, taking $f_\mathrm{bg} > 0.35$ as a lower limit based on the maximum absorption depth found in this band. This approach results in a best-fit velocity dispersion of \SI{90}{\kilo\meter\per\second}, but fails to constrain the background fraction. Initially, we varied the OPR as well, but we fixed it later as we found no significant deviation from the equilibrium value of 3.

Fig.~\ref{fig:H2O_model_spec} shows the best-fit model spectrum for $f_\mathrm{bg} = 0.35$. Only a single component was used in this model. Evidently, the model significantly underestimates the absorption in several $R$-branch lines, while overpredicting most $P$-branch lines. The failure to fit the $P$-branch can be attributed to the $P$-/$R$-branch asymmetry noted above. However, the underpredicted $R$-branch lines at \qtyrange{5.4}{5.8}{\micron} require an explanation. We attempted LTE models with two temperature components, but found that an additional high-excitation component did not improve the fit. This is unsurprising, as the discrepant lines arise from low- to medium-excitation levels; in an LTE model, deep lines from these levels would therefore be accompanied by detectable additional optical depth in the low-$J$ lines.
One possible explanation is a steep negative slope with increasing wavelength of the ``true'' background continuum: if the background fraction varies significantly over the spectral range, long-wavelength lines will be more {diluted} than short-wavelength ones. This scenario is possible if the background source has a temperature of $T_\mathrm{bg} \gtrsim \SI{600}{\kelvin}$.

We further note that, with the high critical densities of its rotational transitions and the wide variety of allowed radiative transitions (in the submillimetre, far-infrared and mid-infrared), \ce{H2O} is highly susceptible to far-infrared radiative excitation, which will strongly affect the population of the rotational levels. Indeed the $4_{3,2}$, $3_{3,0}$ and $3_{2,1}$ levels, all of which have excess population with respect to the model in Fig.~\ref{fig:H2O_model_spec}, are efficiently populated by strong far-infrared radiation, as shown by the Herschel/PACS observations \citet{Gonzalez-Alfonso2012}. Modelling these effects is beyond the scope of our LTE approach, and we therefore do not analyse the \ce{H2O} excitation further in the present paper.

\subsection{Carbon monoxide} \label{sec:res_CO}
\begin{figure*}[ht]
    \centering
    \includegraphics[width=\linewidth]{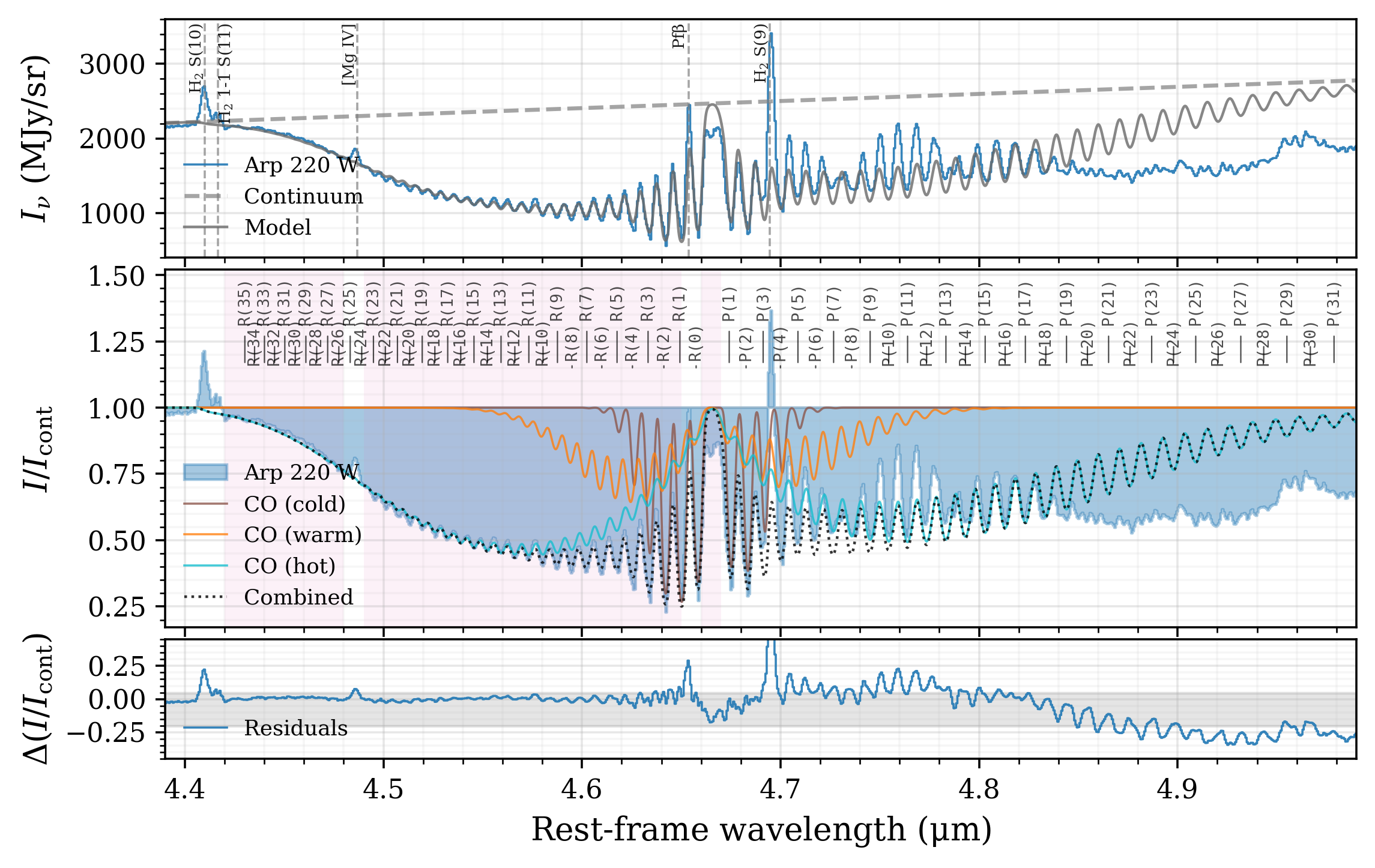}
    \caption{Best-fit model spectrum for CO, extracted from an $r = 0\farcs15$ aperture, with $f_\mathrm{bg} = 0.77$. The top panel shows the total model including the continuum, the middle panel shows the {continuum-normalised} spectrum, and the bottom panel shows the residuals on the {continuum-normalised} spectrum. The grey shaded area in the bottom panel indicates the 68\% interval of the residuals. The pink shaded area indicates the spectral region used in the fit.}
    \label{fig:CO_modelspec}
\end{figure*}

At \SI{4.7}{\micron}, the fundamental band of CO is clearly detected with NIRSpec, with very prominent absorption from the $J < 5$ levels, and higher-excitation tails extending out to $J = 27$, as seen in Fig. \ref{fig:CO_modelspec}. However, the overall shape of the spectrum is peculiar, with a broad absorption feature visible between \SI{4.4}{\micron} and \SI{4.67}{\micron}, and irregular spacing of absorption features in the $P$-branch beyond \SI{4.7}{\micron}. With no ice band known that could produce the broad shape seen in the $R$-branch, we propose that this broad absorption is a pseudocontinuum formed by broad ($\sim \SI{200}{\kilo\meter\per\second}$) gas-phase \ce{CO} lines, with possibly some contribution from \ce{OCN-} ice at \SI{4.6}{\micron}.
Similar \ce{CO}-induced pseudocontinuum formation has been observed in absorption in several U/LIRGs \citep{Onishi2024} and in emission towards the Galactic SNR Cassiopeia A \citep{Rho2024}.

For the continuum estimation, we extend the \SIrange{3.5}{4.4}{\micron} slope out to \SI{5.08}{\micron}. However, on the red side of the CO band, it is unclear where exactly the CO-induced pseudocontinuum ends, in part because the $P$-branch is contaminated by absorption from \ce{^{13}CO} and \ce{C^{18}O}. Furthermore, the CO fundamental band is known to show $P$-/$R$-branch asymmetry \citep[e.g][]{Gonzalez-Alfonso2002, Pereira-Santaella2024, Buiten2024a}, with the $P$-branch preferentially seen in emission and the $R$-branch in absorption. For these reasons, we restrict our spectral fit to the $\qtyrange{4.42}{4.67}{\micron}$ spectral region.

We attempt to fit both a 2-component model and a 3-component model, fixing the background fraction to 1 and 0.77 to determine the minimum and maximum column densities, respectively. The radial velocities and velocity dispersions are fixed in the models; they are estimated visually through experimentation with spectral models, as there are no isolated lines available for direct measurements. We find that the model with three velocity components better represents the data; the best-fit model spectrum for $f_\mathrm{bg} = 0.77$ is shown in Fig. \ref{fig:CO_modelspec}. In this model we assume a \SI{60}{\kilo\meter\per\second} blueshift for the warm component, and a \SI{60}{\kilo\meter\per\second} redshift for the cold component. We note that the relative contributions of the cold and warm component vary strongly with small changes to the assumed kinematics, and therefore the column densities inferred for them may not represent a unique solution.

The exact inferred rotational temperatures and column densities depend on the choice of continuum. Nevertheless, the overall shape of the spectrum on the $R$-branch side, between \qty{4.4}{\micron} and \qty{4.67}{\micron}, can be reproduced well by broad ($\sigma_V \approx \SI{190}{\kilo\meter\per\second}$) lines of CO at a rotational temperature of $T_\mathrm{rot} \approx \SI{700}{\kelvin}$. The obtained constraints on the \ce{CO} properties are summarised in Table~\ref{tab:properties_summary}.

\subsection{\ce{HCN}, \ce{C2H2} and \ce{CH4} at 7 micron}
\subsubsection{The \SI{7.7}{\micron} complex} \label{sec:res_8micron}
\begin{figure*}[ht]
    \centering
    \includegraphics[width=\linewidth]{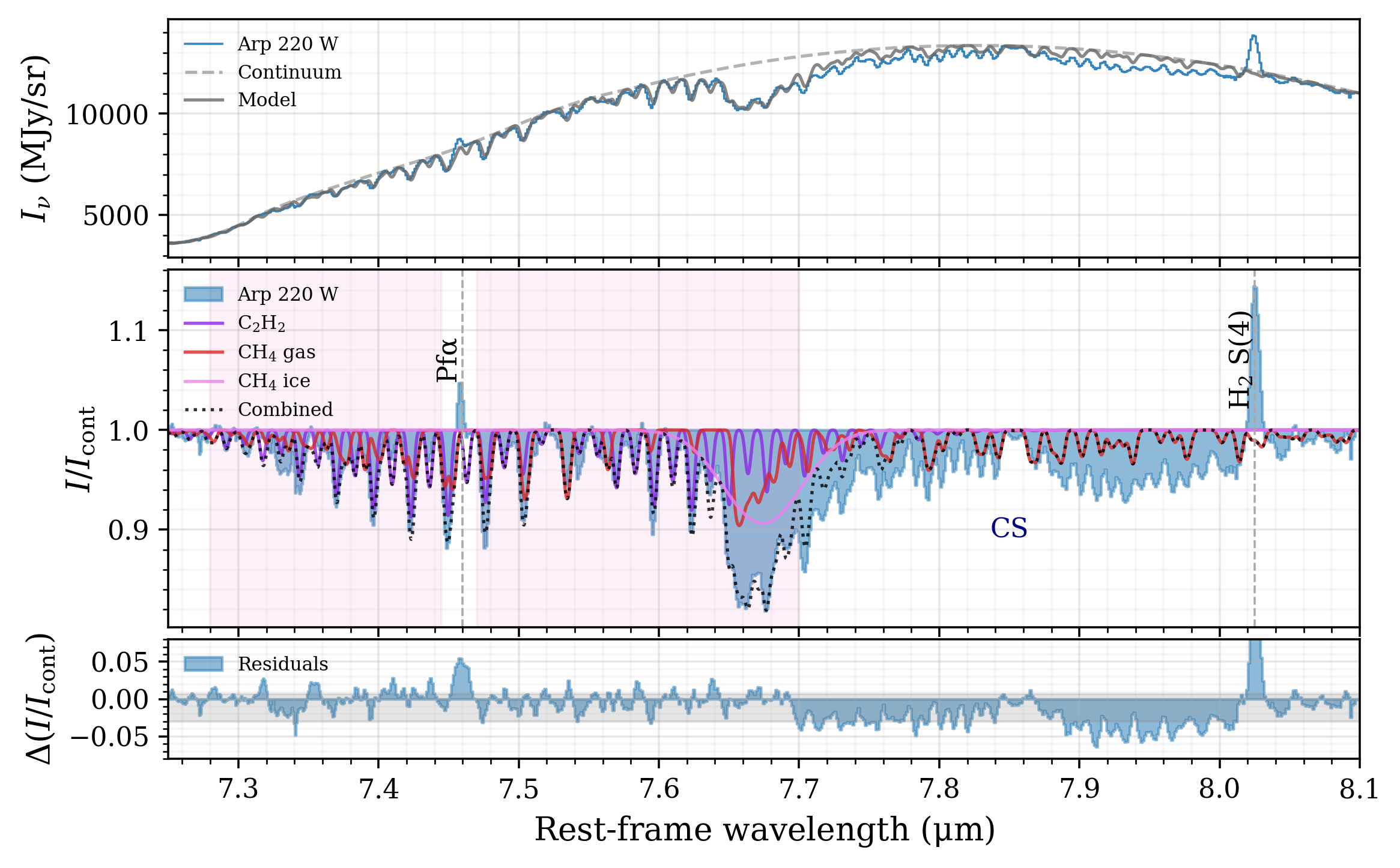}
    \caption{{Continuum-normalised} spectrum in between \SI{7.2}{\micron} and \SI{8.1}{\micron} vs. the best-fit LTE model for $f_\mathrm{bg}(\ce{C2H2}) = 0.2$. The model includes the \ce{C2H2} $\nu_4 + \nu_5$ combination band, \ce{CH4} gas, and \ce{CH4} ice. Only the spectral region indicated in by the pink shaded area was included in the fit in order to avoid the effects of Pf~$\mathrm{\alpha}$ contamination, \ce{CS} absorption, and further systematic uncertainties in the continuum estimation. The grey shaded area in the bottom panel indicates the 68\% interval of the residuals.}
    \label{fig:CH4_C2H2_modelspec}
\end{figure*}

At $\sim 7-\SI{8}{\micron}$, the \SI{7.7}{\micron} PAH complex leads to a bump in the spectrum that smoothly transitions into the broad \SI{9.7}{\micron} silicate absorption band (see Fig. \ref{fig:fullspec}). In the spectrum of the WN, absorption features of several species are imprinted on top of this bump. The tail of the \ce{H2O} $P$-branch extends out to $\sim \SI{7.1}{\micron}$ (Fig. \ref{fig:H2O_model_spec}). Between \SI{6.9}{\micron} and \SI{7.3}{\micron}, we detect the \ce{HCN} 2$\nu_2$ overtone band (see Section \ref{sec:res_HCN_overtone}; Fig. \ref{fig:HCN_overtone_modelspec}). Further redwards, deep absorption lines appear due to the combination of \ce{CH4} and the \ce{C2H2} $\nu_4 + \nu_5$ band, as presented in Fig. \ref{fig:CH4_C2H2_modelspec}. The peculiar, broader feature at \SI{7.7}{\micron} is the $Q$-branch of the gas-phase \ce{CH4} band combined with \ce{CH4} ice absorption, with an additional contribution from the \ce{C2H2} $\nu_4 + \nu_5$ $P$-branch. Finally, the 7.7-\SI{8.0}{\micron} region contains a series of very closely-spaced absorption lines produced by \ce{CS}, with some contribution from the \ce{CH4} $P$-branch.

Our ability to draw strong conclusions on the excitation temperatures, column densities, and covering factors for these spectral bands is severely limited by several complications. First, the overlap of a number of spectral bands leads to strong degeneracies. Second, the spectrum between \SI{7.64}{\micron} and \SI{8.06}{\micron} is most likely affected by pseudocontinuum formation due to \ce{CH4} and \ce{CS} absorption. Third, if a significant portion of the \SI{7.7}{\micron} PAH emission arises in the foreground, the shape of the perceived absorption spectrum is severely distorted, with the \ce{CH4} $Q$-branch being more filled-in than the $R$-branch. The latter two problems lead to considerable systematic uncertainties on the continuum estimate. We attempt to take the pseudocontinuum into account by fitting a very smooth spline to estimate the continuum. Ideally, we would model the combined dust continuum and \qty{7.7}{\micron} PAH complex, and subtract the latter before taking the first as our background continuum. However, this approach would introduce additional degeneracies and uncertainties, and is beyond the scope of this work. We therefore make no explicit correction for foreground PAH emission here.

We use the \SI{7.35}{\micron}-\SI{7.7}{\micron} spectrum to simultaneously fit for \ce{C2H2}, gas-phase \ce{CH4}, and solid \ce{CH4}. For the methane ice band, we use a Gaussian with a fixed central wavelength of \SI{1303}{\centi\meter^{-1}} (\SI{7.674}{\micron}) and a full width at half maximum (FWHM) of \SI{11}{\centi\meter^{-1}} \citep{Boogert2015}, and only fit for the amplitude. For the \ce{C2H2} we assume an OPR of 1.71, as determined from the \SI{14}{\micron} analysis; for \ce{CH4} we assume equilibrium conditions. As the \ce{C2H2} $\nu_4 + \nu_5$ band lacks a $Q$-branch, its background fraction is unconstrained. We therefore choose to fix the \ce{C2H2} background fraction to 1 and 0.2 here, resulting in a lower and upper limit on its column density. Only the \ce{CH4} gas background fraction is left as a free parameter.

The resulting best-fit model spectrum is presented in Fig. \ref{fig:CH4_C2H2_modelspec}. We note that although CS absorption is clearly detected, we do not model it due to large systematic uncertainties beyond \SI{7.7}{\micron}. For the \ce{CH4} ice band, modelled as a simple Gaussian with unity background fraction, we derive a column density of $\sim \SI{e17}{\centi\meter^{-2}}$.

\subsubsection{The \ce{HCN} overtone band} \label{sec:res_HCN_overtone}
\begin{figure*}[ht]
    \centering
    \includegraphics[width=\linewidth]{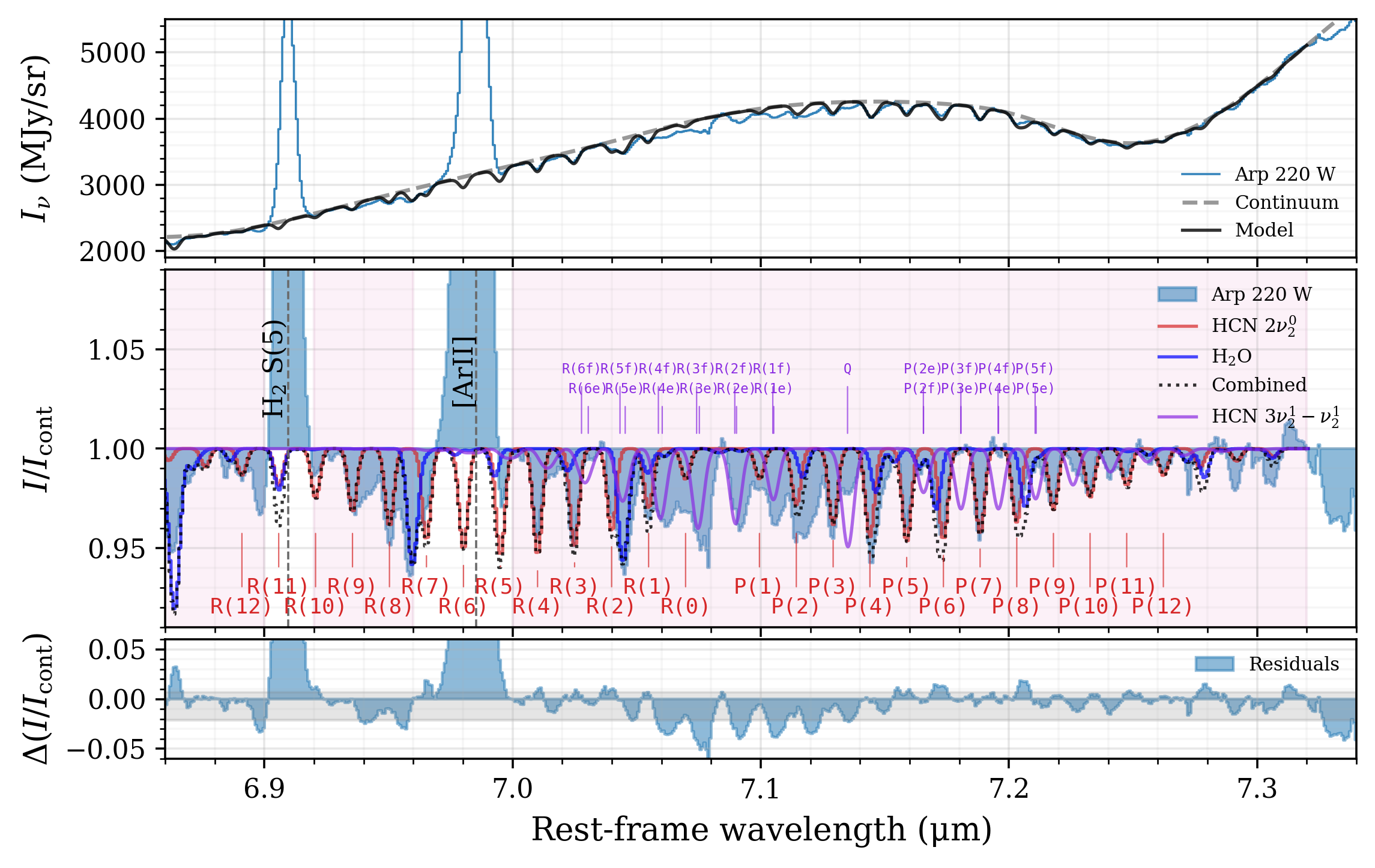}
    \caption{Best-fit LTE model for the \ce{HCN} 2$\nu_2$ overtone band and the interloping \ce{H2O} lines at \SI{7}{\micron}, for $f_\mathrm{bg} = 1$. The pink shaded area indicates the spectral region used in the fit. A simultaneously-fit $\ce{H2O}$ model is also shown. The \ce{HCN} overtone lines are labelled, and the residuals are shown in the bottom panel. A fiducial model of the \ce{HCN} $3\nu_2 - \nu_2$ band, with $N(\ce{HCN}) = \SI{2e18}{\centi\meter^{-2}}$, $T_\mathrm{vib} = \SI{330}{\kelvin}$, $T_\mathrm{rot} = \SI{50}{\kelvin}$, and $f_\mathrm{bg} = 1$ is shown in purple. The grey shaded area in the bottom panel indicates the 68\% interval of the residuals.}
    \label{fig:HCN_overtone_modelspec}
\end{figure*}

The shallow \ce{HCN} $2\nu_2$ overtone band, presented in Fig.~\ref{fig:HCN_overtone_modelspec}, is contaminated by $P$-branch \ce{H2O} lines. To mitigate their effect on the inferred \ce{HCN} column density, we simultaneously fit for \ce{HCN} and \ce{H2O} between \SI{6.8}{\micron} and \SI{7.32}{\micron}, masking out the strong \ce{H2} S(5) and [Ar~II] lines. Here we fix the \ce{H2O} background fraction and velocity dispersion to the values that were derived separately from the $R$-branch (see Section \ref{sec:res_6um}). Similarly to our treatment of the \ce{C2H2} $\nu_4 + \nu_5$ band, we fix the \ce{HCN} $2\nu_2$ background fraction to 1 and $f_\mathrm{bg,min} = 0.1$ in the fit, leading to a lower limit on the column density. The best-fit model spectrum for $f_\mathrm{bg} = 1$ is presented in Fig.~\ref{fig:HCN_overtone_modelspec}.

Some residual lines remain in between the \ce{HCN} overtone lines. We tentatively identify these with the \ce{HCN} $3\nu_2 - \nu_2$ band; that is, the transitions from already vibrationally-excited \ce{HCN}---most likely through the strong \SI{14}{\micron}-pumping---to the third excited state. This is a surprising result, as we do not detect any hot band absorption at \SI{14}{\micron}. To assess the plausibility of this identification, we construct a fiducial model of this band based on ALMA observations of \ce{HCN} and \ce{HCN-vib} lines. \citet{Tunnard2015} analysed isotopologue ratios with large velocity gradient (LVG) modelling, and derived an \ce{HCN} column density of $\SI{2.3e18}{\centi\meter^{-2}}$, corrected for the \SI{90}{\kilo\meter\per\second} velocity dispersion that we measure. From the \ce{HCN}-vib line fluxes obtained by \citet{Martin2016}---who attempted to account for the severe line blending in the WN---and the emission size estimated by \citet{Sakamoto2021b}, we can infer $\nu_2 = 1f$ column densities $\sim\SI{e17}{\centi\meter^{-2}}$. Combining these two column density estimates, we find vibrational temperatures of \qtyrange{300}{500}{\kelvin}.

A fiducial model based on these estimates, with $T_\mathrm{rot} = \SI{50}{\kelvin}$ and $T_\mathrm{vib} = \SI{330}{\kelvin}$, is shown alongside the \ce{HCN} $2\nu_2$ + \ce{H2O} fit in Fig.~\ref{fig:HCN_overtone_modelspec}. Although the exact values for these parameters are uncertain, we find that, for these credible column densities, the \ce{HCN} $3\nu_2 - \nu_2$ $R$-branch lines match well with the strongest residual lines observed, at a small redshift with respect to the $2\nu_2$ lines. However, the $Q$-branch is overpredicted, and the $P$-branch lines do not seem to be present at all. Thus, if the identification with the \ce{HCN} $3\nu_2 - \nu_2$ band is correct, we again observe an asymmetry between the branches, as for \ce{H2O} and CO (see Figures \ref{fig:H2O_model_spec} and \ref{fig:CO_modelspec}). This asymmetry indicates the presence of emission, affecting the $P$-branch most, the $Q$-branch less, and the $R$-branch least. While modelling of this effect requires non-LTE methods, we discuss the possible implications in Section~\ref{sec:discussion_line_emission}.

The inferred column densities and rotational temperatures for \ce{HCN}, \ce{C2H2} and \ce{CH4} in the \SI{7}{\micron} to \SI{8}{\micron} range are listed in Table~\ref{tab:properties_summary}. Even the lower limits inferred here are a factor 1.6-2.0 higher than those derived from the corresponding \SI{14}{\micron} bands. We reflect on this discrepancy in Section~\ref{sec:emission_effects}.

\subsection{\ce{HNC} and \ce{HCO+}} \label{sec:res_HNC}
\begin{figure*}[h]
    \centering
    \includegraphics[width=\linewidth]{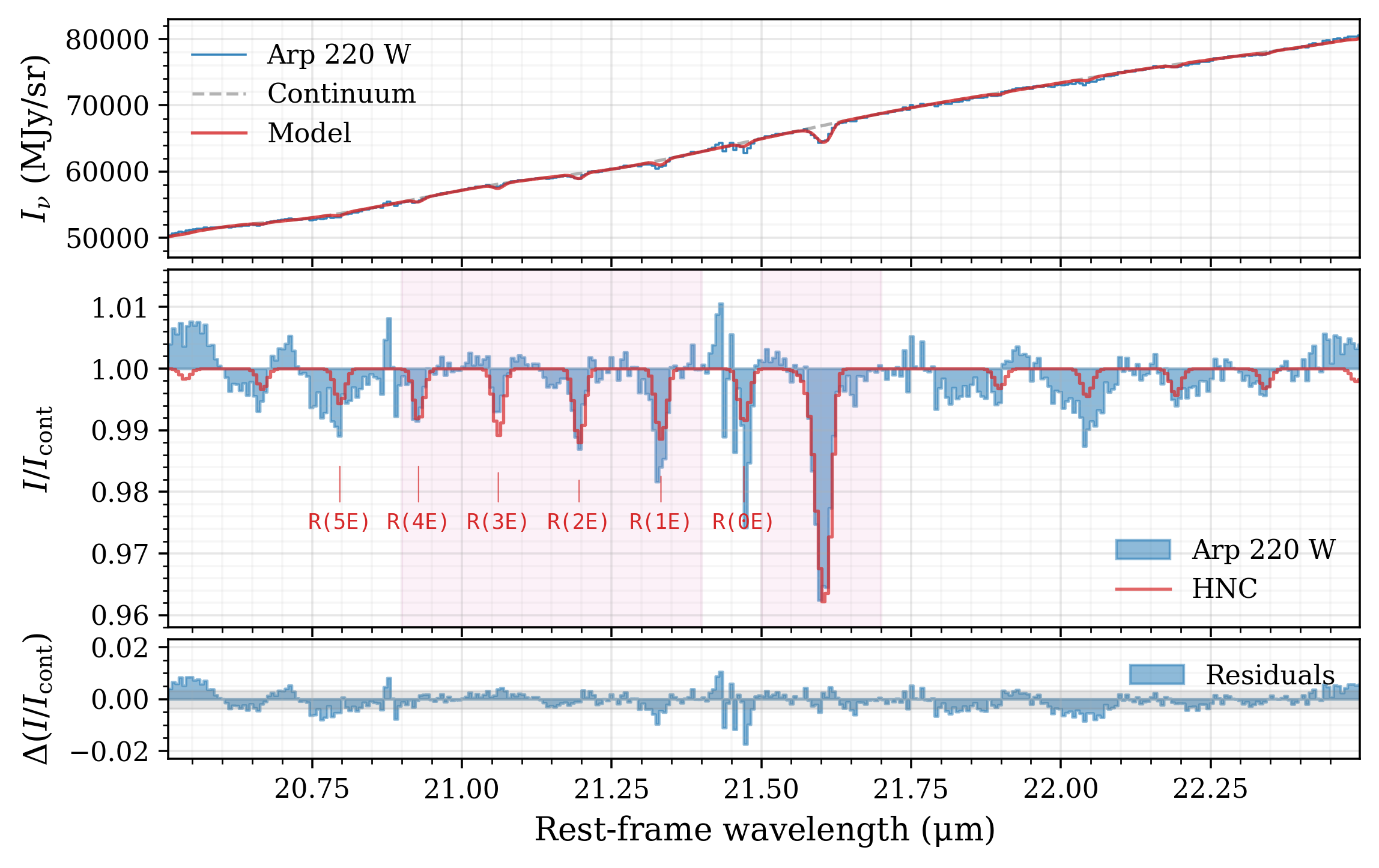}
    \caption{Spectrum at \SI{21}{\micron} (blue) and best-fit model for \ce{HNC} (red). The pink shaded areas indicate the spectral region used in the fit. The residuals are shown in the bottom panel, with the 68\% interval indicated by the grey shaded area.}
    \label{fig:HNC_modelspec}
\end{figure*}

The strong \ce{HCN} absorption in both the $\nu_2$ and $2\nu_2$ bands provides a compelling argument to search for its isomer \ce{HNC}. \ce{HNC} has its $\nu_2$ band at \SI{21.6}{\micron}. At this long wavelength, the MIRI MRS spectra suffer from significant fringing, even after the residual fringe correction in Stage 2 of the reduction pipeline (see Section~\ref{sec:data}). To mitigate this effect, we apply the additional 1D residual fringe correction to the extracted Channel 4B spectrum. The last remaining fringes are taken into account in fitting the continuum by first smoothing the spectrum with a Gaussian filter, and fitting a spline to the smoothed spectrum. The resulting spectrum, shown in Fig.~\ref{fig:HNC_modelspec}, reveals a clear detection of not only the \ce{HNC} $Q$-branch, but 5 $R$-branch lines as well; this combination of branches allows us to constrain the background fraction.

We estimate the radial velocity of the \ce{HNC} gas by direct measurement from the $R$(1)-$R$(4) lines. We find that they are all consistent with the systemic redshift of the WN $z_\mathrm{WN} = 0.01786$ \citep{Sakamoto2009}, except for the $R$(2) line, which appears to be redshifted by $\sim \SI{100}{\kilo\meter\per\second}$. In the spectral models, we fix the radial velocity to zero. The velocity dispersion is included in the fit as a free parameter to account for broadening due to optical depth. We fit a single-component LTE model to a selected spectral region with the clearest HNC features, masking out the noisy $R$(0) line. The best-fit model is presented in Fig.~\ref{fig:HNC_modelspec}. We find a rotational temperature of $T_\mathrm{rot} = \qty{38\pm4}{\kelvin}$, roughly equivalent to the cold \ce{HCN} component, and a remarkably low background fraction of $f_\mathrm{bg} = 0.06 \pm 0.02$. However, we note that the $R$(0e) and $R$(1e) lines are not well-fit by this model. Given the limited data quality at these long wavelengths, we do not attempt to fit for a second component.

The inferred \ce{HNC} column density is \SI{3e16}{\centi\meter^{-2}}. Given the similar rotational temperatures, we assume that the observed \ce{HNC} traces the same gas as the cold \ce{HCN} component observed at \SI{14}{\micron}, resulting in a column density ratio $N(\ce{HCN})/N(\ce{HNC}) \approx 8$. If we instead take the \ce{HCN} column density range derived from the overtone band, we find $N(\ce{HCN})/N(\ce{HNC}) = \numrange{16}{220}$. Regardless of the \ce{HCN} measurement used, we do not find \ce{HNC} enhancement.

\begin{figure*}[h]
    \centering
    \includegraphics[width=\linewidth]{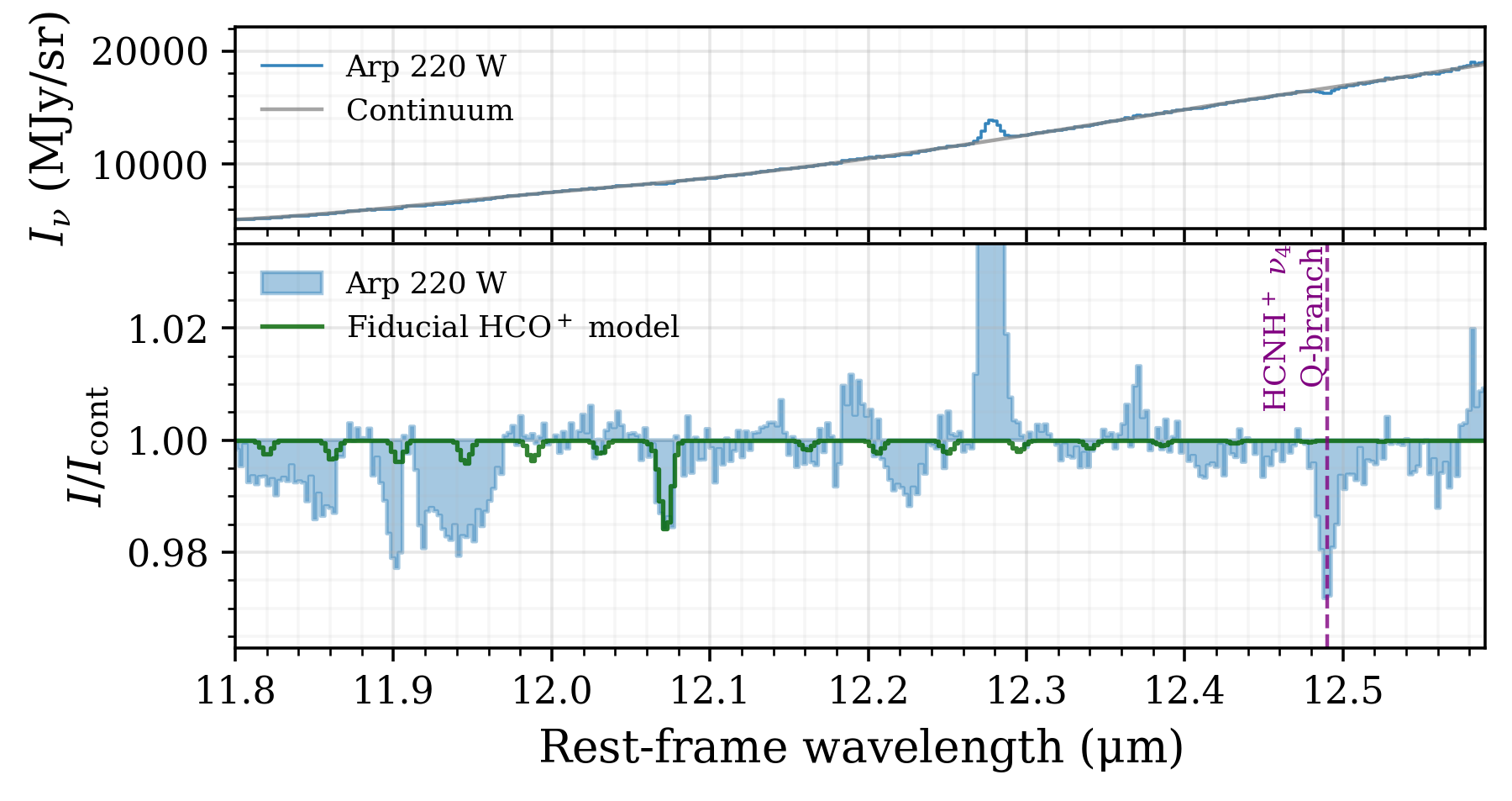}
    \caption{Fiducial model for the tentative \ce{HCO+} band at \SI{12.1}{\micron}. This model has a temperature of $T = \SI{50}{\kelvin}$, a column density of $N(\ce{HCO+}) = \SI{5e15}{\centi\meter^{-2}}$, a velocity dispersion of $\sigma_V = \SI{75}{\kilo\meter\per\second}$, and a background fraction of $f_\mathrm{bg} = 0.28$, matching that found for \ce{HCN} and \ce{C2H2} at \SI{14}{\micron}. The detected $\nu_4$ band of protonated hydrogen cyanide is indicated as well.}
    \label{fig:HCOp_fidmodelspec}
\end{figure*}

Another species of interest is \ce{HCO+}. The submillimetre \ce{HCN}/\ce{HCO+} line ratios have been proposed to be an AGN indicator \citep[e.g.][]{Kohno2001, Krips2008, Imanishi2009, Manohar2017, Butterworth2022, Nishimura2024}. \ce{HCO+} has its strongest rovibrational band at \SI{12.1}{\micron}. We investigate this spectral region, and find a small apparent 1\% dip at the wavelength of the \ce{HCO+} $Q$-branch, as shown in Fig.~\ref{fig:HCOp_fidmodelspec}. However, the absorption depth is similar to that of unexplained noise peaks in this region, making it difficult to solidly identify it with \ce{HCO+}. The \SI{11.8}{\micron}-\SI{12.6}{\micron} is shown alongside a fiducial \ce{HCO+} model in Fig.~\ref{fig:HCOp_fidmodelspec}. We note that the lack of a clear \ce{HCO+} detection implies a relatively low column density compared to that of the dominant \SI{50}{\kelvin} \ce{HCN} component, with an abundance ratio $\ce{HCN}/\ce{HCO+} \gtrsim 50$.

At \SI{12.5}{\micron}, we find a deeper absorption feature, which we attribute to the $Q$-branch of the \ce{HCNH+} $\nu_4$ band \citep{Tanaka1986}. Protonated hydrogen cyanide is a direct precursor of both \ce{HCN} and \ce{HNC}, and its only previous extragalactic detection was through a rotational line in the nearby starburst galaxy NGC 253 \citep{Harada2024}. In the absence of an available line list for the $\nu_4$ band, we cannot derive a column density or temperature from this band, but we note that the lack of a blue wing to the $Q$-branch is suggestive of a relatively low rotational temperature ($T_\mathrm{rot} \lesssim \SI{100}{\kelvin}$).

\subsection{\ce{C2H} and \ce{NO}} \label{sec:res_C2H}
\begin{figure*}[h]
    \centering
    \includegraphics[width=\linewidth]{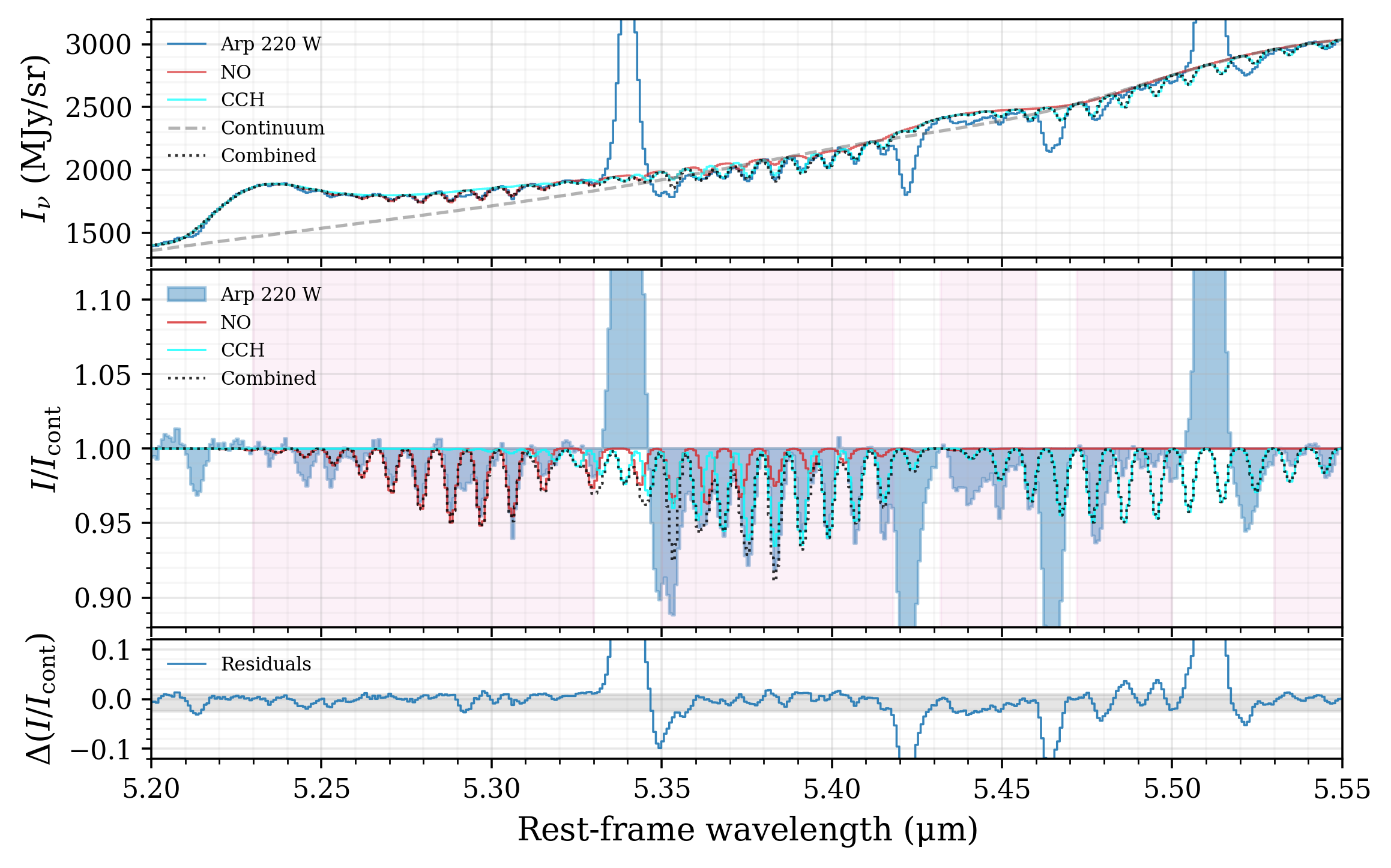}
    \caption{Best-fit LTE model for \ce{NO} and \ce{C2H} at \SI{5}{\micron}, with $f_\mathrm{bg} = 1$. The top panel shows the absorption model with the continuum and PAH emission, the middle panel shows the absorption spectrum, and the bottom panel shows the residuals on the absorption spectrum. Pink shaded areas indicate the spectral regions used in the fit. The grey shaded area in the bottom panel indicates the 68\% interval of the residuals. All model spectra are rebinned onto the spectral axis of the observed spectrum to enable a more precise visual comparison to the data.}
    \label{fig:NO_CCH_modelspec}
\end{figure*}

On the far blue end of the \ce{H2O} band (at \SI{5.4}{\micron}; see Figures \ref{fig:H2O_model_spec} and \ref{fig:NO_CCH_modelspec}), we find a series of regularly spaced absorption lines. We identify these lines with \ce{NO} and \ce{C2H}. As their bands overlap, we fit a model for these two species simultaneously, masking out the strong [Fe~II] and \ce{H2} S(7) emission lines and several \ce{H2O} absorption lines. The continuum is affected by weak PAH emission peaks at \SI{5.25}{\micron} and \SI{5.535}{\micron}; both of these bands have recently also been detected in the Orion Bar \citep{Chown2024}. Assuming that this emission fully arises in the foreground, we subtract it from the observed spectrum before dividing out the continuum. We cannot constrain the background fraction here as the \ce{NO} $Q$-branch completely overlaps with the [Fe~II] line; therefore, we fix the background fraction to 1 and 0.2 to find the lower and upper limits to the column density. For both species, we take a radial velocity of \SI{-50}{\kilo\meter\per\second} and a velocity dispersion of \SI{75}{\kilo\meter\per\second}. The best-fit model for $f_\mathrm{bg} = 1$ is shown in Fig.~\ref{fig:NO_CCH_modelspec}.

\section{Discussion} \label{sec:discussion}
\subsection{Emission effects} \label{sec:emission_effects}
\subsubsection{Interpretation of the background fraction} \label{sec:discussion_fbg}
Although we can only constrain the background fraction $f_\mathrm{bg}$ for bands with clearly saturated features (i.e. \ce{HCN}, \ce{C2H2}, \ce{HNC}, and to a lesser extent \ce{CH4} and \ce{H2O}; see Figures \ref{fig:14micron_model_spec}, \ref{fig:H2O_model_spec}, \ref{fig:CH4_C2H2_modelspec}, and \ref{fig:HNC_modelspec}), the inferred values have interesting implications. First, we note that, where constrained, the values are decidedly below unity, suggesting that overall there is considerable foreground dust adding to the total continuum flux and diluting the detected absorption signal. Second, we find a particularly low background fraction of only $f_\mathrm{bg} = 0.06$ for \ce{HNC} at \SI{21.6}{\micron}, our longest-wavelength band.

We propose the following scenario to explain these two observations: the background continuum against which most of the molecular absorption features arise is that of an embedded hot $T \sim \SIrange{500}{1000}{\kelvin}$ dust component, for which the flux decreases with increasing wavelength at $\lambda \gtrsim \SIrange{3}{6}{\micron}$. The observed continuum, however, rises up to $\lambda \sim \SI{100}{\micron}$, due to the contribution of cooler dust, which becomes dominant for $\lambda \gtrsim \SI{8}{\micron}$ \citep{Armus2007}. The relative contribution of this cool component increases with wavelength, resulting in more dilution of the absorption at longer wavelengths, and hence a lower derived background fraction. This hypothesis also offers a natural explanation for the non-detection of \ce{CO2} ice at \SI{15}{\micron}, despite the strength of the corresponding band at \SI{4.2}{\micron} \citep[][{see also Fig.~\ref{fig:fullspec}}]{Perna2024}.

\begin{figure}[ht]
    \centering
    \includegraphics[width=\linewidth]{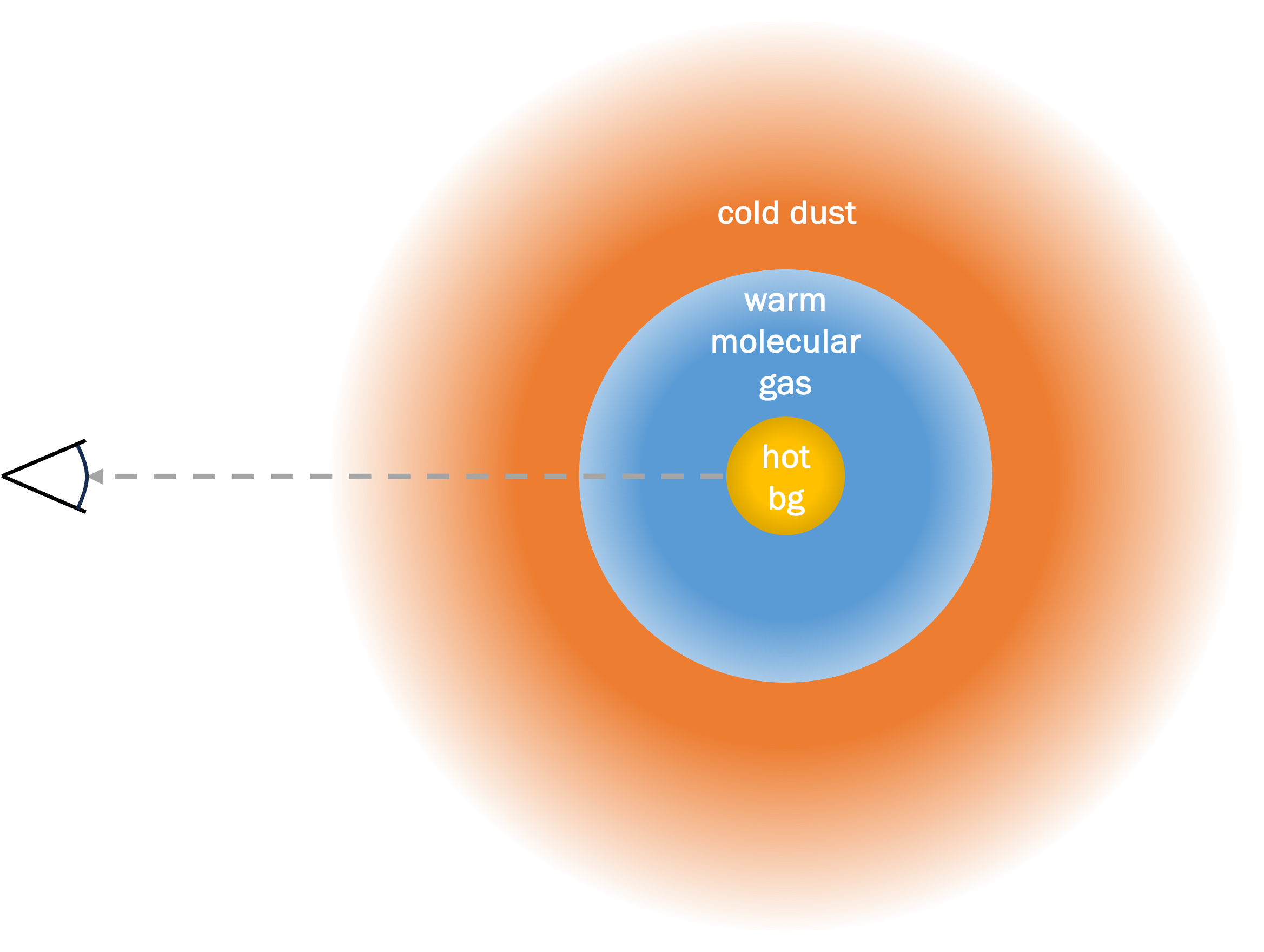}
    \includegraphics[width=\linewidth]{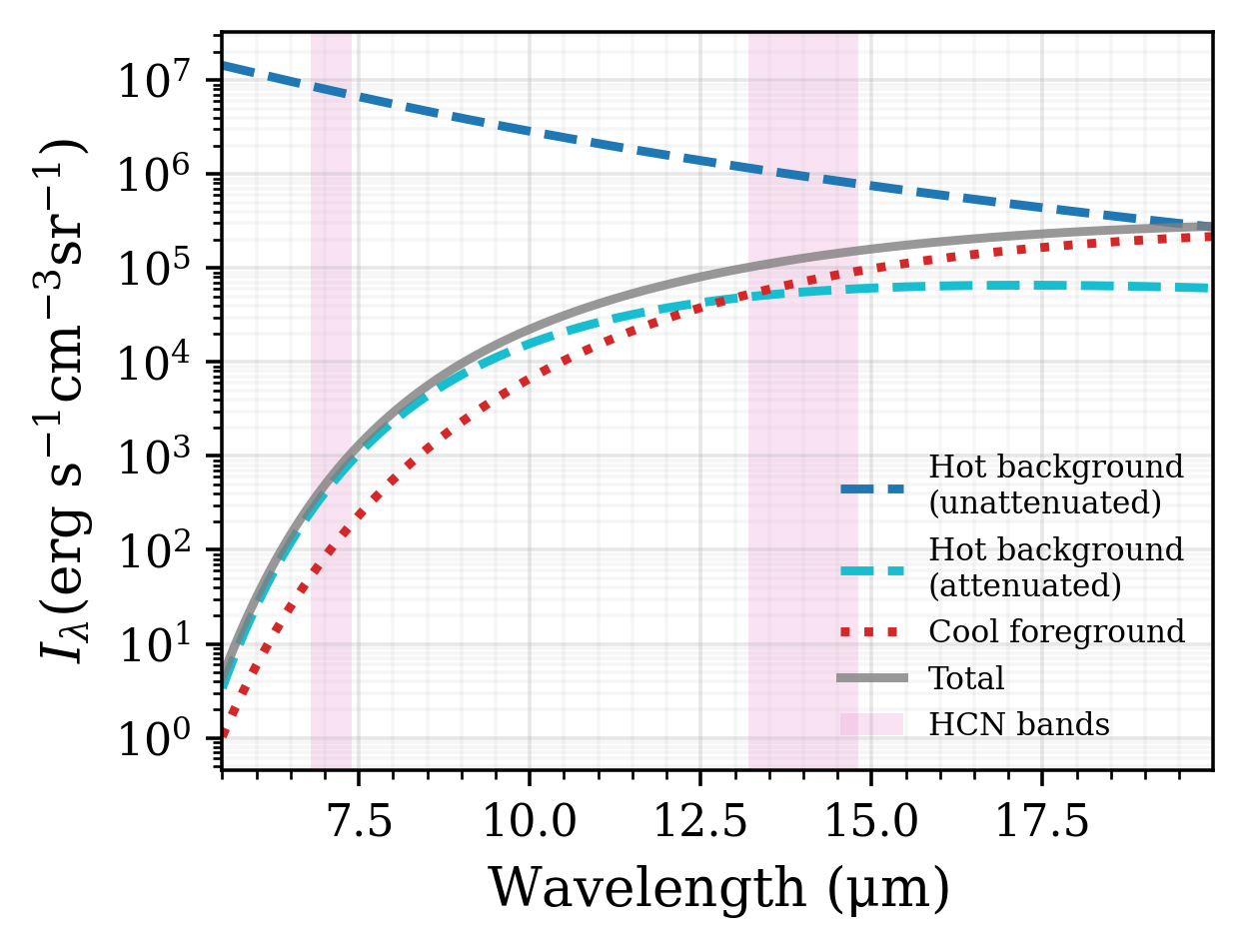}
    \caption{Top: schematic illustration of the proposed line-of-sight geometry. The molecular gas is embedded in cool dust that produces significant foreground emission, filling in the features produced by absorption of hot background photons. Bottom: expected continuum spectrum for a toy model consisting of a \SI{1000}{\kelvin} background blackbody and a \SI{100}{\kelvin} foreground modified blackbody.}
    \label{fig:geometry_los_illustration}
\end{figure}

To illustrate this mechanism, we construct a toy model where the observed continuum consists of a hot \SI{1000}{\kelvin} background blackbody and a cool \SI{100}{\kelvin} foreground modified blackbody. The foreground dust both attenuates the background and produces emission itself. A shell of molecular gas is located in between the hot background and the cool foreground. In the following we consider \ce{HCN} gas in this intermediate layer, adopting a column density of $N(\ce{HCN}_{\nu_2 = 0}) = \SI{2e18}{\centi\meter^{-2}}$, a vibrational temperature of $T_\mathrm{vib} = \SI{440}{\kelvin}$, and a rotational temperature of $T_\mathrm{rot} = \SI{150}{\kelvin}$, based on independent submillimetre studies (see Section \ref{sec:res_HCN_overtone}). The optical depth of the foreground dust is parametrised by a power law $\tau_\mathrm{d} = (\lambda/\lambda_0)^{-\beta}$, with index $\beta = 1.8$ and reference wavelength $\lambda_0 = \qty{25}{\micron}$, where the dust becomes optically thick. This scenario and its associated continuum spectrum are illustrated in Fig.~\ref{fig:geometry_los_illustration}. In this simple toy model, we do not include silicate features or PAH bands.

\begin{figure}[ht]
    \centering
    \includegraphics[width=\linewidth]{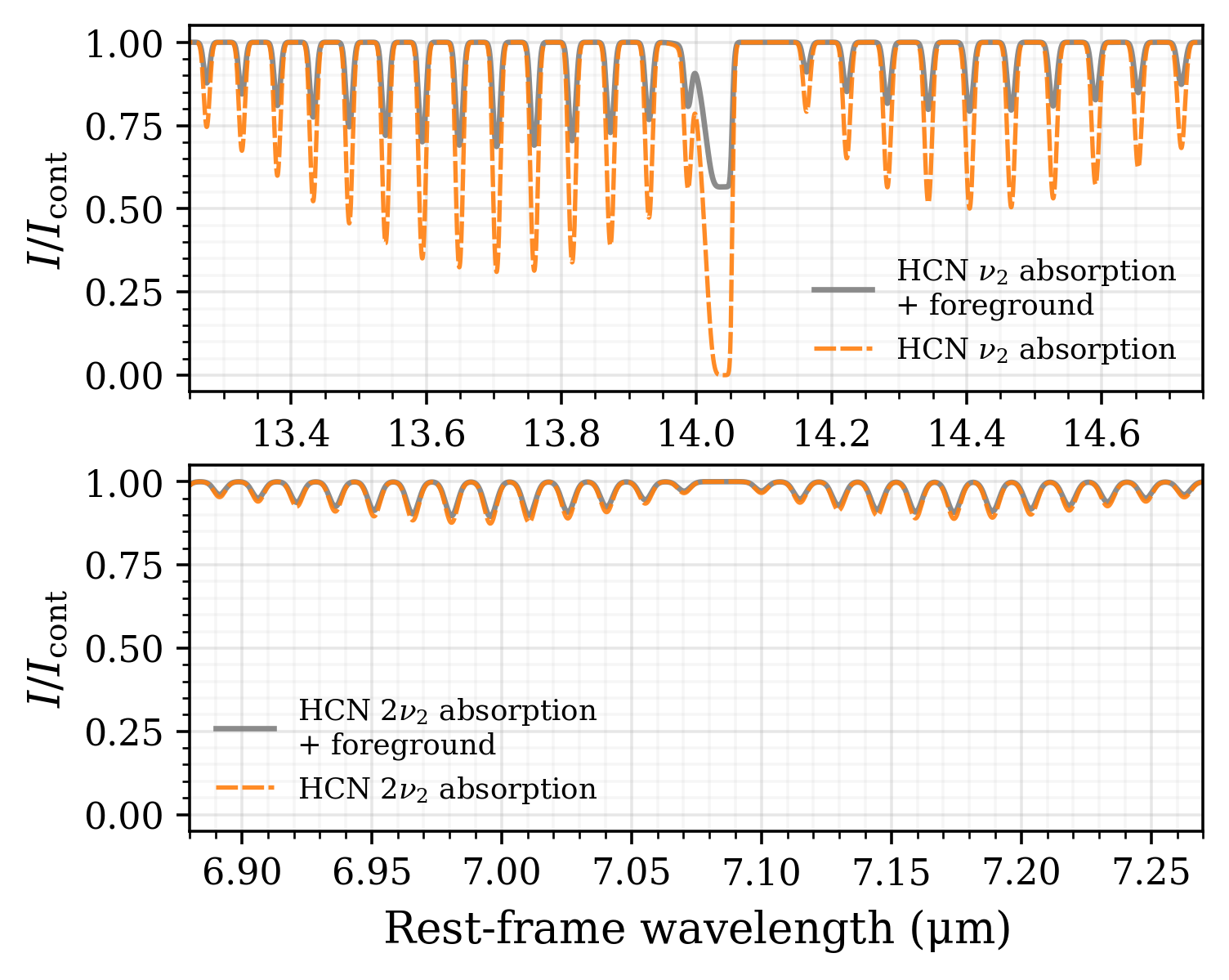}
    \caption{\ce{HCN}$\nu_2$ and $2\nu_2$ absorption bands relative to the observed continuum in the toy model of Fig. \ref{fig:geometry_los_illustration}). A column density of \SI{2e18}{\centi\meter^{-2}} and a rotational temperature of \SI{150}{\kelvin} were used.}
    \label{fig:toy_model_HCN_nu2_absorption}
\end{figure}

As the hot background spectrum is bluer, it can dominate the total observed continuum at \SI{7}{\micron}, while contributing only a small fraction at \SI{14}{\micron}, depending on the exact optical depth spectrum of the foreground dust and the beam filling factors of both dust components. If we now consider the \ce{HCN} $\nu_2$ and $2\nu_2$ bands, we find that, relative to the total observed continuum, the \SI{7}{\micron} $2\nu_2$ band does not noticeably change, while the \SI{14}{\micron} $\nu_2$ band is diluted considerably. This effect is demonstrated in Fig.~\ref{fig:toy_model_HCN_nu2_absorption}. However, even for $f_\mathrm{bg} = 1$, the column densities derived from the \ce{HCN} and \ce{C2H2} bands at \qty{7}{\micron} still exceed those derived from the \qty{14}{\micron} bands by a factor 2 (see Table~\ref{tab:properties_summary} and Sections \ref{sec:res_14um}, \ref{sec:res_8micron} and \ref{sec:res_HCN_overtone}), so foreground dust continuum alone cannot account for this discrepancy.

For similarly obscured starburst galaxies, spectral decompositions using various codes have demonstrated that much of the apparent continuum from \SIrange{5}{9}{\micron} consists of broad overlapping PAH profiles \citep[e.g.][]{Smith2007, Lai2022, Donnan2024}. If this emission arises largely in the foreground, the background fraction we derive is suppressed at these wavelengths. The strong \SI{7.7}{\micron} PAH features further complicate the analysis, as they fill in the \ce{CH4} $Q$-branch much more than the corresponding $R$-branch lines, severely distorting the absorption spectrum and thereby decreasing the inferred background fraction. As such, all properties derived from the \ce{C2H2} $\nu_4 + \nu_5$ and \ce{CH4} bands should be treated with caution.

\subsubsection{Potential molecular line emission} \label{sec:discussion_line_emission}
Throughout our analysis, we have assumed that the lines are pure absorption lines. In the following we consider the effects of possible molecular line emission partly filling in the absorption lines, in the context of the toy model described in Section~\ref{sec:discussion_fbg}.

Both line emission and foreground continuum emission can suppress absorption features; these effects are degenerate unless the line emission overtakes {the background continuum}. All rovibrational bands studied in this work appear in absorption: {across the full spectral range, the incident continuum radiation is more intense than the molecular line emission.} 
{However, if the background continuum is not completely dominant, hidden line emission partially fills in the absorption features. In this case, the background fraction derived from optically thick features, under the assumption of pure absorption, is an underestimate, as the absorption floor would be deeper in the absence of line emission. Nevertheless, the high vibrational temperatures ($T_\mathrm{vib} \gtrsim \qty{200}{\kelvin}$) required to produce appreciable rovibrational emission can only be achieved in the presence of a strong local mid-IR radiation field. Thus, any molecular gas for which line emission affects our analysis, must still be located close to the nucleus.}

Whether a rovibrational line is observed in emission or absorption is determined by the competition between the intensity of a blackbody at the vibrational temperature of the gas on the one hand, and the strength of the true background continuum on the other. If the incoming radiation is that of an unattenuated hot blackbody with $T_\mathrm{bg} > T_\mathrm{vib}$, this competition is more easily won by emission at long wavelengths. Thus, in the proposed scenario ($T_\mathrm{vib} \approx \qty{400}{\kelvin}$; $T_\mathrm{bg} \gtrsim \qty{500}{\kelvin}$), we expect any line emission effects to be {much} greater at \SI{14}{\micron} than at \SI{7}{\micron}. This mechanism could explain the discrepancy in column densities found for \ce{HCN} and \ce{C2H2} between their fundamental and overtone/combination bands: the short-wavelength bands are almost unaffected by emission, while the long-wavelength bands are partially filled in. 
{Additionally, if the vibrational temperature is higher than the rotational temperature, our LTE fits of the absorption bands underestimate the total column density, because the excess population in higher vibrational states is not considered. Both of these effects may play a role in reducing our inferred column densities and producing the apparent discrepancy between \qty{7}{\um} and \qty{14}{um}.}

\begin{figure}
    \centering
    \includegraphics[width=\linewidth]{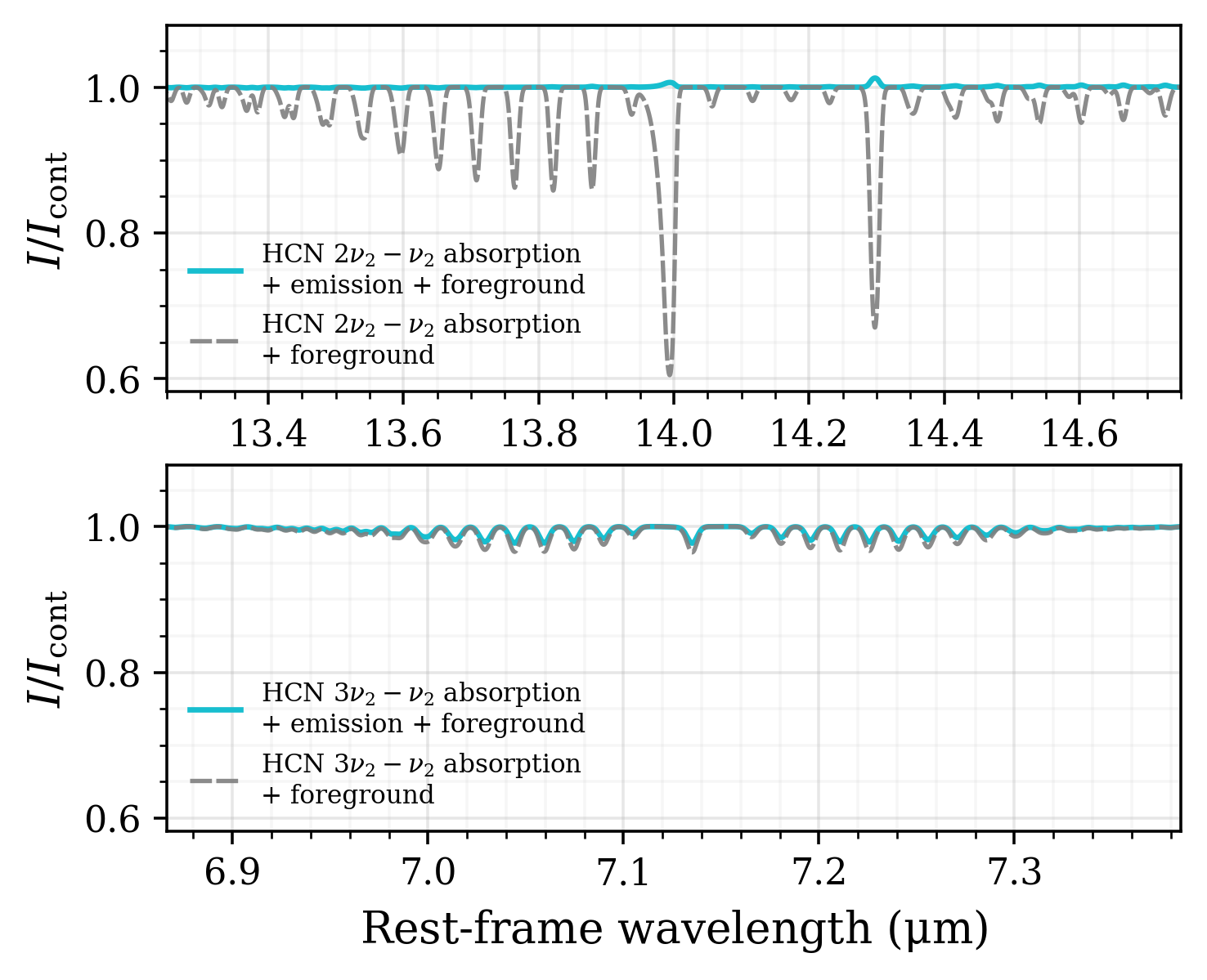}
    \caption{\ce{HCN} $2\nu_2 - \nu_2$ and $3\nu_2 - \nu_2$ hot bands relative to the observed continuum in the toy model of Fig. \ref{fig:geometry_los_illustration}, for both a pure absorption model and a model including both absorption and emission. A column density of \SI{2e18}{\centi\meter^{-2}}, a rotational temperature of \SI{150}{\kelvin}, and a vibrational temperature of \SI{440}{\kelvin} were used. The modelled \SI{14}{\micron} absorption is completely filled in by line emission, while the \SI{7}{\micron} band is almost unchanged.}
    \label{fig:toy_model_HCN_hotbands_abs_em}
\end{figure}

The tentative detection of the \ce{HCN} $3\nu_2 - \nu_2$ hot band (Section \ref{sec:res_HCN_overtone}) can be explained in a similar fashion. Depending on the exact beam filling factors, temperatures and column density, it is possible to fill in all detectable $2\nu_2 - \nu_2$ absorption at \SI{14}{\micron} while leaving the $3\nu_2 - \nu_2$ hot band almost unchanged. An example is shown in Fig. \ref{fig:toy_model_HCN_hotbands_abs_em}. We note that the tentatively detected $3\nu_2-\nu_2$ hot band displays a prominent $R$-branch, weak $Q$-branch and no obvious $P$-branch. As noted in Section~\ref{sec:res_6um}, this branch asymmetry reveals the presence of emission filling in the absorption lines, with the strongest effect on the $P$-branch, weaker effect on the $Q$-branch and the weakest effect on the $R$-branch. Emission from the $\nu_2 = 3$ level (\qty{3047}{\kelvin} above the ground state) could reveal the presence of a very hot radiation field populating the $\nu_2 = 3$ level. However, quantitative modelling of this situation requires a sophisticated non-LTE approach {and a detailed treatment of all emission effects}, which is beyond the scope of the present paper.

In summary, the variety in inferred background fractions, the difference in column densities between the \SI{7}{\micron} and \SI{14}{\micron} bands of \ce{HCN} and \ce{C2H2}, and the potential \ce{HCN} hot band detection at \SI{7}{\micron} without the corresponding \SI{14}{\micron} hot band can all be naturally explained if the background continuum that produces the molecular absorption features is that of a hot central source. The implication is that the bulk of the molecular gas is located deep inside the nuclear region, embedded in the cooler dust that produces most of the bolometric luminosity. This physical concept is illustrated in Fig.~\ref{fig:geometry_los_illustration}. {We further emphasise that, despite our treatment of the background fraction, our LTE pure absorption analysis may still underestimate the column densities if the vibrational temperatures are high.}

\subsection{Properties of the embedded molecular gas}   \label{sec:categories}

We detect a number of species in the gas phase, for which we derive different rotational temperatures, as summarised in Table~\ref{tab:properties_summary}. As discussed in Section~\ref{sec:emission_effects}, the inferred background fraction is below unity wherever it is constrained, suggesting that the bulk of the molecular gas is located behind large amounts of cooler dust, and must therefore be close to the nucleus itself. We stress that for this nuclear gas, the inferred rotational temperatures may not reflect the actual gas temperatures. Both subthermal excitation and radiative excitation may play a role, particularly for species with high (rotational) critical densities. Nevertheless, the rotational temperature offers a useful way to categorise our detections.

We distinguish four groups based on rotational temperature: cold $T_\mathrm{rot} \lesssim \SI{60}{\kelvin}$ gas (\ce{CO}, \ce{HCN}, \ce{CO2}, \ce{NO}, \ce{HNC}); moderately warm $T_\mathrm{rot} \sim \SI{150}{\kelvin}$ gas (\ce{CO}, \ce{C2H2}, \ce{HCN}, \ce{C2H}); and warm $T_\mathrm{rot} \sim \SI{300}{\kelvin}$ gas (\ce{H2O}, \ce{CH4}, \ce{HCN}). The \ce{CO} fundamental band is dominated by a hot \SI{650}{\kelvin} component, and it is the only band for which we derive such a high temperature. Given the combination of a high temperature and broad ($\sigma_V \approx \SI{190}{\kilo\meter\per\second}$) line widths, and the similarity to the \ce{CO} band detected towards the Galactic SNR Cas~A \citep{Rho2024}, we posit that this gas is most likely shocked. We note that the contributions of the two other \ce{CO} components are largely degenerate, and therefore the column densities and excitation temperatures derived for the cold and warm \ce{CO} suffer from large systematic uncertainties.

All species detected with cold rotational temperatures are simple molecules that can form efficiently in the gas and/or sublimate from ices. \ce{CO} is ubiquitous in the nuclear region of Arp~220, as confirmed by its large emission size in high-resolution ALMA studies \citep{Sakamoto2021b}, and can trace molecular gas in almost any physical environment. For \ce{HCN} and \ce{HNC}, the low-$J$ levels are possibly subthermally excited, as they have high critical densities ($n_\mathrm{crit}(\ce{HCN}) \sim 10^{6}-\SI{e7}{\centi\meter^{-3}}$) and do not couple to the radiation field as easily as the high-$J$ levels.

The molecular ions detected in this work---\ce{N2H+}, \ce{HCO+} and \ce{HCNH+}---do not have well-constrained rotational temperatures. However, their $Q$-branches appear to be symmetric, and we therefore tentatively group them under the cold components. The cosmic ray ionisation rate (CRIR) in the WN of Arp~220 is orders of magnitude above the Milky Way average \citep{Pereira-Santaella2024b}, so fairly large columns of molecular ions are not unexpected.

Several species are detected with a moderately warm $T \sim \SI{150}{\kelvin}$ rotational temperature, without a clear colder component. \ce{C2H2} in particular has a well-constrained excitation temperature of $\qtyrange{150}{180}{\kelvin}$ in both detected bands. Its direct parent species \ce{C2H} is found at a similar rotational temperature. \ce{C2H2} has no allowed rotational transitions, so its rotational temperature probes either the gas temperature or the local $\qtyrange{7}{14}{\micron}$ radiation field. \ce{C2H} has rotational critical densities of $\gtrsim \SI{e6}{\centi\meter^{-2}}$ \citep{Nagy2015}, so we consider it likely that radiative excitation plays an important role for this molecule.

A moderately warm \SI{110}{\kelvin} component is found for \ce{HCN} in the $2\nu_2$ overtone band at \SI{7.1}{\micron}. Given the many processes capable of forming \ce{HCN}, its appearance in a moderately warm phase is not surprising, and the \ce{HCN} overtone band may trace the same gas as the \ce{C2H2}. Alternatively, the \SI{50}{\kelvin} and \SI{330}{\kelvin} \ce{HCN} components identified at \SI{14}{\micron} could be blended into one apparent \SI{110}{\kelvin} component in the overtone band, depending on their true column densities and background fractions at \SI{7}{\micron}. Due to the limited absorption depth and the contamination in this band, we did not attempt to fit for more than one \ce{HCN} component here.

The warm components of \ce{H2O}, \ce{CH4} and \ce{HCN} stand out. For \ce{H2O}, LTE models could not sufficiently reproduce the observed spectrum, and the derived excitation temperature should be interpreted with caution; for \ce{HCN}, however, a \SI{330}{\kelvin} component with $N(\ce{HCN}) \geq \SI{1e17}{\centi\meter^{-2}}$ is decidedly needed to account for the blue wing of the $Q$-branch. These high-$J$ levels ($J \gtrsim 10$) are most likely radiatively excited. Unlike \ce{HCN}, \ce{CH4} only appears in a warm $T_\mathrm{rot} \approx \qty{300}{\kelvin}$ component, although its exact rotational temperature is uncertain due to the strongly peaked PAH feature its band lies on top of. Like \ce{C2H2}, methane lacks a permanent dipole moment and must therefore probe either the kinetic gas temperature or the local radiation temperature at \SI{7.7}{\micron}.

\subsection{A multi-wavelength view of \ce{HCN} in Arp~220} \label{sec:discussion_HCN_comparison}

The molecular gas in Arp~220 has been extensively studied, from centimetre wavelengths to the mid- and near-infrared. In general, emission line studies with single-dish telescopes derive lower column densities than we do, {due to their large beam sizes}. Even with interferometers and at the highest resolutions, it is challenging to separate relatively cool, more extended gas from the warm gas in the inner nucleus through rotational emission lines. Absorption studies, on the other hand, benefit from the pencil beam effect, restricting detection of gas to a narrow line of sight towards the compact effective background source. In the following, we therefore compare our derived molecular gas properties to other works that were able to probe the compact inner region. The focus of this discussion is on \ce{HCN}, as this species has been especially well-targeted in previous observations of Arp~220.

The \ce{C2H2}, \ce{HCN} and \ce{CO2} absorption bands at $\sim \SI{14}{\micron}$ were previously detected in the Spitzer IRS spectrum of Arp~220 \citep{Lahuis2007}. These observations could not separate the two nuclei, and the spectral resolution was too low to detect the individual $P$- and $R$-branch lines. The $Q$-branches of all three species were analysed together, assuming LTE conditions, a single common excitation temperature, and a unity covering factor. \citet{Lahuis2007} derived an excitation temperature of \SI{250}{\kelvin}, with an uncertainty of up to 30\%. They found column densities of \SI{1.7e16}{\centi\meter^{-2}}, \SI{2.9e16}{\centi\meter^{-2}} and \SI{0.7e16}{\centi\meter^{-2}} for \ce{C2H2}, \ce{HCN} and \ce{CO2} respectively.

The column densities inferred here for the WN are an order of magnitude higher than previously reported. This difference is caused by a combination of effects. First, with the MIRI MRS we can now resolve out the two nuclei of Arp~220, and thus remove contaminating continuum emission from the eastern nucleus. Second, the clear detection of $P$- and $R$-branch lines of both \ce{C2H2} and \ce{HCN}, and the saturation of the $Q$-branches, now lifts the degeneracy between column density and covering factor/background fraction. Despite the removal of the eastern nucleus, we still find a background fraction of only $f_\mathrm{bg} = 0.28$, and so the true optical depth of the lines is much higher than previously thought. Using the same methods as \citet{Lahuis2007}, we now derive abundances of $\sim 10^{-6}$ for \ce{C2H2} and \ce{HCN}, and $\sim 10^{-7}$ for \ce{CO2}.

The excitation temperature derived by \citet{Lahuis2007} was highly uncertain due to the limited spectral resolution of the Spitzer IRS. With MIRI MRS, however, the improved spatial and spectral resolving power allows us to decouple the rotational temperatures of the three species, and puts much tighter constraints on them. The velocity dispersion and radial velocity can also be directly measured from the individual lines, removing a source of systematic uncertainty in the modelling. We find that lower rotational temperatures dominate the signal (\SI{50}{\kelvin} for \ce{HCN} and \SI{150}{\kelvin} for \ce{C2H2}), but with a necessary additional \SI{330}{\kelvin} \ce{HCN} component.

This rotational temperature was also derived from high-$J$ rotational lines seen in absorption with \textit{Herschel}/SPIRE \citep{Rangwala2011}, although they inferred a much lower column density of $N(\ce{HCN}) = \SI{2e15}{\centi\meter^{-2}}$. Given the two \ce{HCN} components we find at \SI{14}{\micron}, the high-$J$ lines probed by \citet{Rangwala2011} should be completely dominated by the $T_\mathrm{rot} \sim \SI{300}{\kelvin}$ component, and therefore their rotational temperature is entirely consistent with our results. Their lower derived column density is likely a result of contaminating emission from both the eastern nucleus and cool foreground dust. Indeed, using a more complex model involving multiple continuum sources, \citet{Gonzalez-Alfonso2012} infer much higher column densities of $N(\ce{HCN}) \sim \SI{e17}{}-\SI{e18}{\centi\meter^{-2}}$, consistent with our findings.

\ce{HCN} absorption has also been seen in direct l-type transitions in the $\nu_2 = 1f$ $J = 4$ and $J = 6$ levels at centimetre wavelengths. These suggest a rotational temperature of $T_\mathrm{rot} \sim \SI{150}{\kelvin}$ \citep{Salter2008}. However, these lines show evidence for several velocity components that are not taken into account, and the two nuclei were blended together in their Arecibo observations.

Vibrationally excited \ce{HCN} has been proposed as a way to isolate the inner nucleus in emission studies. A strong mid-infrared radiation field is required to populate these levels, and the subsequent \ce{HCN}-vib rotational emission lines therefore only arise in the most compact, dense and/or warm regions \citep[e.g.][]{Aalto2015, Martin2016}. Towards Arp~220 such lines have been detected for \ce{HCN} and \ce{HC3N}, and \citet{Sakamoto2021b} used a high-resolution ALMA spectral scan to confirm the relatively compact emission sizes of vibrationally excited species. ALMA observations of the \ce{HCN} $\nu_2 = 1f$ $J=3-2$ and $J=4-3$ emission lines indicate a rotational temperature of \SI{38}{\kelvin} in the WN \citep{Martin2016}. For vibrationally excited \ce{HC3N}, of which many more transitions have been detected, \citet{Martin2011} derived a rotational temperature $T_\mathrm{rot} \sim \SI{350}{\kelvin}$, closer to the warm \ce{HCN} component identified in the present work.

In the \SI{14}{\micron} $Q$-branch, the \SI{330}{\kelvin} component is dominant for $J \geq 10$. For these high levels, the rotational critical densities exceed \SI{e8}{\centi\meter^{-3}}. Although the effective critical density will be reduced through photon trapping, for these high values it is unlikely that the levels are collisionally excited. It is more probable that the \SI{330}{\kelvin} rotational temperature is driven by the local radiation field, either through the rotational transitions or through the \SI{14}{\micron} pumping. \citet{Rangwala2011} argue for mid-infrared pumping, requiring a local radiation temperature of $T_\mathrm{rad,\SI{14}{\micron}} > \SI{350}{\kelvin}$. \citet{Gonzalez-Alfonso2012} instead find FIR radiation to dominate the high-$J$ levels, but make note of the dependence on their choice of dust and gas distribution.

The importance of the vibrational and rotational radiative excitation processes for populating the levels depends on the product of the Einstein A coefficient and the photon occupation number $n_\gamma$ at the transition frequency (see \citet{Buiten2024a} for details). For the rovibrational lines at \SI{14}{\micron}, the Einstein A coefficients are $A_\mathrm{rovib} \sim \SI{1}{\per\second}$, and the product $A_\mathrm{rovib} \times n_\gamma(\SI{14}{\micron})$ depends only on the radiation temperature and the dilution factor; it is almost constant across $J$-levels. The rotational transitions, however, lie at a wide range of frequencies, and the strength of the rotational radiative excitation therefore varies between levels. Their Einstein A values range from \SI{2e-5}{\per\second} for $J = 1-0$ to \SI{0.5}{\per\second} for $J = 25-24$.

Considering again the structure of the toy model sketched in Fig. \ref{fig:geometry_los_illustration}, suppose the local radiation field at the position of the warm molecular gas is entirely dominated by the geometrically diluted but unattenuated hot $T_\mathrm{d} \gtrsim \SI{500}{\kelvin}$ background component identified by \citet{Sakamoto2017}. For such a blackbody, the rotational pumping rate will be higher than the vibrational pumping rate for $J \geq 9$; we thus expect rotational radiative pumping to dominate the rotational temperature of high-$J$ levels. In that case, we can derive the dilution factor $f_\mathrm{D}$ directly from the rotational temperature $T_\mathrm{rot}$ and the background temperature $T_\mathrm{bg}$:

\begin{equation}
    f_\mathrm{D} = \frac{e^{E_{ul}/{k_BT_\mathrm{bg}}} - 1}{e^{E_{ul}/{k_BT_\mathrm{rot}}} - 1}
\end{equation}

Here $E_{ul}$ is the energy difference between the levels and $k_B$ is the Boltzmann constant. For $T_\mathrm{rot} = \SI{330}{\kelvin}$ and $\SI{600}{\kelvin} < T_\mathrm{bg} < \SI{1500}{\kelvin}$, we find dilution factors $f_\mathrm{D} = 0.2-0.5$. This result is consistent with the picture that the molecular gas seen in absorption is located at or near the edge of the optically thick background source. The gas cannot be fully embedded in the hot dust itself, as the dilution factor is decidedly below unity, but it must be close to it for the radiation field to still set the rotational temperature to a relatively high \SI{330}{\kelvin}. We therefore propose that the absorbing molecular gas we detect in the mid-infrared is located either in an outer layer of the $\sim \qty{10}{\parsec}$ nucleus identified by \citet{Sakamoto2017}, or in the inner regions of the starburst disk (see Section~\ref{sec:discussion_geometry}).

\subsection{A hot-core-like chemistry} \label{sec:discussion_chemistry}
The wide variety of molecular species detected in this work and the high signal-to-noise ratios on their features provide several new insights into the chemistry of the WN. In the following we characterise this chemistry. Given the extraordinarily high optical depths derived for the WN of Arp~220, as well as the layered structure of the dust continuum, it is unclear whether the \SI{9.8}{\micron} silicate depth is a good indicator of the total \ce{H2} column density traced by the various molecular absorption bands. We therefore refrain from presenting abundances with respect to \ce{H2} for these species. Instead, we consider column density ratios between the species studied in this work, and compare these to abundance ratios predicted by chemical models. However, we cautiously note that the medium is not uniform along the line of sight, and thus the column densities of different species may be dominated by different physical environments. We further emphasise that the derived rotational temperatures do not necessarily reflect the temperature of the gas.

We find the highest column densities for \ce{CO} and \ce{H2O}: both have components with column densities $\sim \SI{e19}{\centi\meter^{-2}}$. \ce{CH4}, \ce{HCN}, \ce{C2H2}, \ce{C2H} and \ce{NO} are all found with column densities 10-100 times smaller. This implies elevated abundances: in cold gas, \ce{HCN} and \ce{C2H2} are typically $10^{-4}$ times as abundant as \ce{CO}. We find considerably lower column densities for \ce{HNC}, \ce{CO2}, \ce{N2H+}, and our tentative \ce{HCO+} and \ce{HC3N} detections, with ratios to \ce{CO} of $\sim \numlist[list-exponents=individual]{e-3;e-3;e-4;e-4;e-3}$ respectively.

The chemical footprint of the WN is remarkably similar to that of Galactic hot cores; this resemblance is a common feature of compact obscured nuclei \citep[CONs; e.g.][]{Lahuis2007, Aalto2015, Falstad2015, Costagliola2015, Gorski2023}. In hot core envelopes, ice evaporation and warm gas-phase chemistry can efficiently produce \ce{C2H2}, \ce{HCN} and \ce{CH4}, enhancing their gas-phase abundances by orders of magnitude at temperatures $T > \SI{200}{\kelvin}$ \citep[e.g][]{Doty2002}. While Galactic massive protostellar objects often show particularly elevated \ce{CO2} abundances due to its efficient formation through a reaction between \ce{CO} and \ce{OH} \citep[e.g.][]{Boonman2003b, Francis2024}, the \ce{CO2} column density derived here for the Arp~220 WN is an order of magnitude lower than that of \ce{HCN} and \ce{C2H2}. This suggests that the gas temperature is $T > \SI{250}{\kelvin}$, at which temperature \ce{OH} is rapidly driven into \ce{H2O}, blocking the main gas-phase formation route of \ce{CO2} \citep[e.g][]{vanDishoeck2023}. The large columns of \ce{H2O} detected in this work are further evidence of this scenario.

The temperatures inferred, however, are lower than those required to efficiently produce the derived density ratios. \ce{C2H2}, which is expected to be collisionally excited as it has no radiative pure rotational transitions, is found at only $T_\mathrm{rot} \approx \qty{150}{\kelvin}$. We do derive $T_\mathrm{rot} \approx \qty{300}{\kelvin}$ for \ce{CH4}, but we do not consider this measurement reliable due to the overlap of this band with the \qty{7.7}{\micron} PAH complex (see Sections \ref{sec:res_8micron} and \ref{sec:discussion_fbg}). \citet{Lahuis2007} suggested that the \ce{HCN}, \ce{C2H2} and \ce{CO2} abundances could be produced in hotter ($\gtrsim \qty{800}{\kelvin}$) gas, which is then diffused to cooler layers by winds or turbulent motions. This mechanism requires the chemical timescale to be longer than the mixing timescale. However, under the greatly enhanced CRIR measured for the Arp~220 WN \citep{Pereira-Santaella2024}, the chemistry is expected to be rapid. Furthermore, the higher column densities derived in the present work translate to a larger path length than estimated by \citet{Lahuis2007}. Thus, we consider it unlikely that the hot gas can be transported to cooler regions sufficiently rapidly to fully explain the observed chemical signature.

Shocks could also play a role in decreasing the \ce{CO2} abundance. \citet{CharnleyKaufman2000} demonstrated that C-type shocks can efficiently destroy \ce{CO2} to form \ce{CO}, \ce{H2O} and \ce{OH}. In these models, the $[\ce{H2O}]/[\ce{CO}]$ abundance ratio stabilises at $\sim 1$ in the post-shock gas, similar to the column density ratio we derive for the Arp~220 WN. If this shocked gas subsequently collapses to form protostars and undergoes hot core chemistry, the \ce{CO2} abundance will increase again, but will not reach the high levels predicted for pure hot core chemistry. In a merger, where large amounts of gas are driven into the nucleus, this scenario is not unexpected. Evidence for shocks in Arp~220 is plentiful: several lines of \ce{SiO} have been detected towards the WN \citep[e.g.][]{Wheeler2020, Sakamoto2021b}, the starburst disks contain numerous SNRs, and we attribute the $T_\mathrm{rot} \approx \qty{700}{\kelvin}$, $\sigma_V \approx \qty{190}{\kilo\meter\per\second}$ \ce{CO} component found in this work to shocks as well. We therefore conclude that this combination of shocks and high-mass star formation may be responsible for the comparative lack of \ce{CO2} found in Arp~220 and other U/LIRGs \citep{Lahuis2007}.

Another striking similarity to hot cores is found in the ortho-to-para ratio of \ce{C2H2}, for which we have tight constraints due to the high-SNR detection of individual $R$- and $P$-branch lines in the $\nu_5$ band. This OPR of $1.71 \pm 0.05$ is decidedly below the value of 3 expected for gas in equilibrium at high temperatures, but it coincides with those found for the Orion IRc2 and AFGL~2136~IRS1 hot cores \citep{Rangwala2018, Barr2020}. In Orion IRc2, the low OPR was proposed to be a remnant of an earlier cold phase, as gas-phase reactions are slow to change the OPR.

Contrary to \citet{Aalto2007}, we find an \ce{HCN}/\ce{HNC} ratio of $10-10^2$, well above unity. This result suggests that the reported overluminous \ce{HNC} is an excitation effect rather than a sign of X-ray driven chemistry. For the center of the starburst galaxy NGC~253, \citet{Behrens2022} derived \ce{HCN}/\ce{HNC} ratios much closer to unity, which they attributed to the efficient formation of both species from \ce{HCNH+} at very high cosmic ray ionisation rates (CRIR). As the CRIR in the WN of Arp~220 is similarly high \citep{Pereira-Santaella2024b}, the high \ce{HCN}/\ce{HNC} could indicate that mechanical heating plays a role in the chemistry \citep[e.g.][]{Meijerink2011}. This picture is consistent with our hypothesis that both shocks and high-mass star formation are relevant chemical drivers in the WN of Arp~220.

\subsection{The geometry of the western nucleus} \label{sec:discussion_geometry}

High-resolution radio and (sub)millimetre studies have led to a detailed a picture of the structure of the WN. SNe, dust continuum, and molecular lines all trace an inclined, rotating starburst disk with a major axis size (FWHM) of $\sim \qty{100}{\parsec}$ \citep{Varenius2019, BarcosMunoz2015, Scoville2017, Sakamoto2017, Sakamoto2021a}. At the center of the disk, molecular rotational lines exhibit narrow absorption features, and the continuum shows a more symmetric hot dust peak with a FWHM size of $\sim \qty{20}{\parsec}$ and significant opacity up to wavelengths of $\sim \qty{3}{\milli\meter}$ \citep{Scoville2017, Sakamoto2017}. We interpret this additional component as a distinct opaque inner nucleus. A fast molecular outflow, spanning $\sim \qty{200}{\parsec}$ across in projection, was identified by \citet{BarcosMunoz2018}; it has a blueshifted lobe to the south of the nucleus and a redshifted lobe to the north. This biconical outflow appears to be normal to the disk in the plane of the sky.

\begin{figure}[h]
    \centering
    \includegraphics[width=\linewidth]{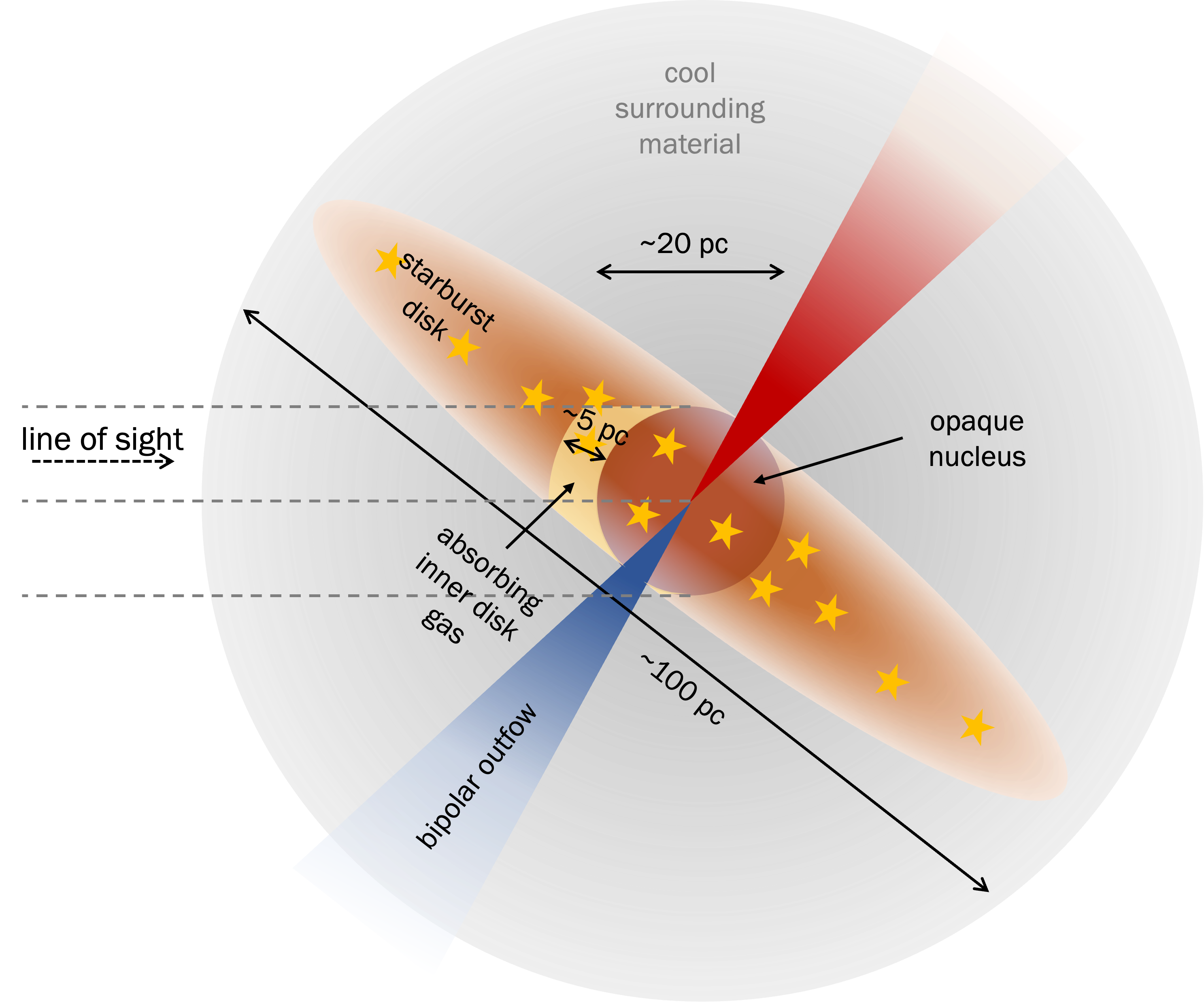}
    \caption{Schematic illustration of the proposed geometry of the WN along the line of sight. Following the interpretation of \citet{Sakamoto2021b}, we assume that the south side of the disk is facing us with an inclination of $i \approx \ang{60}$, and that the biconical outflow is normal to the disk. The direction of the tilt is debated \citep[][Van der Werf et al. in prep]{Scoville1998}, but we note that it does not affect the interpretation of the absorption bands in this work.}
    \label{fig:WN_geometry_cartoon}
\end{figure}

The ratio of the major and minor axis of the disk suggests an inclination of $\gtrsim \ang{52}$ \citep{BarcosMunoz2015, Sakamoto2017}. However, there is no clear consensus as to the direction of the disk tilt: infrared observations suggest that the north side of the disk is facing us \citep[][Van der Werf et al. in prep]{Scoville1998}, while millimetre studies find the opposite \citep{Tunnard2015, Sakamoto2021b}. Likewise, the inclination of the outflow is not well-constrained, and it is unclear whether it is normal to the disk or slanted. In the following discussion, we adopt the interpretation of \citet{Sakamoto2021b}, who argue for an outflow orientation perpendicular to the disk, with the south side of the disk facing us. This configuration is illustrated in Fig.~\ref{fig:WN_geometry_cartoon}.

Estimates of the geometric thickness of the disk are limited. \citet{Scoville2017} use kinematic modelling of \ce{CO} (1-0) emission to calculate the disk thickness as a function of radius for the case of a fully self-gravitating disk. However, they assume an inclination of only \qty{30}{\degree}, underestimating the effect of outflowing material on the measurement of the minor axis size. Correcting for the inclination, we obtain a mean disk thickness of $\sim \qty{20}{\parsec}$ (FWHM)---approximately the size of the inner nucleus. This estimate suggests that the inner core is (nearly) entirely embedded in the surrounding starburst disk, as sketched in Fig.~\ref{fig:WN_geometry_cartoon}.

Under the assumption that the opaque nucleus is the dominant background source for our MIR absorption bands, Fig. \ref{fig:WN_geometry_cartoon} illustrates that only a small fraction of the disk falls within our detection beam. We estimate the path length along which we detect our warm molecular gas through from the derived \ce{HCN} column density of \qtyrange[range-exponents=individual]{e17}{e18}{\centi\meter^{-2}}. In the warm gas, the \ce{HCN} abundance is likely elevated; \citet{Gonzalez-Alfonso2012} derive an abundance $X_{\ce{HCN}} \sim 10^{-6}$, and this abundance is also commonly found in Galactic hot cores \citep[e.g.][]{Boonman2001, Barr2020, VanGelder2024}. Our measured \ce{HCN} column density then translates to an \ce{H2} column density of \qtyrange[range-exponents=individual]{e23}{e24}{\centi\meter^{-2}}. Though high, this falls short of the highest column densities of \qty{e26}{\centi\meter^{-2}} estimated for the opaque nucleus by two orders of magnitude, corroborating the picture that we do not see into the nucleus. The gas density is estimated at \qtyrange[range-exponents=individual]{e5}{e6}{\centi\meter^{-3}} \citep{Scoville2015, Tunnard2015}; the implied average path length is then \qtyrange{0.1}{10}{\parsec}.

We propose the following picture for the origin of the observed MIR absorption features: the ``moderately warm'' and ``warm'' molecular gas is located in the inner regions of the starburst disk and externally irradiated by the $T_\mathrm{d} \gtrsim \qty{500}{\kelvin}$ nucleus, causing significant radiative excitation in the case of \ce{HCN} and \ce{H2O}. In these inner regions, \ce{CO2} is so depleted through previous shocks that we only observe it in colder, shielded regions in the midplane of the disk, where the detected ices are also protected from evaporation. The detected $T_\mathrm{rot} \approx \qty{700}{\kelvin}$ \ce{CO}, with high velocity dispersion, is located in shocked regions of the disk affected by SNRs.

The geometry of the WN as sketched in Fig.~\ref{fig:WN_geometry_cartoon} also demonstrates an explanation for the non-detection of the outflow in our MIR spectrum. The mass and size presented for the southern lobe by \citet{BarcosMunoz2018} suggest an average \ce{HCN} column density of \qty{6e16}{\centi\meter^{-2}}---slightly above the detection limit indicated by our spectral models. However, the outflow covers only a small fraction of the opaque nucleus in the plane of the sky, suppressing any absorption lines through continuum emission from the uncovered part of background source. From the observation that the extreme blueshift is only seen south of the nucleus, the covering factor is at most $f_\mathrm{cov} < 0.5$, even if the opening angle is wide or the outflow is slanted. As a result, combined with the effect of foreground dust emission described in Section \ref{sec:emission_effects}, the MIR absorption of this fast outflow is not detected.

\section{Summary and conclusions} \label{sec:conclusions}

In this work, we studied the WN of Arp~220 using MIRI/MRS and NIRSpec/IFU spectroscopy. Owing to the leap in sensitivity and spatial and spectral resolution brought by JWST, we were able to detect rovibrational absorption bands of 14 distinct molecular species in a narrow line of sight towards the opaque $\qty{20}{\parsec}$ nucleus, and perform a detailed analysis of the strongest features. Using LTE spectral models, we characterised the column densities and excitation of this molecular gas, and reached the following conclusions:

\begin{enumerate}
    \item Optically thick features of \ce{C2H2}, \ce{HCN} and \ce{HNC} indicate that a considerable fraction of the observed continuum emission does not pass through the absorbing molecular gas, particularly for \ce{HNC} at \qty{21}{\micron}. The implication is that the bulk of this gas must be hidden behind large columns of cooler dust, with the associated foreground continuum emission diluting the absorption signal. When we account for this effect, we derive \ce{HCN}, \ce{C2H2} and \ce{CO2} column densities an order of magnitude above the previous mid-IR estimates \citep[][Sections~\ref{sec:res_14um}, \ref{sec:discussion_fbg}]{Lahuis2007}. {We note that these column densities may still be underestimates if the vibrational temperatures are higher than the derived rotational temperatures (Section~\ref{sec:discussion_line_emission}).}
    \item We find two temperature components for \ce{HCN}. The warm \SI{330}{\kelvin} component is radiatively excited, most likely through its far-IR rotational transitions, and probes the local radiation field. {This excitation can be produced by a postulated unattenuated \qtyrange{600}{1500}{\kelvin} blackbody under a dilution factor \numrange{0.5}{0.2}.}
    The implied \ce{H2} column density is of the order of \qty{e24}{\centi\meter^{-2}}. We therefore propose that the detected \ce{HCN} is located in the inner regions of the \qty{100}{\parsec} circumnuclear starburst disk, directly surrounding the optically thick \qty{20}{\parsec}, $T_\mathrm{d} \gtrsim \SI{500}{\kelvin}$ background source identified by \citet{Sakamoto2017} (Sections~\ref{sec:discussion_HCN_comparison} and \ref{sec:discussion_geometry}).
    \item The fast collimated molecular outflow found in submillimetre studies \citep{BarcosMunoz2018} is not detected in the mid-infrared, despite the fact that the implied average column density is above our detection limit. We attribute this non-detection to a dilution of the absorption signal, as the blueshifted southern lobe covers only a small fraction of the opaque nucleus in the plane of the sky (Section~\ref{sec:discussion_geometry}).
    \item The molecular footprint probed by the MIR absorption bands resembles that of a hot core with additional evidence for shocks. In particular, the \ce{HCN}/\ce{HNC} ratio is well above unity and favours a hot core chemistry over X-ray-driven chemistry. We also obtain the first extragalactic measurement of the \ce{C2H2} ortho-to-para ratio, finding a decidedly non-equilibrium value of $1.71 \pm 0.05$. This value is remarkably consistent with measurements in hot cores \citep[][Section~\ref{sec:discussion_chemistry}]{Rangwala2018, Barr2020}.
    \item We detect large columns of \ce{CO} at a rotational temperature of $T_\mathrm{rot} \approx \qty{700}{\kelvin}$ and with extremely broad lines ($\sigma_V \approx \qty{190}{\kilo\meter\per\second}$. \ce{CO} has the highest rotational temperature and velocity dispersion in our sample of rovibrational bands, and we attribute this to shocks induced by SNe in the starburst disk (Sections~\ref{sec:res_CO} and \ref{sec:discussion_geometry}).
    \item Although we find evidence for considerable radiative excitation and line emission from vibrationally excited states, the inferred radiation temperatures do not evidently require the presence of an AGN. The detected molecular gas is located outside the central $T_\mathrm{d} \gtrsim \qty{500}{\kelvin}$ radiation source, and so we do not pierce deeply into the inner nucleus. The chemical footprint of the detected gas also shows no evidence of being influenced by X-rays from an obscured AGN, {despite the proximity of this gas to the hot central source.}
\end{enumerate}

The JWST IFU observations of the WN of Arp~220 demonstrate the immense power of mid-IR molecular absorption bands to probe and characterise the central concentrations of molecular gas in compact obscured nuclei. With its exceptional data quality, the extracted Arp~220 spectrum presented here may serve as a guide for similar studies of other U/LIRGs. Our analysis has shed new light on both the chemistry and excitation of the gas, as well as its embedded location in the dust distribution. We have found no unambiguous evidence for the presence of an AGN, and the absorbing molecular gas appears to trace an extreme, deeply buried nuclear starburst. Future detailed modelling of the chemistry and the radiative excitation is needed to determine whether all observed characteristics presented in this work can be achieved by pure star formation in the opaque \qty{20}{\parsec} background source.

\begin{acknowledgements}
We thank the anonymous reviewer for their helpful suggestions.
We are grateful to Santiago Arribas for help with the NIRSpec data. We thank Susanne Aalto, Loreto Barcos-Muñoz, John Black, George Rieke, Nick Scoville, and Henrik Spoon for helpful discussions.
S.V. acknowledges support from the European Research Council (ERC) Advanced Grant MOPPEX 833460.
AAH and LHM acknowledge support from grant PID2021-124665NB-I00 funded by MCIN/AEI/10.13039/501100011033 and by ERDF A way of making Europe.
LC acknowledges support by grant PIB2021-725127718NB-100 from the Spanish Ministry of Science and Innovation/State Agency of Research MCIN/AEI/10.13039/501100011033.
G\"O acknowledges support from the Swedish National Space Agency (SNSA).
TRG acknowledges funding from the Cosmic Dawn Center (DAWN), funded by the Danish National Research Foundation (DNRF) under grant DNRF140. TRG is also grateful for support from the Carlsberg Foundation via grant No. CF20-0534.
MP acknowledges grant PID2021-127718NB-I00 funded by the Spanish Ministry of Science and Innovation/State Agency of Research (MICIN/AEI/ 10.13039/501100011033), and the grant RYC2023-044853-I, funded by  MICIU/AEI/10.13039/501100011033 and European Social Fund Plus (FSE+).

This work is based on observations made with the NASA/ESA/CSA James Webb Space Telescope. The data were obtained from the Mikulski Archive for Space Telescopes at the Space Telescope Science Institute, which is operated by the Association of Universities for Research in Astronomy, Inc., under NASA contract NAS 5-03127 for JWST. These observations are associated with program \#1267, {and are available via \href{http://dx.doi.org/10.17909/t6c5-ks25}{DOI 10.17909/t6c5-ks25}.}
This research has made use of the NASA/IPAC Extragalactic Database (NED) which is operated by the Jet Propulsion Laboratory, California Institute of Technology, under contract with the National Aeronautics and Space Administration. This research made use of Astropy, a community-developed core Python package for Astronomy \citep{2018AJ....156..123A, 2013A&A...558A..33A}. This research made use of SciPy \citep{Virtanen_2020}. This research made use of matplotlib, a Python library for publication quality graphics \citep{Hunter:2007}. This research made use of NumPy \citep{harris2020array}. This research made use of spectralcube, a library for astronomical spectral data cubes. This research made use of APLpy, an open-source plotting package for Python \citep{aplpy}.
\end{acknowledgements}

\bibliographystyle{aa}
\bibliography{biblio.bib}

\begin{thebibliography}{108}
\expandafter\ifx\csname natexlab\endcsname\relax\def\natexlab#1{#1}\fi

\bibitem[{{Aalto} {et~al.}(2015){Aalto}, {Mart{\'\i}n}, {Costagliola}, {Gonz{\'a}lez-Alfonso}, {Muller}, {Sakamoto}, {Fuller}, {Garc{\'\i}a-Burillo}, {van der Werf}, {Neri}, {Spaans}, {Combes}, {Viti}, {M{\"u}hle}, {Armus}, {Evans}, {Sturm}, {Cernicharo}, {Henkel}, \& {Greve}}]{Aalto2015}
{Aalto}, S., {Mart{\'\i}n}, S., {Costagliola}, F., {et~al.} 2015, \aap, 584, A42

\bibitem[{{Aalto} {et~al.}(2007){Aalto}, {Spaans}, {Wiedner}, \& {H{\"u}ttemeister}}]{Aalto2007}
{Aalto}, S., {Spaans}, M., {Wiedner}, M.~C., \& {H{\"u}ttemeister}, S. 2007, \aap, 464, 193

\bibitem[{{Aladro} {et~al.}(2015){Aladro}, {Mart{\'\i}n}, {Riquelme}, {Henkel}, {Mauersberger}, {Mart{\'\i}n-Pintado}, {Wei{\ss}}, {Lefevre}, {Kramer}, {Requena-Torres}, \& {Armijos-Abenda{\~n}o}}]{Aladro2015}
{Aladro}, R., {Mart{\'\i}n}, S., {Riquelme}, D., {et~al.} 2015, \aap, 579, A101

\bibitem[{{Alonso Herrero} {et~al.}(2024){Alonso Herrero}, {Hermosa Mu{\~n}oz}, {Labiano}, {Guillard}, {Buiten}, {Dicken}, {van der Werf}, {{\'A}lvarez-M{\'a}rquez}, {B{\"o}ker}, {Colina}, {Eckart}, {Garc{\'\i}a-Mar{\'\i}n}, {Jones}, {Pantoni}, {P{\'e}rez-Gonz{\'a}lez}, {Rouan}, {Ward}, {Baes}, {{\"O}stlin}, {Royer}, {Wright}, {G{\"u}del}, {Henning}, {Lagage}, \& {van Dishoeck}}]{AlonsoHerrero2024}
{Alonso Herrero}, A., {Hermosa Mu{\~n}oz}, L., {Labiano}, A., {et~al.} 2024, \aap, 690, A95

\bibitem[{{Araya} {et~al.}(2004){Araya}, {Baan}, \& {Hofner}}]{Araya2004}
{Araya}, E., {Baan}, W.~A., \& {Hofner}, P. 2004, \apjs, 154, 541

\bibitem[{{Argyriou} {et~al.}(2023){Argyriou}, {Glasse}, {Law}, {Labiano}, {{\'A}lvarez-M{\'a}rquez}, {Patapis}, {Kavanagh}, {Gasman}, {Mueller}, {Larson}, {Vandenbussche}, {Glauser}, {Royer}, {Dicken}, {Harkett}, {Sargent}, {Engesser}, {Jones}, {Kendrew}, {Noriega-Crespo}, {Brandl}, {Rieke}, {Wright}, {Lee}, \& {Wells}}]{Argyriou2023}
{Argyriou}, I., {Glasse}, A., {Law}, D.~R., {et~al.} 2023, \aap, 675, A111

\bibitem[{{Armus} {et~al.}(2007){Armus}, {Charmandaris}, {Bernard-Salas}, {Spoon}, {Marshall}, {Higdon}, {Desai}, {Teplitz}, {Hao}, {Devost}, {Brandl}, {Wu}, {Sloan}, {Soifer}, {Houck}, \& {Herter}}]{Armus2007}
{Armus}, L., {Charmandaris}, V., {Bernard-Salas}, J., {et~al.} 2007, \apj, 656, 148

\bibitem[{{Astropy Collaboration} {et~al.}(2018){Astropy Collaboration}, {Price-Whelan}, {Sip{\H o}cz}, {G{\"u}nther}, {Lim}, {Crawford}, {Conseil}, {Shupe}, {Craig}, {Dencheva}, {Ginsburg}, {VanderPlas}, {Bradley}, {P{\'e}rez-Su{\'a}rez}, {de Val-Borro}, {Aldcroft}, {Cruz}, {Robitaille}, {Tollerud}, {Ardelean}, {Babej}, {Bach}, {Bachetti}, {Bakanov}, {Bamford}, {Barentsen}, {Barmby}, {Baumbach}, {Berry}, {Biscani}, {Boquien}, {Bostroem}, {Bouma}, {Brammer}, {Bray}, {Breytenbach}, {Buddelmeijer}, {Burke}, {Calderone}, {Cano Rodr{\'{\i}}guez}, {Cara}, {Cardoso}, {Cheedella}, {Copin}, {Corrales}, {Crichton}, {D'Avella}, {Deil}, {Depagne}, {Dietrich}, {Donath}, {Droettboom}, {Earl}, {Erben}, {Fabbro}, {Ferreira}, {Finethy}, {Fox}, {Garrison}, {Gibbons}, {Goldstein}, {Gommers}, {Greco}, {Greenfield}, {Groener}, {Grollier}, {Hagen}, {Hirst}, {Homeier}, {Horton}, {Hosseinzadeh}, {Hu}, {Hunkeler}, {Ivezi{\'c}}, {Jain}, {Jenness}, {Kanarek}, {Kendrew}, {Kern}, {Kerzendorf}, {Khvalko}, {King}, {Kirkby}, {Kulkarni},
  {Kumar}, {Lee}, {Lenz}, {Littlefair}, {Ma}, {Macleod}, {Mastropietro}, {McCully}, {Montagnac}, {Morris}, {Mueller}, {Mumford}, {Muna}, {Murphy}, {Nelson}, {Nguyen}, {Ninan}, {N{\"o}the}, {Ogaz}, {Oh}, {Parejko}, {Parley}, {Pascual}, {Patil}, {Patil}, {Plunkett}, {Prochaska}, {Rastogi}, {Reddy Janga}, {Sabater}, {Sakurikar}, {Seifert}, {Sherbert}, {Sherwood-Taylor}, {Shih}, {Sick}, {Silbiger}, {Singanamalla}, {Singer}, {Sladen}, {Sooley}, {Sornarajah}, {Streicher}, {Teuben}, {Thomas}, {Tremblay}, {Turner}, {Terr{\'o}n}, {van Kerkwijk}, {de la Vega}, {Watkins}, {Weaver}, {Whitmore}, {Woillez}, {Zabalza}, \& {Astropy Contributors}}]{2018AJ....156..123A}
{Astropy Collaboration}, {Price-Whelan}, A.~M., {Sip{\H o}cz}, B.~M., {et~al.} 2018, \aj, 156, 123

\bibitem[{{Astropy Collaboration} {et~al.}(2013){Astropy Collaboration}, {Robitaille}, {Tollerud}, {Greenfield}, {Droettboom}, {Bray}, {Aldcroft}, {Davis}, {Ginsburg}, {Price-Whelan}, {Kerzendorf}, {Conley}, {Crighton}, {Barbary}, {Muna}, {Ferguson}, {Grollier}, {Parikh}, {Nair}, {Unther}, {Deil}, {Woillez}, {Conseil}, {Kramer}, {Turner}, {Singer}, {Fox}, {Weaver}, {Zabalza}, {Edwards}, {Azalee Bostroem}, {Burke}, {Casey}, {Crawford}, {Dencheva}, {Ely}, {Jenness}, {Labrie}, {Lim}, {Pierfederici}, {Pontzen}, {Ptak}, {Refsdal}, {Servillat}, \& {Streicher}}]{2013A&A...558A..33A}
{Astropy Collaboration}, {Robitaille}, T.~P., {Tollerud}, E.~J., {et~al.} 2013, \aap, 558, A33

\bibitem[{{Baan} {et~al.}(2023){Baan}, {Aditya}, {An}, \& {Kl{\"o}ckner}}]{Baan2023}
{Baan}, W.~A., {Aditya}, J.~N.~H.~S., {An}, T., \& {Kl{\"o}ckner}, H.-R. 2023, \mnras, 523, 5487

\bibitem[{{Baan} \& {Haschick}(1984)}]{BaanHaschick1984}
{Baan}, W.~A. \& {Haschick}, A.~D. 1984, \apj, 279, 541

\bibitem[{{Baan} {et~al.}(1989){Baan}, {Haschick}, \& {Henkel}}]{Baan1989}
{Baan}, W.~A., {Haschick}, A.~D., \& {Henkel}, C. 1989, \apj, 346, 680

\bibitem[{{Baan} {et~al.}(1982){Baan}, {Wood}, \& {Haschick}}]{Baan1982}
{Baan}, W.~A., {Wood}, P.~A.~D., \& {Haschick}, A.~D. 1982, \apjl, 260, L49

\bibitem[{{Barcos-Mu{\~n}oz} {et~al.}(2018){Barcos-Mu{\~n}oz}, {Aalto}, {Thompson}, {Sakamoto}, {Mart{\'\i}n}, {Leroy}, {Privon}, {Evans}, \& {Kepley}}]{BarcosMunoz2018}
{Barcos-Mu{\~n}oz}, L., {Aalto}, S., {Thompson}, T.~A., {et~al.} 2018, \apjl, 853, L28

\bibitem[{{Barcos-Mu{\~n}oz} {et~al.}(2015){Barcos-Mu{\~n}oz}, {Leroy}, {Evans}, {Privon}, {Armus}, {Condon}, {Mazzarella}, {Meier}, {Momjian}, {Murphy}, {Ott}, {Reichardt}, {Sakamoto}, {Sanders}, {Schinnerer}, {Stierwalt}, {Surace}, {Thompson}, \& {Walter}}]{BarcosMunoz2015}
{Barcos-Mu{\~n}oz}, L., {Leroy}, A.~K., {Evans}, A.~S., {et~al.} 2015, \apj, 799, 10

\bibitem[{{Barr} {et~al.}(2020){Barr}, {Boogert}, {DeWitt}, {Montiel}, {Richter}, {Lacy}, {Neufeld}, {Indriolo}, {Pendleton}, {Chiar}, \& {Tielens}}]{Barr2020}
{Barr}, A.~G., {Boogert}, A., {DeWitt}, C.~N., {et~al.} 2020, \apj, 900, 104

\bibitem[{{Behrens} {et~al.}(2022){Behrens}, {Mangum}, {Holdship}, {Viti}, {Harada}, {Mart{\'\i}n}, {Sakamoto}, {Muller}, {Tanaka}, {Nakanishi}, {Herrero-Illana}, {Yoshimura}, {Aladro}, {Colzi}, {Emig}, {Henkel}, {Huang}, {Humire}, {Meier}, {Rivilla}, {van der Werf}, \& {Alma Comprehensive High-Resolution Extragalactic Molecular Inventory (Alchemi) Collaboration}}]{Behrens2022}
{Behrens}, E., {Mangum}, J.~G., {Holdship}, J., {et~al.} 2022, \apj, 939, 119

\bibitem[{{B{\"o}ker} {et~al.}(2023){B{\"o}ker}, {Beck}, {Birkmann}, {Giardino}, {Keyes}, {Kumari}, {Muzerolle}, {Rawle}, {Zeidler}, {Abul-Huda}, {Alves de Oliveira}, {Arribas}, {Bechtold}, {Bhatawdekar}, {Bonaventura}, {Bunker}, {Cameron}, {Carniani}, {Charlot}, {Curti}, {Espinoza}, {Ferruit}, {Franx}, {Jakobsen}, {Karakla}, {L{\'o}pez-Caniego}, {L{\"u}tzgendorf}, {Maiolino}, {Manjavacas}, {Marston}, {Moseley}, {Ogle}, {Perna}, {Pe{\~n}a-Guerrero}, {Pirzkal}, {Plesha}, {Proffitt}, {Rauscher}, {Rix}, {Rodr{\'\i}guez del Pino}, {Rustamkulov}, {Sabbi}, {Sing}, {Sirianni}, {te Plate}, {{\'U}beda}, {Wahlgren}, {Wislowski}, {Wu}, \& {Willott}}]{Boeker2023}
{B{\"o}ker}, T., {Beck}, T.~L., {Birkmann}, S.~M., {et~al.} 2023, \pasp, 135, 038001

\bibitem[{{Boogert} {et~al.}(2015){Boogert}, {Gerakines}, \& {Whittet}}]{Boogert2015}
{Boogert}, A.~C.~A., {Gerakines}, P.~A., \& {Whittet}, D. C.~B. 2015, \araa, 53, 541

\bibitem[{{Boonman} {et~al.}(2001){Boonman}, {Stark}, {van der Tak}, {van Dishoeck}, {van der Wal}, {Sch{\"a}fer}, {de Lange}, \& {Laauwen}}]{Boonman2001}
{Boonman}, A.~M.~S., {Stark}, R., {van der Tak}, F.~F.~S., {et~al.} 2001, \apjl, 553, L63

\bibitem[{{Boonman} {et~al.}(2003){Boonman}, {van Dishoeck}, {Lahuis}, \& {Doty}}]{Boonman2003b}
{Boonman}, A.~M.~S., {van Dishoeck}, E.~F., {Lahuis}, F., \& {Doty}, S.~D. 2003, \aap, 399, 1063

\bibitem[{{Buiten} {et~al.}(2024){Buiten}, {van der Werf}, {Viti}, {Armus}, {Barr}, {Barcos-Mu{\~n}oz}, {Evans}, {Inami}, {Linden}, {Privon}, {Song}, {Rich}, {Aalto}, {Appleton}, {B{\"o}ker}, {Charmandaris}, {Diaz-Santos}, {Hayward}, {Lai}, {Medling}, {Ricci}, \& {U}}]{Buiten2024a}
{Buiten}, V.~A., {van der Werf}, P.~P., {Viti}, S., {et~al.} 2024, \apj, 966, 166

\bibitem[{{Butterworth} {et~al.}(2022){Butterworth}, {Holdship}, {Viti}, \& {Garc{\'\i}a-Burillo}}]{Butterworth2022}
{Butterworth}, J., {Holdship}, J., {Viti}, S., \& {Garc{\'\i}a-Burillo}, S. 2022, \aap, 667, A131

\bibitem[{{Charnley} \& {Kaufman}(2000)}]{CharnleyKaufman2000}
{Charnley}, S.~B. \& {Kaufman}, M.~J. 2000, \apjl, 529, L111

\bibitem[{{Chown} {et~al.}(2024){Chown}, {Sidhu}, {Peeters}, {Tielens}, {Cami}, {Bern{\'e}}, {Habart}, {Alarc{\'o}n}, {Canin}, {Schroetter}, {Trahin}, {Van De Putte}, {Abergel}, {Bergin}, {Bernard-Salas}, {Boersma}, {Bron}, {Cuadrado}, {Dartois}, {Dicken}, {El-Yajouri}, {Fuente}, {Goicoechea}, {Gordon}, {Issa}, {Joblin}, {Kannavou}, {Khan}, {Lacinbala}, {Languignon}, {Le Gal}, {Maragkoudakis}, {Meshaka}, {Okada}, {Onaka}, {Pasquini}, {Pound}, {Robberto}, {R{\"o}llig}, {Schefter}, {Schirmer}, {Vicente}, {Wolfire}, {Zannese}, {Aleman}, {Allamandola}, {Auchettl}, {Baratta}, {Bejaoui}, {Bera}, {Black}, {Boulanger}, {Bouwman}, {Brandl}, {Brechignac}, {Br{\"u}nken}, {Buragohain}, {Burkhardt}, {Candian}, {Cazaux}, {Cernicharo}, {Chabot}, {Chakraborty}, {Champion}, {Colgan}, {Cooke}, {Coutens}, {Cox}, {Demyk}, {Meyer}, {Foschino}, {Garc{\'\i}a-Lario}, {Gavilan}, {Gerin}, {Gottlieb}, {Guillard}, {Gusdorf}, {Hartigan}, {He}, {Herbst}, {Hornekaer}, {J{\"a}ger}, {Janot-Pacheco}, {Kaufman}, {Kemper}, {Kendrew},
  {Kirsanova}, {Klaassen}, {Kwok}, {Labiano}, {Lai}, {Lee}, {Lefloch}, {Le Petit}, {Li}, {Linz}, {Mackie}, {Madden}, {Mascetti}, {McGuire}, {Merino}, {Micelotta}, {Misselt}, {Morse}, {Mulas}, {Neelamkodan}, {Ohsawa}, {Omont}, {Paladini}, {Palumbo}, {Pathak}, {Pendleton}, {Petrignani}, {Pino}, {Puga}, {Rangwala}, {Rapacioli}, {Ricca}, {Roman-Duval}, {Roser}, {Roueff}, {Rouill{\'e}}, {Salama}, {Sales}, {Sandstrom}, {Sarre}, {Sciamma-O'Brien}, {Sellgren}, {Shenoy}, {Teyssier}, {Thomas}, {Togi}, {Verstraete}, {Witt}, {Wootten}, {Zettergren}, {Zhang}, {Zhang}, \& {Zhen}}]{Chown2024}
{Chown}, R., {Sidhu}, A., {Peeters}, E., {et~al.} 2024, \aap, 685, A75

\bibitem[{{Costagliola} {et~al.}(2015){Costagliola}, {Sakamoto}, {Muller}, {Mart{\'\i}n}, {Aalto}, {Harada}, {van der Werf}, {Viti}, {Garcia-Burillo}, \& {Spaans}}]{Costagliola2015}
{Costagliola}, F., {Sakamoto}, K., {Muller}, S., {et~al.} 2015, \aap, 582, A91

\bibitem[{{Delahaye} {et~al.}(2021){Delahaye}, {Armante}, {Scott}, {Jacquinet-Husson}, {Ch{\'e}din}, {Cr{\'e}peau}, {Crevoisier}, {Douet}, {Perrin}, {Barbe}, {Boudon}, {Campargue}, {Coudert}, {Ebert}, {Flaud}, {Gamache}, {Jacquemart}, {Jolly}, {Kwabia Tchana}, {Kyuberis}, {Li}, {Lyulin}, {Manceron}, {Mikhailenko}, {Moazzen-Ahmadi}, {M{\"u}ller}, {Naumenko}, {Nikitin}, {Perevalov}, {Richard}, {Starikova}, {Tashkun}, {Tyuterev}, {Vander Auwera}, {Vispoel}, {Yachmenev}, \& {Yurchenko}}]{GEISA}
{Delahaye}, T., {Armante}, R., {Scott}, N.~A., {et~al.} 2021, Journal of Molecular Spectroscopy, 380, 111510

\bibitem[{{D'Eugenio} {et~al.}(2024){D'Eugenio}, {P{\'e}rez-Gonz{\'a}lez}, {Maiolino}, {Scholtz}, {Perna}, {Circosta}, {{\"U}bler}, {Arribas}, {B{\"o}ker}, {Bunker}, {Carniani}, {Charlot}, {Chevallard}, {Cresci}, {Curtis-Lake}, {Jones}, {Kumari}, {Lamperti}, {Looser}, {Parlanti}, {Rix}, {Robertson}, {Rodr{\'\i}guez Del Pino}, {Tacchella}, {Venturi}, \& {Willott}}]{D'Eugenio2024}
{D'Eugenio}, F., {P{\'e}rez-Gonz{\'a}lez}, P.~G., {Maiolino}, R., {et~al.} 2024, Nature Astronomy, 8, 1443

\bibitem[{{Donnan} {et~al.}(2024){Donnan}, {Garc{\'\i}a-Bernete}, {Rigopoulou}, {Pereira-Santaella}, {Roche}, \& {Alonso-Herrero}}]{Donnan2024}
{Donnan}, F.~R., {Garc{\'\i}a-Bernete}, I., {Rigopoulou}, D., {et~al.} 2024, \mnras, 529, 1386

\bibitem[{{Doty} {et~al.}(2002){Doty}, {van Dishoeck}, {van der Tak}, \& {Boonman}}]{Doty2002}
{Doty}, S.~D., {van Dishoeck}, E.~F., {van der Tak}, F.~F.~S., \& {Boonman}, A.~M.~S. 2002, \aap, 389, 446

\bibitem[{{Endres} {et~al.}(2016){Endres}, {Schlemmer}, {Schilke}, {Stutzki}, \& {M{\"u}ller}}]{CDMS}
{Endres}, C.~P., {Schlemmer}, S., {Schilke}, P., {Stutzki}, J., \& {M{\"u}ller}, H. S.~P. 2016, Journal of Molecular Spectroscopy, 327, 95

\bibitem[{{Falstad} {et~al.}(2015){Falstad}, {Gonz{\'a}lez-Alfonso}, {Aalto}, {van der Werf}, {Fischer}, {Veilleux}, {Mel{\'e}ndez}, {Farrah}, \& {Smith}}]{Falstad2015}
{Falstad}, N., {Gonz{\'a}lez-Alfonso}, E., {Aalto}, S., {et~al.} 2015, \aap, 580, A52

\bibitem[{{Foreman-Mackey} {et~al.}(2013){Foreman-Mackey}, {Hogg}, {Lang}, \& {Goodman}}]{emcee}
{Foreman-Mackey}, D., {Hogg}, D.~W., {Lang}, D., \& {Goodman}, J. 2013, PASP, 125, 306

\bibitem[{{Francis} {et~al.}(2024){Francis}, {van Gelder}, {van Dishoeck}, {Gieser}, {Beuther}, {Tychoniec}, {Perotti}, {Caratti o Garatti}, {Kavanagh}, {Ray}, {Klaassen}, {Justtanont}, {Linnartz}, {Rocha}, {Slavicinska}, {G{\"u}del}, {Henning}, {Lagage}, \& {{\"O}stlin}}]{Francis2024}
{Francis}, L., {van Gelder}, M.~L., {van Dishoeck}, E.~F., {et~al.} 2024, \aap, 683, A249

\bibitem[{{Gaia Collaboration} {et~al.}(2023){Gaia Collaboration}, {Vallenari}, {Brown}, {Prusti}, {de Bruijne}, {Arenou}, {Babusiaux}, {Biermann}, {Creevey}, {Ducourant}, {Evans}, {Eyer}, {Guerra}, {Hutton}, {Jordi}, {Klioner}, {Lammers}, {Lindegren}, {Luri}, {Mignard}, {Panem}, {Pourbaix}, {Randich}, {Sartoretti}, {Soubiran}, {Tanga}, {Walton}, {Bailer-Jones}, {Bastian}, {Drimmel}, {Jansen}, {Katz}, {Lattanzi}, {van Leeuwen}, {Bakker}, {Cacciari}, {Casta{\~n}eda}, {De Angeli}, {Fabricius}, {Fouesneau}, {Fr{\'e}mat}, {Galluccio}, {Guerrier}, {Heiter}, {Masana}, {Messineo}, {Mowlavi}, {Nicolas}, {Nienartowicz}, {Pailler}, {Panuzzo}, {Riclet}, {Roux}, {Seabroke}, {Sordo}, {Th{\'e}venin}, {Gracia-Abril}, {Portell}, {Teyssier}, {Altmann}, {Andrae}, {Audard}, {Bellas-Velidis}, {Benson}, {Berthier}, {Blomme}, {Burgess}, {Busonero}, {Busso}, {C{\'a}novas}, {Carry}, {Cellino}, {Cheek}, {Clementini}, {Damerdji}, {Davidson}, {de Teodoro}, {Nu{\~n}ez Campos}, {Delchambre}, {Dell'Oro}, {Esquej},
  {Fern{\'a}ndez-Hern{\'a}ndez}, {Fraile}, {Garabato}, {Garc{\'\i}a-Lario}, {Gosset}, {Haigron}, {Halbwachs}, {Hambly}, {Harrison}, {Hern{\'a}ndez}, {Hestroffer}, {Hodgkin}, {Holl}, {Jan{\ss}en}, {Jevardat de Fombelle}, {Jordan}, {Krone-Martins}, {Lanzafame}, {L{\"o}ffler}, {Marchal}, {Marrese}, {Moitinho}, {Muinonen}, {Osborne}, {Pancino}, {Pauwels}, {Recio-Blanco}, {Reyl{\'e}}, {Riello}, {Rimoldini}, {Roegiers}, {Rybizki}, {Sarro}, {Siopis}, {Smith}, {Sozzetti}, {Utrilla}, {van Leeuwen}, {Abbas}, {{\'A}brah{\'a}m}, {Abreu Aramburu}, {Aerts}, {Aguado}, {Ajaj}, {Aldea-Montero}, {Altavilla}, {{\'A}lvarez}, {Alves}, {Anders}, {Anderson}, {Anglada Varela}, {Antoja}, {Baines}, {Baker}, {Balaguer-N{\'u}{\~n}ez}, {Balbinot}, {Balog}, {Barache}, {Barbato}, {Barros}, {Barstow}, {Bartolom{\'e}}, {Bassilana}, {Bauchet}, {Becciani}, {Bellazzini}, {Berihuete}, {Bernet}, {Bertone}, {Bianchi}, {Binnenfeld}, {Blanco-Cuaresma}, {Blazere}, {Boch}, {Bombrun}, {Bossini}, {Bouquillon}, {Bragaglia}, {Bramante}, {Breedt},
  {Bressan}, {Brouillet}, {Brugaletta}, {Bucciarelli}, {Burlacu}, {Butkevich}, {Buzzi}, {Caffau}, {Cancelliere}, {Cantat-Gaudin}, {Carballo}, {Carlucci}, {Carnerero}, {Carrasco}, {Casamiquela}, {Castellani}, {Castro-Ginard}, {Chaoul}, {Charlot}, {Chemin}, {Chiaramida}, {Chiavassa}, {Chornay}, {Comoretto}, {Contursi}, {Cooper}, {Cornez}, {Cowell}, {Crifo}, {Cropper}, {Crosta}, {Crowley}, {Dafonte}, {Dapergolas}, {David}, {David}, {de Laverny}, {De Luise}, \& {De March}}]{GaiaDR3}
{Gaia Collaboration}, {Vallenari}, A., {Brown}, A.~G.~A., {et~al.} 2023, \aap, 674, A1

\bibitem[{{Garc{\'\i}a-Bernete} {et~al.}(2024){Garc{\'\i}a-Bernete}, {Pereira-Santaella}, {Gonz{\'a}lez-Alfonso}, {Rigopoulou}, {Efstathiou}, {Donnan}, \& {Thatte}}]{Garcia-Bernete2024}
{Garc{\'\i}a-Bernete}, I., {Pereira-Santaella}, M., {Gonz{\'a}lez-Alfonso}, E., {et~al.} 2024, \aap, 682, L5

\bibitem[{{Goldberg} {et~al.}(2024){Goldberg}, {Buiten}, {Rieke}, {Alonso-Herrero}, {Paggi}, {van der Werf}, {Stone}, {Morrison}, {Alberts}, {Dicken}, \& {Wright}}]{Goldberg2024}
{Goldberg}, C.~E., {Buiten}, V.~A., {Rieke}, G.~H., {et~al.} 2024, \apj, 977, 55

\bibitem[{{Goldsmith} \& {Langer}(1999)}]{GoldsmithLanger1999}
{Goldsmith}, P.~F. \& {Langer}, W.~D. 1999, \apj, 517, 209

\bibitem[{{Gonz{\'a}lez-Alfonso} {et~al.}(2013){Gonz{\'a}lez-Alfonso}, {Fischer}, {Bruderer}, {M{\"u}ller}, {Graci{\'a}-Carpio}, {Sturm}, {Lutz}, {Poglitsch}, {Feuchtgruber}, {Veilleux}, {Contursi}, {Sternberg}, {Hailey-Dunsheath}, {Verma}, {Christopher}, {Davies}, {Genzel}, \& {Tacconi}}]{Gonzalez-Alfonso2013}
{Gonz{\'a}lez-Alfonso}, E., {Fischer}, J., {Bruderer}, S., {et~al.} 2013, \aap, 550, A25

\bibitem[{{Gonz{\'a}lez-Alfonso} {et~al.}(2012){Gonz{\'a}lez-Alfonso}, {Fischer}, {Graci{\'a}-Carpio}, {Sturm}, {Hailey-Dunsheath}, {Lutz}, {Poglitsch}, {Contursi}, {Feuchtgruber}, {Veilleux}, {Spoon}, {Verma}, {Christopher}, {Davies}, {Sternberg}, {Genzel}, \& {Tacconi}}]{Gonzalez-Alfonso2012}
{Gonz{\'a}lez-Alfonso}, E., {Fischer}, J., {Graci{\'a}-Carpio}, J., {et~al.} 2012, \aap, 541, A4

\bibitem[{{Gonz{\'a}lez-Alfonso} {et~al.}(2024){Gonz{\'a}lez-Alfonso}, {Garc{\'\i}a-Bernete}, {Pereira-Santaella}, {Neufeld}, {Fischer}, \& {Donnan}}]{Gonzalez-Alfonso2024}
{Gonz{\'a}lez-Alfonso}, E., {Garc{\'\i}a-Bernete}, I., {Pereira-Santaella}, M., {et~al.} 2024, \aap, 682, A182

\bibitem[{{Gonz{\'a}lez-Alfonso} {et~al.}(2002){Gonz{\'a}lez-Alfonso}, {Wright}, {Cernicharo}, {Rosenthal}, {Boonman}, \& {van Dishoeck}}]{Gonzalez-Alfonso2002}
{Gonz{\'a}lez-Alfonso}, E., {Wright}, C.~M., {Cernicharo}, J., {et~al.} 2002, \aap, 386, 1074

\bibitem[{{Gordon} {et~al.}(2022){Gordon}, {Rothman}, {Hargreaves}, {Hashemi}, {Karlovets}, {Skinner}, {Conway}, {Hill}, {Kochanov}, {Tan}, {Wcis{\l}o}, {Finenko}, {Nelson}, {Bernath}, {Birk}, {Boudon}, {Campargue}, {Chance}, {Coustenis}, {Drouin}, {Flaud}, {Gamache}, {Hodges}, {Jacquemart}, {Mlawer}, {Nikitin}, {Perevalov}, {Rotger}, {Tennyson}, {Toon}, {Tran}, {Tyuterev}, {Adkins}, {Baker}, {Barbe}, {Can{\`e}}, {Cs{\'a}sz{\'a}r}, {Dudaryonok}, {Egorov}, {Fleisher}, {Fleurbaey}, {Foltynowicz}, {Furtenbacher}, {Harrison}, {Hartmann}, {Horneman}, {Huang}, {Karman}, {Karns}, {Kassi}, {Kleiner}, {Kofman}, {Kwabia-Tchana}, {Lavrentieva}, {Lee}, {Long}, {Lukashevskaya}, {Lyulin}, {Makhnev}, {Matt}, {Massie}, {Melosso}, {Mikhailenko}, {Mondelain}, {M{\"u}ller}, {Naumenko}, {Perrin}, {Polyansky}, {Raddaoui}, {Raston}, {Reed}, {Rey}, {Richard}, {T{\'o}bi{\'a}s}, {Sadiek}, {Schwenke}, {Starikova}, {Sung}, {Tamassia}, {Tashkun}, {Vander Auwera}, {Vasilenko}, {Vigasin}, {Villanueva}, {Vispoel}, {Wagner}, {Yachmenev}, \&
  {Yurchenko}}]{HITRAN}
{Gordon}, I.~E., {Rothman}, L.~S., {Hargreaves}, R.~J., {et~al.} 2022, \jqsrt, 277, 107949

\bibitem[{{Gorski} {et~al.}(2023){Gorski}, {Aalto}, {K{\"o}nig}, {Wethers}, {Yang}, {Muller}, {Viti}, {Black}, {Onishi}, \& {Sato}}]{Gorski2023}
{Gorski}, M.~D., {Aalto}, S., {K{\"o}nig}, S., {et~al.} 2023, \aap, 670, A70

\bibitem[{{Graham} {et~al.}(1990){Graham}, {Carico}, {Matthews}, {Neugebauer}, {Soifer}, \& {Wilson}}]{Graham1990}
{Graham}, J.~R., {Carico}, D.~P., {Matthews}, K., {et~al.} 1990, \apjl, 354, L5

\bibitem[{{Harada} {et~al.}(2024){Harada}, {Meier}, {Mart{\'\i}n}, {Muller}, {Sakamoto}, {Saito}, {Gorski}, {Henkel}, {Tanaka}, {Mangum}, {Aalto}, {Aladro}, {Bouvier}, {Colzi}, {Emig}, {Herrero-Illana}, {Huang}, {Kohno}, {K{\"o}nig}, {Nakanishi}, {Nishimura}, {Takano}, {Rivilla}, {Viti}, {Watanabe}, {van der Werf}, \& {Yoshimura}}]{Harada2024}
{Harada}, N., {Meier}, D.~S., {Mart{\'\i}n}, S., {et~al.} 2024, \apjs, 271, 38

\bibitem[{Harris {et~al.}(2020)Harris, Millman, van~der Walt, Gommers, Virtanen, Cournapeau, Wieser, Taylor, Berg, Smith, Kern, Picus, Hoyer, van Kerkwijk, Brett, Haldane, del R{'{\i}}o, Wiebe, Peterson, G{'{e}}rard-Marchant, Sheppard, Reddy, Weckesser, Abbasi, Gohlke, \& Oliphant}]{harris2020array}
Harris, C.~R., Millman, K.~J., van~der Walt, S.~J., {et~al.} 2020, Nature, 585, 357

\bibitem[{{Hermosa Mu{\~n}oz} {et~al.}(2025){Hermosa Mu{\~n}oz}, {Alonso-Herrero}, {Labiano}, {Guillard}, {Pantoni}, {Buiten}, {Dicken}, {Baes}, {B{\"o}ker}, {Colina}, {Donnan}, {Garc{\'\i}a-Bernete}, {{\"O}stlin}, {van der Werf}, {Ward}, {Brandl}, {Walter}, {Wright}, {G{\"u}del}, {Henning}, {Lagage}, \& {Ray}}]{HermosaMunoz2024}
{Hermosa Mu{\~n}oz}, L., {Alonso-Herrero}, A., {Labiano}, A., {et~al.} 2025, \aap, 693, A321

\bibitem[{{Hibbard} {et~al.}(2000){Hibbard}, {Vacca}, \& {Yun}}]{Hibbard2000}
{Hibbard}, J.~E., {Vacca}, W.~D., \& {Yun}, M.~S. 2000, \aj, 119, 1130

\bibitem[{Hunter(2007)}]{Hunter:2007}
Hunter, J.~D. 2007, Computing In Science \& Engineering, 9, 90

\bibitem[{{Imanishi} {et~al.}(2009){Imanishi}, {Nakanishi}, {Tamura}, \& {Peng}}]{Imanishi2009}
{Imanishi}, M., {Nakanishi}, K., {Tamura}, Y., \& {Peng}, C.-H. 2009, \aj, 137, 3581

\bibitem[{{Jones} {et~al.}(2023){Jones}, {{\'A}lvarez-M{\'a}rquez}, {Sloan}, {Kavanagh}, {Argyriou}, {Law}, {Labiano}, {Patapis}, {Mueller}, {Larson}, {Bright}, {Klaassen}, {Fox}, {Gasman}, {Geers}, {Glauser}, {Guillard}, {Nayak}, {Noriega-Crespo}, {Ressler}, {Sargent}, {Temim}, {Vandenbussche}, \& {Garc{\'\i}a Mar{\'\i}n}}]{Jones2023}
{Jones}, O.~C., {{\'A}lvarez-M{\'a}rquez}, J., {Sloan}, G.~C., {et~al.} 2023, \mnras, 523, 2519

\bibitem[{{Joseph} \& {Wright}(1985)}]{JosephWright1985}
{Joseph}, R.~D. \& {Wright}, G.~S. 1985, \mnras, 214, 87

\bibitem[{{Kohno} {et~al.}(2001){Kohno}, {Matsushita}, {Vila-Vilar{\'o}}, {Okumura}, {Shibatsuka}, {Okiura}, {Ishizuki}, \& {Kawabe}}]{Kohno2001}
{Kohno}, K., {Matsushita}, S., {Vila-Vilar{\'o}}, B., {et~al.} 2001, in Astronomical Society of the Pacific Conference Series, Vol. 249, The Central Kiloparsec of Starbursts and AGN: The La Palma Connection, ed. J.~H. {Knapen}, J.~E. {Beckman}, I.~{Shlosman}, \& T.~J. {Mahoney}, 672

\bibitem[{{Krips} {et~al.}(2008){Krips}, {Neri}, {Garc{\'\i}a-Burillo}, {Mart{\'\i}n}, {Combes}, {Graci{\'a}-Carpio}, \& {Eckart}}]{Krips2008}
{Krips}, M., {Neri}, R., {Garc{\'\i}a-Burillo}, S., {et~al.} 2008, \apj, 677, 262

\bibitem[{{Lahuis} {et~al.}(2007){Lahuis}, {Spoon}, {Tielens}, {Doty}, {Armus}, {Charmandaris}, {Houck}, {St{\"a}uber}, \& {van Dishoeck}}]{Lahuis2007}
{Lahuis}, F., {Spoon}, H.~W.~W., {Tielens}, A.~G.~G.~M., {et~al.} 2007, \apj, 659, 296

\bibitem[{{Lahuis} \& {van Dishoeck}(2000)}]{Lahuis2000}
{Lahuis}, F. \& {van Dishoeck}, E.~F. 2000, \aap, 355, 699

\bibitem[{{Lai} {et~al.}(2022){Lai}, {Armus}, {U}, {D{\'\i}az-Santos}, {Larson}, {Evans}, {Malkan}, {Appleton}, {Rich}, {M{\"u}ller-S{\'a}nchez}, {Inami}, {Bohn}, {McKinney}, {Finnerty}, {Law}, {Linden}, {Medling}, {Privon}, {Song}, {Stierwalt}, {van der Werf}, {Barcos-Mu{\~n}oz}, {Smith}, {Togi}, {Aalto}, {B{\"o}ker}, {Charmandaris}, {Howell}, {Iwasawa}, {Kemper}, {Mazzarella}, {Murphy}, {Brown}, {Hayward}, {Marshall}, {Sanders}, \& {Surace}}]{Lai2022}
{Lai}, T. S.~Y., {Armus}, L., {U}, V., {et~al.} 2022, \apjl, 941, L36

\bibitem[{{Law} {et~al.}(2023){Law}, {E. Morrison}, {Argyriou}, {Patapis}, {{\'A}lvarez-M{\'a}rquez}, {Labiano}, \& {Vandenbussche}}]{Law2023}
{Law}, D.~R., {E. Morrison}, J., {Argyriou}, I., {et~al.} 2023, \aj, 166, 45

\bibitem[{{Lonsdale} {et~al.}(2006){Lonsdale}, {Diamond}, {Thrall}, {Smith}, \& {Lonsdale}}]{Lonsdale2006}
{Lonsdale}, C.~J., {Diamond}, P.~J., {Thrall}, H., {Smith}, H.~E., \& {Lonsdale}, C.~J. 2006, \apj, 647, 185

\bibitem[{{Manohar} \& {Scoville}(2017)}]{Manohar2017}
{Manohar}, S. \& {Scoville}, N. 2017, \apj, 835, 127

\bibitem[{{Mart{\'\i}n} {et~al.}(2016){Mart{\'\i}n}, {Aalto}, {Sakamoto}, {Gonz{\'a}lez-Alfonso}, {Muller}, {Henkel}, {Garc{\'\i}a-Burillo}, {Aladro}, {Costagliola}, {Harada}, {Krips}, {Mart{\'\i}n-Pintado}, {M{\"u}hle}, {van der Werf}, \& {Viti}}]{Martin2016}
{Mart{\'\i}n}, S., {Aalto}, S., {Sakamoto}, K., {et~al.} 2016, \aap, 590, A25

\bibitem[{{Mart{\'\i}n} {et~al.}(2011){Mart{\'\i}n}, {Krips}, {Mart{\'\i}n-Pintado}, {Aalto}, {Zhao}, {Peck}, {Petitpas}, {Monje}, {Greve}, \& {An}}]{Martin2011}
{Mart{\'\i}n}, S., {Krips}, M., {Mart{\'\i}n-Pintado}, J., {et~al.} 2011, \aap, 527, A36

\bibitem[{{Meijerink} {et~al.}(2011){Meijerink}, {Spaans}, {Loenen}, \& {van der Werf}}]{Meijerink2011}
{Meijerink}, R., {Spaans}, M., {Loenen}, A.~F., \& {van der Werf}, P.~P. 2011, \aap, 525, A119

\bibitem[{{Nagy} {et~al.}(2015){Nagy}, {Ossenkopf}, {Van der Tak}, {Faure}, {Makai}, \& {Bergin}}]{Nagy2015}
{Nagy}, Z., {Ossenkopf}, V., {Van der Tak}, F.~F.~S., {et~al.} 2015, \aap, 578, A124

\bibitem[{{Nishimura} {et~al.}(2024){Nishimura}, {Aalto}, {Gorski}, {K{\"o}nig}, {Onishi}, {Wethers}, {Yang}, {Barcos-Mu{\~n}oz}, {Combes}, {D{\'\i}az-Santos}, {Gallagher}, {Garc{\'\i}a-Burillo}, {Gonz{\'a}lez-Alfonso}, {Greve}, {Harada}, {Henkel}, {Imanishi}, {Kohno}, {Linden}, {Mangum}, {Mart{\'\i}n}, {Muller}, {Privon}, {Ricci}, {Stanley}, {van der Werf}, \& {Viti}}]{Nishimura2024}
{Nishimura}, Y., {Aalto}, S., {Gorski}, M.~D., {et~al.} 2024, \aap, 686, A48

\bibitem[{{Norris}(1988)}]{Norris1988}
{Norris}, R.~P. 1988, \mnras, 230, 345

\bibitem[{{Onishi} {et~al.}(2024){Onishi}, {Nakagawa}, {Baba}, {Matsumoto}, {Isobe}, {Shirahata}, {Terada}, {Usuda}, \& {Oyabu}}]{Onishi2024}
{Onishi}, S., {Nakagawa}, T., {Baba}, S., {et~al.} 2024, \apj, 976, 106

\bibitem[{{Ott} {et~al.}(2011){Ott}, {Henkel}, {Braatz}, \& {Wei{\ss}}}]{Ott2011}
{Ott}, J., {Henkel}, C., {Braatz}, J.~A., \& {Wei{\ss}}, A. 2011, \apj, 742, 95

\bibitem[{{Pereira-Santaella} {et~al.}(2024{\natexlab{a}}){Pereira-Santaella}, {Gonz{\'a}lez-Alfonso}, {Garc{\'\i}a-Bernete}, {Donnan}, {Santa-Maria}, {Goicoechea}, {Lamperti}, {Perna}, \& {Rigopoulou}}]{Pereira-Santaella2024b}
{Pereira-Santaella}, M., {Gonz{\'a}lez-Alfonso}, E., {Garc{\'\i}a-Bernete}, I., {et~al.} 2024{\natexlab{a}}, \aap, 689, L12

\bibitem[{{Pereira-Santaella} {et~al.}(2024{\natexlab{b}}){Pereira-Santaella}, {Gonz{\'a}lez-Alfonso}, {Garc{\'\i}a-Bernete}, {Garc{\'\i}a-Burillo}, \& {Rigopoulou}}]{Pereira-Santaella2024}
{Pereira-Santaella}, M., {Gonz{\'a}lez-Alfonso}, E., {Garc{\'\i}a-Bernete}, I., {Garc{\'\i}a-Burillo}, S., \& {Rigopoulou}, D. 2024{\natexlab{b}}, \aap, 681, A117

\bibitem[{{Perna} {et~al.}(2020){Perna}, {Arribas}, {Catal{\'a}n-Torrecilla}, {Colina}, {Bellocchi}, {Fluetsch}, {Maiolino}, {Cazzoli}, {Hern{\'a}n Caballero}, {Pereira Santaella}, {Piqueras L{\'o}pez}, \& {Rodr{\'\i}guez del Pino}}]{Perna2020}
{Perna}, M., {Arribas}, S., {Catal{\'a}n-Torrecilla}, C., {et~al.} 2020, \aap, 643, A139

\bibitem[{{Perna} {et~al.}(2024){Perna}, {Arribas}, {Lamperti}, {Pereira-Santaella}, {Ulivi}, {B{\"o}ker}, {Maiolino}, {Bunker}, {Charlot}, {Cresci}, {Rodr{\'\i}guez Del Pino}, {D'Eugenio}, {{\"U}bler}, {Fahrion}, \& {Ceci}}]{Perna2024}
{Perna}, M., {Arribas}, S., {Lamperti}, I., {et~al.} 2024, \aap, 690, A171

\bibitem[{{Planck Collaboration} {et~al.}(2020){Planck Collaboration}, {Aghanim}, {Akrami}, {Ashdown}, {Aumont}, {Baccigalupi}, {Ballardini}, {Banday}, {Barreiro}, {Bartolo}, {Basak}, {Battye}, {Benabed}, {Bernard}, {Bersanelli}, {Bielewicz}, {Bock}, {Bond}, {Borrill}, {Bouchet}, {Boulanger}, {Bucher}, {Burigana}, {Butler}, {Calabrese}, {Cardoso}, {Carron}, {Challinor}, {Chiang}, {Chluba}, {Colombo}, {Combet}, {Contreras}, {Crill}, {Cuttaia}, {de Bernardis}, {de Zotti}, {Delabrouille}, {Delouis}, {Di Valentino}, {Diego}, {Dor{\'e}}, {Douspis}, {Ducout}, {Dupac}, {Dusini}, {Efstathiou}, {Elsner}, {En{\ss}lin}, {Eriksen}, {Fantaye}, {Farhang}, {Fergusson}, {Fernandez-Cobos}, {Finelli}, {Forastieri}, {Frailis}, {Fraisse}, {Franceschi}, {Frolov}, {Galeotta}, {Galli}, {Ganga}, {G{\'e}nova-Santos}, {Gerbino}, {Ghosh}, {Gonz{\'a}lez-Nuevo}, {G{\'o}rski}, {Gratton}, {Gruppuso}, {Gudmundsson}, {Hamann}, {Handley}, {Hansen}, {Herranz}, {Hildebrandt}, {Hivon}, {Huang}, {Jaffe}, {Jones}, {Karakci}, {Keih{\"a}nen},
  {Keskitalo}, {Kiiveri}, {Kim}, {Kisner}, {Knox}, {Krachmalnicoff}, {Kunz}, {Kurki-Suonio}, {Lagache}, {Lamarre}, {Lasenby}, {Lattanzi}, {Lawrence}, {Le Jeune}, {Lemos}, {Lesgourgues}, {Levrier}, {Lewis}, {Liguori}, {Lilje}, {Lilley}, {Lindholm}, {L{\'o}pez-Caniego}, {Lubin}, {Ma}, {Mac{\'\i}as-P{\'e}rez}, {Maggio}, {Maino}, {Mandolesi}, {Mangilli}, {Marcos-Caballero}, {Maris}, {Martin}, {Martinelli}, {Mart{\'\i}nez-Gonz{\'a}lez}, {Matarrese}, {Mauri}, {McEwen}, {Meinhold}, {Melchiorri}, {Mennella}, {Migliaccio}, {Millea}, {Mitra}, {Miville-Desch{\^e}nes}, {Molinari}, {Montier}, {Morgante}, {Moss}, {Natoli}, {N{\o}rgaard-Nielsen}, {Pagano}, {Paoletti}, {Partridge}, {Patanchon}, {Peiris}, {Perrotta}, {Pettorino}, {Piacentini}, {Polastri}, {Polenta}, {Puget}, {Rachen}, {Reinecke}, {Remazeilles}, {Renzi}, {Rocha}, {Rosset}, {Roudier}, {Rubi{\~n}o-Mart{\'\i}n}, {Ruiz-Granados}, {Salvati}, {Sandri}, {Savelainen}, {Scott}, {Shellard}, {Sirignano}, {Sirri}, {Spencer}, {Sunyaev}, {Suur-Uski}, {Tauber}, {Tavagnacco},
  {Tenti}, {Toffolatti}, {Tomasi}, {Trombetti}, {Valenziano}, {Valiviita}, {Van Tent}, {Vibert}, {Vielva}, {Villa}, {Vittorio}, {Wandelt}, {Wehus}, {White}, {White}, {Zacchei}, \& {Zonca}}]{planck2020}
{Planck Collaboration}, {Aghanim}, N., {Akrami}, Y., {et~al.} 2020, \aap, 641, A6

\bibitem[{{Rangwala} {et~al.}(2018){Rangwala}, {Colgan}, {Le Gal}, {Acharyya}, {Huang}, {Lee}, {Herbst}, {deWitt}, {Richter}, {Boogert}, \& {McKelvey}}]{Rangwala2018}
{Rangwala}, N., {Colgan}, S. W.~J., {Le Gal}, R., {et~al.} 2018, \apj, 856, 9

\bibitem[{{Rangwala} {et~al.}(2011){Rangwala}, {Maloney}, {Glenn}, {Wilson}, {Rykala}, {Isaak}, {Baes}, {Bendo}, {Boselli}, {Bradford}, {Clements}, {Cooray}, {Fulton}, {Imhof}, {Kamenetzky}, {Madden}, {Mentuch}, {Sacchi}, {Sauvage}, {Schirm}, {Smith}, {Spinoglio}, \& {Wolfire}}]{Rangwala2011}
{Rangwala}, N., {Maloney}, P.~R., {Glenn}, J., {et~al.} 2011, \apj, 743, 94

\bibitem[{{Rho} {et~al.}(2024){Rho}, {Park}, {Arendt}, {Matsuura}, {Milisavljevic}, {Temim}, {De Looze}, {Blair}, {Rest}, {Fox}, {Ravi}, {Koo}, {Barlow}, {Burrows}, {Chevalier}, {Clayton}, {Fesen}, {Fransson}, {Fryer}, {Gomez}, {Janka}, {Kirchschlager}, {Laming}, {Orlando}, {Patnaude}, {Pavlov}, {Plucinsky}, {Posselt}, {Priestley}, {Raymond}, {Sartorio}, {Schmidt}, {Slane}, {Smith}, {Sravan}, {Vink}, {Weil}, {Wheeler}, \& {Yoon}}]{Rho2024}
{Rho}, J., {Park}, S.~H., {Arendt}, R., {et~al.} 2024, \apjl, 969, L9

\bibitem[{{Robitaille} \& {Bressert}(2012)}]{aplpy}
{Robitaille}, T. \& {Bressert}, E. 2012, {APLpy: Astronomical Plotting Library in Python}, Astrophysics Source Code Library, record ascl:1208.017

\bibitem[{{Sakamoto} {et~al.}(2017){Sakamoto}, {Aalto}, {Barcos-Mu{\~n}oz}, {Costagliola}, {Evans}, {Harada}, {Mart{\'\i}n}, {Wiedner}, \& {Wilner}}]{Sakamoto2017}
{Sakamoto}, K., {Aalto}, S., {Barcos-Mu{\~n}oz}, L., {et~al.} 2017, \apj, 849, 14

\bibitem[{{Sakamoto} {et~al.}(2010){Sakamoto}, {Aalto}, {Evans}, {Wiedner}, \& {Wilner}}]{Sakamoto2010}
{Sakamoto}, K., {Aalto}, S., {Evans}, A.~S., {Wiedner}, M.~C., \& {Wilner}, D.~J. 2010, \apjl, 725, L228

\bibitem[{{Sakamoto} {et~al.}(2009){Sakamoto}, {Aalto}, {Wilner}, {Black}, {Conway}, {Costagliola}, {Peck}, {Spaans}, {Wang}, \& {Wiedner}}]{Sakamoto2009}
{Sakamoto}, K., {Aalto}, S., {Wilner}, D.~J., {et~al.} 2009, \apjl, 700, L104

\bibitem[{{Sakamoto} {et~al.}(2021{\natexlab{a}}){Sakamoto}, {Gonz{\'a}lez-Alfonso}, {Mart{\'\i}n}, {Wilner}, {Aalto}, {Evans}, \& {Harada}}]{Sakamoto2021a}
{Sakamoto}, K., {Gonz{\'a}lez-Alfonso}, E., {Mart{\'\i}n}, S., {et~al.} 2021{\natexlab{a}}, \apj, 923, 206

\bibitem[{{Sakamoto} {et~al.}(2021{\natexlab{b}}){Sakamoto}, {Mart{\'\i}n}, {Wilner}, {Aalto}, {Evans}, \& {Harada}}]{Sakamoto2021b}
{Sakamoto}, K., {Mart{\'\i}n}, S., {Wilner}, D.~J., {et~al.} 2021{\natexlab{b}}, \apj, 923, 240

\bibitem[{{Sakamoto} {et~al.}(1999){Sakamoto}, {Scoville}, {Yun}, {Crosas}, {Genzel}, \& {Tacconi}}]{Sakamoto1999}
{Sakamoto}, K., {Scoville}, N.~Z., {Yun}, M.~S., {et~al.} 1999, \apj, 514, 68

\bibitem[{{Salter} {et~al.}(2008){Salter}, {Ghosh}, {Catinella}, {Lebron}, {Lerner}, {Minchin}, \& {Momjian}}]{Salter2008}
{Salter}, C.~J., {Ghosh}, T., {Catinella}, B., {et~al.} 2008, \aj, 136, 389

\bibitem[{{Sanders} {et~al.}(1988){Sanders}, {Soifer}, {Elias}, {Madore}, {Matthews}, {Neugebauer}, \& {Scoville}}]{Sanders1988}
{Sanders}, D.~B., {Soifer}, B.~T., {Elias}, J.~H., {et~al.} 1988, \apj, 325, 74

\bibitem[{{Scoville} {et~al.}(2017){Scoville}, {Murchikova}, {Walter}, {Vlahakis}, {Koda}, {Vanden Bout}, {Barnes}, {Hernquist}, {Sheth}, {Yun}, {Sanders}, {Armus}, {Cox}, {Thompson}, {Robertson}, {Zschaechner}, {Tacconi}, {Torrey}, {Hayward}, {Genzel}, {Hopkins}, {van der Werf}, \& {Decarli}}]{Scoville2017}
{Scoville}, N., {Murchikova}, L., {Walter}, F., {et~al.} 2017, \apj, 836, 66

\bibitem[{{Scoville} {et~al.}(2015){Scoville}, {Sheth}, {Walter}, {Manohar}, {Zschaechner}, {Yun}, {Koda}, {Sanders}, {Murchikova}, {Thompson}, {Robertson}, {Genzel}, {Hernquist}, {Tacconi}, {Brown}, {Narayanan}, {Hayward}, {Barnes}, {Kartaltepe}, {Davies}, {van der Werf}, \& {Fomalont}}]{Scoville2015}
{Scoville}, N., {Sheth}, K., {Walter}, F., {et~al.} 2015, \apj, 800, 70

\bibitem[{{Scoville} {et~al.}(1998){Scoville}, {Evans}, {Dinshaw}, {Thompson}, {Rieke}, {Schneider}, {Low}, {Hines}, {Stobie}, {Becklin}, \& {Epps}}]{Scoville1998}
{Scoville}, N.~Z., {Evans}, A.~S., {Dinshaw}, N., {et~al.} 1998, \apjl, 492, L107

\bibitem[{{Scoville} {et~al.}(1986){Scoville}, {Sanders}, {Sargent}, {Soifer}, {Scott}, \& {Lo}}]{Scoville1986}
{Scoville}, N.~Z., {Sanders}, D.~B., {Sargent}, A.~I., {et~al.} 1986, \apjl, 311, L47

\bibitem[{{Smith} {et~al.}(1998){Smith}, {Lonsdale}, {Lonsdale}, \& {Diamond}}]{Smith1998}
{Smith}, H.~E., {Lonsdale}, C.~J., {Lonsdale}, C.~J., \& {Diamond}, P.~J. 1998, \apjl, 493, L17

\bibitem[{{Smith} {et~al.}(2007){Smith}, {Draine}, {Dale}, {Moustakas}, {Kennicutt}, {Helou}, {Armus}, {Roussel}, {Sheth}, {Bendo}, {Buckalew}, {Calzetti}, {Engelbracht}, {Gordon}, {Hollenbach}, {Li}, {Malhotra}, {Murphy}, \& {Walter}}]{Smith2007}
{Smith}, J.~D.~T., {Draine}, B.~T., {Dale}, D.~A., {et~al.} 2007, \apj, 656, 770

\bibitem[{{Soifer} {et~al.}(1984){Soifer}, {Helou}, {Lonsdale}, {Neugebauer}, {Hacking}, {Houck}, {Low}, {Rice}, \& {Rowan-Robinson}}]{Soifer1984}
{Soifer}, B.~T., {Helou}, G., {Lonsdale}, C.~J., {et~al.} 1984, \apjl, 283, L1

\bibitem[{{Soifer} {et~al.}(1999){Soifer}, {Neugebauer}, {Matthews}, {Becklin}, {Ressler}, {Werner}, {Weinberger}, \& {Egami}}]{Soifer1999}
{Soifer}, B.~T., {Neugebauer}, G., {Matthews}, K., {et~al.} 1999, \apj, 513, 207

\bibitem[{{Sonnentrucker} {et~al.}(2007){Sonnentrucker}, {Gonz{\'a}lez-Alfonso}, \& {Neufeld}}]{Sonnentrucker2007}
{Sonnentrucker}, P., {Gonz{\'a}lez-Alfonso}, E., \& {Neufeld}, D.~A. 2007, \apjl, 671, L37

\bibitem[{{Spoon} {et~al.}(2004){Spoon}, {Moorwood}, {Lutz}, {Tielens}, {Siebenmorgen}, \& {Keane}}]{Spoon2004}
{Spoon}, H.~W.~W., {Moorwood}, A.~F.~M., {Lutz}, D., {et~al.} 2004, \aap, 414, 873

\bibitem[{{Takano} {et~al.}(2005){Takano}, {Nakanishi}, {Nakai}, \& {Takano}}]{Takano2005}
{Takano}, S., {Nakanishi}, K., {Nakai}, N., \& {Takano}, T. 2005, \pasj, 57, L29

\bibitem[{{Tanaka} {et~al.}(1986){Tanaka}, {Kawaguchi}, \& {Hirota}}]{Tanaka1986}
{Tanaka}, K., {Kawaguchi}, K., \& {Hirota}, E. 1986, Journal of Molecular Spectroscopy, 117, 408

\bibitem[{{Teng} {et~al.}(2015){Teng}, {Rigby}, {Stern}, {Ptak}, {Alexander}, {Bauer}, {Boggs}, {Brandt}, {Christensen}, {Comastri}, {Craig}, {Farrah}, {Gandhi}, {Hailey}, {Harrison}, {Hickox}, {Koss}, {Luo}, {Treister}, \& {Zhang}}]{Teng2015}
{Teng}, S.~H., {Rigby}, J.~R., {Stern}, D., {et~al.} 2015, \apj, 814, 56

\bibitem[{{Tunnard} {et~al.}(2015){Tunnard}, {Greve}, {Garcia-Burillo}, {Graci{\'a} Carpio}, {Fischer}, {Fuente}, {Gonz{\'a}lez-Alfonso}, {Hailey-Dunsheath}, {Neri}, {Sturm}, {Usero}, \& {Planesas}}]{Tunnard2015}
{Tunnard}, R., {Greve}, T.~R., {Garcia-Burillo}, S., {et~al.} 2015, \apj, 800, 25

\bibitem[{{Ulivi} {et~al.}(2025){Ulivi}, {Perna}, {Lamperti}, {Arribas}, {Cresci}, {Marconcini}, {Rodr{\'\i}guez Del Pino}, {B{\"o}ker}, {Bunker}, {Ceci}, {Charlot}, {D'Eugenio}, {Fahrion}, {Maiolino}, {Marconi}, \& {Pereira-Santaella}}]{Ulivi2024}
{Ulivi}, L., {Perna}, M., {Lamperti}, I., {et~al.} 2025, \aap, 693, A36

\bibitem[{{van der Tak} {et~al.}(1999){van der Tak}, {van Dishoeck}, {Evans}, {Bakker}, \& {Blake}}]{VanDerTak1999}
{van der Tak}, F. F.~S., {van Dishoeck}, E.~F., {Evans}, II, N.~J., {Bakker}, E.~J., \& {Blake}, G.~A. 1999, \apj, 522, 991

\bibitem[{{van Dishoeck} {et~al.}(2023){van Dishoeck}, {Grant}, {Tabone}, {van Gelder}, {Francis}, {Tychoniec}, {Bettoni}, {Arabhavi}, {Gasman}, {Nazari}, {Vlasblom}, {Kavanagh}, {Christiaens}, {Klaassen}, {Beuther}, {Henning}, \& {Kamp}}]{vanDishoeck2023}
{van Dishoeck}, E.~F., {Grant}, S., {Tabone}, B., {et~al.} 2023, Faraday Discussions, 245, 52

\bibitem[{{van Gelder} {et~al.}(2024){van Gelder}, {Francis}, {van Dishoeck}, {Tychoniec}, {Ray}, {Beuther}, {Caratti o Garatti}, {Chen}, {Devaraj}, {Gieser}, {Justtanont}, {Kavanagh}, {Nazari}, {Reyes}, {Rocha}, {Slavicinska}, {G{\"u}del}, {Henning}, {Lagage}, \& {Wright}}]{VanGelder2024}
{van Gelder}, M.~L., {Francis}, L., {van Dishoeck}, E.~F., {et~al.} 2024, \aap, 692, A197

\bibitem[{{Varenius} {et~al.}(2019){Varenius}, {Conway}, {Batejat}, {Mart{\'\i}-Vidal}, {P{\'e}rez-Torres}, {Aalto}, {Alberdi}, {Lonsdale}, \& {Diamond}}]{Varenius2019}
{Varenius}, E., {Conway}, J.~E., {Batejat}, F., {et~al.} 2019, \aap, 623, A173

\bibitem[{{Virtanen} {et~al.}(2020){Virtanen}, {Gommers}, {Oliphant}, {Haberland}, {Reddy}, {Cournapeau}, {Burovski}, {Peterson}, {Weckesser}, {Bright}, {van der Walt}, {Brett}, {Wilson}, {Jarrod Millman}, {Mayorov}, {Nelson}, {Jones}, {Kern}, {Larson}, {Carey}, {Polat}, {Feng}, {Moore}, {Vand erPlas}, {Laxalde}, {Perktold}, {Cimrman}, {Henriksen}, {Quintero}, {Harris}, {Archibald}, {Ribeiro}, {Pedregosa}, {van Mulbregt}, \& {Contributors}}]{Virtanen_2020}
{Virtanen}, P., {Gommers}, R., {Oliphant}, T.~E., {et~al.} 2020, Nature Methods, 17, 261

\bibitem[{{Wheeler} {et~al.}(2020){Wheeler}, {Glenn}, {Rangwala}, \& {Fyhrie}}]{Wheeler2020}
{Wheeler}, J., {Glenn}, J., {Rangwala}, N., \& {Fyhrie}, A. 2020, \apj, 896, 43

\bibitem[{{Wright} {et~al.}(2023){Wright}, {Rieke}, {Glasse}, {Ressler}, {Garc{\'\i}a Mar{\'\i}n}, {Aguilar}, {Alberts}, {{\'A}lvarez-M{\'a}rquez}, {Argyriou}, {Banks}, {Baudoz}, {Boccaletti}, {Bouchet}, {Bouwman}, {Brandl}, {Breda}, {Bright}, {Cale}, {Colina}, {Cossou}, {Coulais}, {Cracraft}, {De Meester}, {Dicken}, {Engesser}, {Etxaluze}, {Fox}, {Friedman}, {Fu}, {Gasman}, {G{\'a}sp{\'a}r}, {Gastaud}, {Geers}, {Glauser}, {Gordon}, {Greene}, {Greve}, {Grundy}, {G{\"u}del}, {Guillard}, {Haderlein}, {Hashimoto}, {Henning}, {Hines}, {Holler}, {Detre}, {Jahromi}, {James}, {Jones}, {Justtanont}, {Kavanagh}, {Kendrew}, {Klaassen}, {Krause}, {Labiano}, {Lagage}, {Lambros}, {Larson}, {Law}, {Lee}, {Libralato}, {Lorenzo Alverez}, {Meixner}, {Morrison}, {Mueller}, {Murray}, {Mycroft}, {Myers}, {Nayak}, {Naylor}, {Nickson}, {Noriega-Crespo}, {{\"O}stlin}, {O'Sullivan}, {Ottens}, {Patapis}, {Penanen}, {Pietraszkiewicz}, {Ray}, {Regan}, {Roteliuk}, {Royer}, {Samara-Ratna}, {Samuelson}, {Sargent}, {Scheithauer},
  {Schneider}, {Schreiber}, {Shaughnessy}, {Sheehan}, {Shivaei}, {Sloan}, {Tamas}, {Teague}, {Temim}, {Tikkanen}, {Tustain}, {van Dishoeck}, {Vandenbussche}, {Weilert}, {Whitehouse}, \& {Wolff}}]{Wright2023}
{Wright}, G.~S., {Rieke}, G.~H., {Glasse}, A., {et~al.} 2023, \pasp, 135, 048003

\end{thebibliography}

\end{document}